\documentclass[11pt%,fleqn
]{article}
\usepackage{geometry,epsf,amsmath,amsfonts,cite,microtype,mathtools,pdflscape,hyperref,dsfont}
\hypersetup{colorlinks=true,linkcolor=blue,urlcolor=blue,citecolor=blue}

\usepackage{tikz}
\usetikzlibrary{arrows,calc,patterns,decorations.markings,decorations.pathreplacing,plotmarks,shapes.arrows,decorations.pathmorphing,backgrounds}
\tikzset{snake it/.style={decorate, decoration=snake}}

\usepackage{bbm}
\usepackage{parskip}
\usepackage{amssymb}
\usepackage{braket}
\usepackage{url}
\usepackage[all]{xy}
\usepackage{subfig}
\usepackage[normalem]{ulem}
\usepackage{rotating}
\usepackage{rotfloat}
\usepackage{array}
\numberwithin{equation}{section}
\numberwithin{table}{section}
\newcommand{\pb}[1]{\phi_{#1}}
\newcommand{\psib}{{\bar\psi}}

\newcommand{\sigmaB}{{\sigma^{}_{\!B}}}
\newcommand{\muB}{{\mu^{}_{\!B}}}

\newcommand{\nn}{\nonumber}

\def\SVIR2{{SVIR}$_3{}^{\otimes 2}$}

\newcommand{\kett}[1]{{| #1 \rangle\!\rangle }}

\newcommand{\blank}[1]{}%

\newcommand{\be}{\begin{equation}}
\newcommand{\ee}{\end{equation}}

\newcommand{\bc}{\begin{cases}}
\newcommand{\ec}{\end{cases}}

\newcommand{\eps}{{\epsilon}}

\newcommand{\fxed}{{\text{fixed}}}
\newcommand{\fre}{{\text{free}}}

\newcommand{\cB}{{\cal B}}
\newcommand{\cD}{{\cal D}}

\newcommand{\cH}{{\cal H}}
\newcommand{\cI}{{\cal I}}

\newcommand{\FM}{{\ensuremath{\mathit{FM}}}}
\newcommand{\wtFM}{{\ensuremath{\widetilde{\mathit{FM}}}}}

\newcommand{\WB}{{\ensuremath{\mathit{WB}}}}

\newcommand{\SC}{\hat C}
\newcommand{\SCvar}{\check C}

\begin{document}
\begin{flushright}  {~} \\[-12mm]
{\sf KCL-MTH-20-01}
\end{flushright} 

\thispagestyle{empty}

\begin{center} \vskip 24mm
{\Large\bf Fermionic CFTs and classifying algebras}\\[10mm] 
{\large 
Ingo Runkel\,${}^{1,2}$
and
G\'erard M.\ T.\ Watts\,${}^{2}$
}
\\[5mm]
${}^1$ Fachbereich Mathematik, Universit\"at Hamburg,\\
Bundesstra\ss e 55, 20146 Hamburg, Germany
\\[5mm]
${}^2$ Department of Mathematics, King's College London,\\
Strand, London WC2R\;2LS, UK
\\[5mm]

\vskip 4mm
\end{center}

\begin{quote}{\bf Abstract}\\[1mm]
We study fermionic conformal field theories on surfaces with spin
structure in the presence of boundaries, defects, and interfaces. We
obtain the relevant crossing relations, taking particular care with
parity signs and signs arising from the change of spin structure in
different limits. 
We define fermionic classifying algebras for boundaries, defects, and
interfaces, which allow one to read off the elementary boundary
conditions, etc.  

As examples, we define fermionic extensions of Virasoro minimal models
and give explicit solutions for the spectrum and bulk structure
constants. We show how the $A$- and $D$-type fermionic Virasoro
minimal models are related by a parity-shift operation which we
define in general.
We study the boundaries, defects, and interfaces in several
examples, in particular in the fermionic Ising model, i.e.\ the free
fermion, in the fermionic tri-critical Ising model, i.e.\ the
first unitary
$N=1$ superconformal minimal model, and in the supersymmetric Lee-Yang model, of which there are two distinct versions that are related by parity-shift.

\end{quote}

\newpage

\setcounter{tocdepth}{2}
\tableofcontents

\newpage

\section{Introduction}
\label{sec:intro}

In this paper we set out a way to define conformal field theories
with fermions and analyse their conformal boundary conditions, defects and
interfaces. 
Conformal field theories with fermions have been studied for a very
long time, as have their boundary conditions, but in this paper we
take an algebraic approach to the description of fermionic theories
and spin structures. This makes an algebraic analysis of boundary
conditions and defects tractable and has revealed new relations
between models that had not been understood before.

The first objective is to define correlation functions of fields in
fermionic conformal field theories unambiguously on surfaces with spin
structures; we do this using defects as proposed in
\cite{Novak:2015ela}. In this paper we outline this method and state
our results; we will give more details and proofs in \cite{RSW-prep}. 

Our description of fermionic CFTs has two immediate
implications. Firstly, the state space is a super-vector space, 
divided by the grading into fields of even and odd parity.	
Secondly, we need
to consider the different spin structures separately as each spin
structure defines a different way to put fermions on a surface
consistently, so that each circular
boundary, defect, and interface
will have a separate
description for the two possible spin structures in its neighbourhood.
We obtain
a consistent set of sewing constraints for theories
including fermions, which incorporates the signs that arise from
re-ordering of products of fermionic fields and -- importantly --
extra signs that
arise from putting the spin structure back in a standard form in
different limits. 
We describe this in detail in section \ref{sec:fermbulk}.
One particularly interesting result is that given a fermionic CFT, it
is possible to define another fermionic
CFT in which the parities (odd/even) are
swapped in the Ramond sector. This can either result in the original
CFT again or in a new CFT. 

Having defined the bulk theories, we are able to consider their
conformal boundary conditions, defects and interfaces. 
We show in section \ref{sec:ca} how the bulk-boundary, bulk-defect and
bulk-interface 
structure constants define
super-algebras, which we refer to as fermionic classifying algebras, and
which then
allow one to identify the fundamental boundary conditions, etc.,
algebraically from these
algebras alone. 
For purely bosonic theories the corresponding classifying algebras
were introduced in \cite{acafbc,FSS07}.  
One consequence is the natural occurrence of fermionic weight zero
fields on boundaries, defects and interfaces, which are required for
their consistent description.  
These have been known for a long time, 
for example they have been used in the coupling of the Ising model to
boundary magnetic fields \cite{GZ,Chatterjee:1993ca}, see also~\cite{Toth:2006tj,Konechny:2018ujl} for more recent applications to boundary renormalisation group flows.
Our analysis shows they are a necessary and integral part of the description of fermionic
CFTs with boundaries, defects and interfaces.

We illustrate these ideas in the concrete cases of fermionic
extensions of Virasoro minimal models which we define in section
\ref{sec:fermVir}, and for which we give an explicit solution for the bulk structure constants.
As is well known, for some values of the central charge there are
two or more different Virasoro minimal models labelled by pairs of Lie
algebras~\cite{Cappelli:1987xt}.
We show that, remarkably, the fermionic extensions of the $(A_m,A_{4n-1})$ and $(A_m,D_{2n+1})$ bosonic minimal models -- and in particular their bulk
structure constants \cite{runkel,Runkel:1999dz}-- are related by the parity-shift operation.

The fermionic extensions of the Virasoro minimal models 
include such important examples as the fermionic Ising model, i.e.\ the free fermion, and
the fermionic tri-critical Ising model, i.e.\ the first unitary
$N=1$ superconformal minimal model, as well as many  other
theories with extended symmetries. Sections~\ref{sec:Ising} and~\ref{sec:furtherVir}
of the paper are taken up with
exploring our results in these situations and comparing our findings
with the discussions already in the literature, in particular \cite{Nepomechie2001,Nepomechie2002,mw}.

\section{Bulk fields in fermionic CFT}
\label{sec:fermbulk}

In this section we explain how to describe fields of a fermionic CFT.
In order to include the effects of the spin structure and the
bosonic/fermionic nature of fields, we introduce a special type of
topological line defect; {\em all} bulk fields are then connected to one of
these defects, so that we think of them as disorder fields which sit
at the starting point of the specific topological defect. 

We start by describing the relevant properties of the topological defect and then use these to define OPE coefficients and to obtain the crossing constraint they have to satisfy. We show that, given one solution to the crossing constraint, one can obtain another solution by shifting the Ramond sector parity and modifying the given solution by signs.

\subsection{The topological defect $F$}\label{sec:topdefF}

The spin structure on the worldsheet of a fermionic CFT is encoded
by a topological defect which we call $F$. 
The technical details of this procedure are given in \cite{Novak:2014oca,Novak:2015ela}. Here we do not need the full formalism and just state the properties we will use below.

\begin{figure}[p]

a) \begin{tikzpicture}[baseline=2em]
\coordinate (v1) at (0,0);
\coordinate (v2) at (-2,0);

\begin{scope}[very thick,blue!80!black,decoration={markings,mark=at position 0.6 with {\arrow{>}}}]
\draw[postaction={decorate}] (v1) -- node[above,black] {\small $F$} (v2);
\draw[dashed] (v2) -- ++(-0.7,0);
\end{scope}
\draw[very thick,black,fill=black] (v1) circle (0.060);

\node[above] at (v1) {\small $\phi$};
\end{tikzpicture}
\hspace{4.5em}
b)\raisebox{-2em}{
	\begin{tikzpicture}
	\coordinate (v1) at (0,0);
	\coordinate (v2) at (-1.5,0);
	\coordinate (v3) at (-2.5,0);
	
	\begin{scope}[very thick,blue!80!black,decoration={markings,mark=at position 0.6 with {\arrow{>}}}]
	\draw[postaction={decorate}] (v1) -- (v2);
	\draw[postaction={decorate}] (v2) -- (v3);
	\draw[dashed] (v3) -- ++(-0.7,0);
	\end{scope}
	
	\draw[very thick,black,fill=black] (v1) circle (0.060);
	\draw[very thick,blue!80!black,fill=blue!80!black] (v2) circle (0.060);
	
	\node[above] at (v1) {\small $\phi$};
	\node[above] at (v2) {\small $\pi$};
	\end{tikzpicture}
	~=~~
	\begin{tikzpicture}
	\coordinate (v1) at (0,0);
	\coordinate (v2) at (-1.5,0);
	
	\begin{scope}[very thick,blue!80!black,decoration={markings,mark=at position 0.6 with {\arrow{>}}}]
	\draw[postaction={decorate}] (v1) -- (v2);
	\draw[dashed] (v2) -- ++(-0.7,0);
	\end{scope}
	\draw[very thick,black,fill=black] (v1) circle (0.060);
	
	\node[above] at (v1) {\small $(-1)^{\pb{}}\phi$};
	\end{tikzpicture}
}
\\[2em]
c)
\begin{tikzpicture}[baseline=3em]
\coordinate (v1) at (-1,0);
\coordinate (v2) at (-2.5,0);
\coordinate (v3) at (0,0.6);
\coordinate (v4) at (0,-0.6);

\begin{scope}[very thick,blue!80!black,decoration={markings,mark=at position 0.6 with {\arrow{>}}}]
\draw[postaction={decorate}] (v1) -- (v2);
\draw[postaction={decorate}] (v3) -- (v1);
\draw[postaction={decorate}] (v4) -- (v1);
\draw[dashed] (v2) -- ++(-0.7,0);
\draw[dashed] (v3) -- ++(+0.5,0.3);
\draw[dashed] (v4) -- ++(+0.5,-0.3);
\end{scope}
\draw[very thick,blue!80!black,fill=blue!80!black] (v1) circle (0.060);
\end{tikzpicture}
\hspace{3em}
d)\raisebox{-3em}{
	\begin{tikzpicture}[baseline=-1]
	\coordinate (v1) at (-1,0);
	\coordinate (v2) at (-2.5,0);
	\coordinate (v3) at (0,0.6);
	\coordinate (v4) at (0,-0.6);
	\coordinate (pi1) at (-2.0,0);
	\coordinate (pi2) at (-0.4,{(1-0.4)*0.6});
	\coordinate (pi3) at (-0.4,{-(1-0.4)*0.6});
	
	\begin{scope}[very thick,blue!80!black,decoration={markings,mark=at position 0.4 with {\arrow{>}}}]
	\draw[postaction={decorate}] (v1) -- (v2);
	\draw[postaction={decorate}] (v3) -- (v1);
	\draw[postaction={decorate}] (v4) -- (v1);
	\draw[dashed] (v2) -- ++(-0.7,0);
	\draw[dashed] (v3) -- ++(+0.5,0.3);
	\draw[dashed] (v4) -- ++(+0.5,-0.3);
	\end{scope}
	\draw[very thick,blue!80!black,fill=blue!80!black] (v1) circle (0.060);
	\draw[very thick,blue!80!black,fill=blue!80!black] (pi1) circle (0.060);
	\node[above] at (pi1) {\small $\pi$};
	\end{tikzpicture}
	=~~
	\begin{tikzpicture}[baseline=-1]
	\coordinate (v1) at (-1,0);
	\coordinate (v2) at (-2.5,0);
	\coordinate (v3) at (0,0.6);
	\coordinate (v4) at (0,-0.6);
	\coordinate (pi1) at (-2.0,0);
	\coordinate (pi2) at (-0.2,{(1-0.2)*0.6});
	\coordinate (pi3) at (-0.2,{-(1-0.2)*0.6});
	
	\begin{scope}[very thick,blue!80!black,decoration={markings,mark=at position 0.7 with {\arrow{>}}}]
	\draw[postaction={decorate}] (v1) -- (v2);
	\draw[postaction={decorate}] (v3) -- (v1);
	\draw[postaction={decorate}] (v4) -- (v1);
	\draw[dashed] (v2) -- ++(-0.7,0);
	\draw[dashed] (v3) -- ++(+0.5,0.3);
	\draw[dashed] (v4) -- ++(+0.5,-0.3);
	\end{scope}
	\draw[very thick,blue!80!black,fill=blue!80!black] (v1) circle (0.060);
	\draw[very thick,blue!80!black,fill=blue!80!black] (pi2) circle (0.060);
	\draw[very thick,blue!80!black,fill=blue!80!black] (pi3) circle (0.060);
	\node[above] at (pi2) {\small $\pi$};
	\node[above] at (pi3) {\small $\pi$};
	\end{tikzpicture}
}
\\[3em]
e)
\raisebox{-4em}{
\begin{tikzpicture}[baseline=-0.2em]
\coordinate (v1) at (-1,0);
\coordinate (v2) at (-2.5,0);
\coordinate (v3) at (0,0.6);
\coordinate (v4) at (0,-0.6);
\coordinate (v5) at (+1,1.2);
\coordinate (v6) at (+1,-1.2);
\coordinate (j) at (+1,0);
\begin{scope}[very thick,blue!80!black,decoration={markings,mark=at position 0.6 with {\arrow{>}}}]
\draw[postaction={decorate}] (v1) -- (v2);
\draw[postaction={decorate}] (v3) -- (v1);
\draw (v4) -- (v1);
\draw[postaction={decorate}] (v5) -- (v3);
\draw[postaction={decorate}] (v6) -- (v4);
\draw[postaction={decorate}] (j) .. controls ++(-0.5,0) and ++(+0.5,-0.5) .. (v3);
\draw[dashed] (v2) -- ++(-0.7,0);
\draw[dashed] (j) -- ++(+0.7,0);
\draw[dashed] (v5) -- ++(+0.5,0.3);
\draw[dashed] (v6) -- ++(+0.5,-0.3);
\end{scope}
\draw[very thick,blue!80!black,fill=blue!80!black] (v1) circle (0.060);
\draw[very thick,blue!80!black,fill=blue!80!black] (v3) circle (0.060);
\end{tikzpicture}
\quad=\quad
\begin{tikzpicture}[baseline=-0.2em]
\coordinate (v1) at (-1,0);
\coordinate (v2) at (-2.5,0);
\coordinate (v3) at (0,0.6);
\coordinate (v4) at (0,-0.6);
\coordinate (v5) at (+1,1.2);
\coordinate (v6) at (+1,-1.2);
\coordinate (j) at (+1,0);
\begin{scope}[very thick,blue!80!black,decoration={markings,mark=at position 0.6 with {\arrow{>}}}]
\draw[postaction={decorate}] (v1) -- (v2);
\draw (v3) -- (v1);
\draw[postaction={decorate}] (v4) -- (v1);
\draw[postaction={decorate}] (v5) -- (v3);
\draw[postaction={decorate}] (v6) -- (v4);
\draw[postaction={decorate}] (j) .. controls ++(-0.5,0) and ++(0.5,0.5) .. (v4);
\draw[dashed] (v2) -- ++(-0.7,0);
\draw[dashed] (j) -- ++(+0.7,0);
\draw[dashed] (v5) -- ++(+0.5,0.3);
\draw[dashed] (v6) -- ++(+0.5,-0.3);
\end{scope}
\draw[very thick,blue!80!black,fill=blue!80!black] (v1) circle (0.060);
\draw[very thick,blue!80!black,fill=blue!80!black] (v4) circle (0.060);
\end{tikzpicture}
}
\\[3em]
f)\raisebox{-2em}{
	\begin{tikzpicture}[baseline=0]
	\coordinate (l) at (+2,0);
	\coordinate (r) at (-2,0);
	\coordinate (b) at (0,-1);
	\coordinate (t) at (0,1);
	\coordinate (c) at (0.5,0);
	
	\begin{scope}[very thick,blue!80!black,decoration={markings,mark=at position 0.6 with {\arrow{>}}}]
	\draw[postaction={decorate}] (l) .. controls ++(-1,0) and ++(+1,0) .. (t);
	\draw[postaction={decorate}] (t) .. controls ++(-1,0) and ++(+1,0) .. (r);
	\draw[postaction={decorate}] (c) .. controls ++(-0.5,0) and ++(0,-0.5) .. (t);
	\draw[dashed] (l) -- ++(+0.7,0);
	\draw[dashed] (r) -- ++(-0.7,0);
	\end{scope}
	\draw[very thick,blue!80!black,fill=blue!80!black] (t) circle (0.060);
	\draw[very thick,black,fill=black] (c) circle (0.060);
	\node[right] at (c) {\small $\phi$};
	\end{tikzpicture}
	~~=~~
	\begin{tikzpicture}[baseline=0]
	\coordinate (l) at (+2,0);
	\coordinate (r) at (-2,0);
	\coordinate (b) at (0,-1);
	\coordinate (t) at (0,1);
	\coordinate (c) at (+0.5,0);
	\coordinate (pi) at (+0.7,-0.8);
	
	\begin{scope}[very thick,blue!80!black,decoration={markings,mark=at position 0.4 with {\arrow{>}}}]
	\draw[postaction={decorate}] (l) .. controls ++(-1,0) and ++(+1,0) .. (b);
	\draw[postaction={decorate}] (b) .. controls ++(-1,0) and ++(+1,0) .. (r);
	\draw[postaction={decorate}] (c) .. controls ++(-0.5,0) and ++(0,0.5) .. (b);
	\draw[dashed] (l) -- ++(+0.7,0);
	\draw[dashed] (r) -- ++(-0.7,0);
	\end{scope}
	\draw[very thick,blue!80!black,fill=blue!80!black] (b) circle (0.060);
	\draw[very thick,blue!80!black,fill=blue!80!black] (pi) circle (0.060);
	\draw[very thick,black,fill=black] (c) circle (0.060);
	\node at ([shift={(0.2,-0.25)}]pi) {\small $\pi^{\nu_\phi}$};
	\node[right] at (c) {\small $\phi$};
	\end{tikzpicture}
}
\\[2em]
g)\raisebox{-2em}{$
e^{2 \pi i (L_0 - \bar L_0)} \phi 
~~=~~ 
\begin{tikzpicture}[baseline=0]
\coordinate (v1) at (0,0);
\coordinate (v2) at (0,0.7);
\coordinate (v3) at (0,-0.9);
\coordinate (v4) at (-1.5,0);
\coordinate (v5) at (-2,0);

\begin{scope}[very thick,blue!80!black,decoration={markings,mark=at position 0.4 with {\arrow{>}}}]
\draw (v1) .. controls ++(-0.5,0) and ++(-0.5,0) .. (v2);
\draw[postaction={decorate}] (v2) .. controls ++(+1,0) and ++(+1,0) .. (v3);
\draw (v3) .. controls ++(-1,0) and ++(+1,0) .. (v4);
\draw[postaction={decorate}] (v4) -- (v5);
\draw[dashed] (v5) -- ++(-0.7,0);
\end{scope}
\draw[very thick,black,fill=black] (v1) circle (0.060);
\node[right] at (v1) {\small $\phi$};
\end{tikzpicture}
~~=~~
\begin{tikzpicture}[baseline=0]
\coordinate (v1) at (0,0);
\coordinate (v2) at (-2,0);
\coordinate (pi) at (-1,0);

\begin{scope}[very thick,blue!80!black,decoration={markings,mark=at position 0.8 with {\arrow{>}}}]
\draw[postaction={decorate}] (v1) -- (v2);
\draw[dashed] (v2) -- ++(-0.7,0);
\end{scope}
\draw[very thick,black,fill=black] (v1) circle (0.060);
\draw[very thick,blue!80!black,fill=blue!80!black] (pi) circle (0.060);
\node[above] at (v1) {\small $\phi$};
\node at ([shift={(0.2,0.3)}]pi) {\small $\pi^{\nu_\phi + 1}$};
\end{tikzpicture}
~~=~~(-1)^{\pb{}(\nu_\phi + 1)}~~
\begin{tikzpicture}[baseline=0]
\coordinate (v1) at (0,0);
\coordinate (v2) at (-1,0);

\begin{scope}[very thick,blue!80!black,decoration={markings,mark=at position 0.8 with {\arrow{>}}}]
\draw[postaction={decorate}] (v1) -- (v2);
\draw[dashed] (v2) -- ++(-0.7,0);
\end{scope}
\draw[very thick,black,fill=black] (v1) circle (0.060);
\node[above] at (v1) {\small $\phi$};
\end{tikzpicture}
$
}

\caption{Properties of the topological line defect $F$. 
a)~A field $\phi \in \cH_F$ sits at the start of $F$. 
b)~The OPE of the weight zero defect field $\pi$ and a field $\phi$ multiplies $\phi$ by $\pm 1$, depending on its parity.
c)~The weight zero junction joining two $T$ defects into one.
d)~Pushing the defect field $\pi$ through the defect junction.
e)~Associativity relation for the junction field. 
f)~Dragging the $F$ defect through a field $\phi$ of spin grade $\nu_\phi$ inserts $\pi^{\nu_\phi}$. 
g)~Rotating $\phi \in \cH_F^{\nu_\phi}$ by $2\pi$ can be traded for an insertion of $\pi^{\nu_\phi+1}$. In particular, one cannot just unwind an $F$-defect around $\phi$, instead the tangent at the insertion point of $\phi$ has to remain fixed.}
\label{fig:F-properties}
\end{figure}
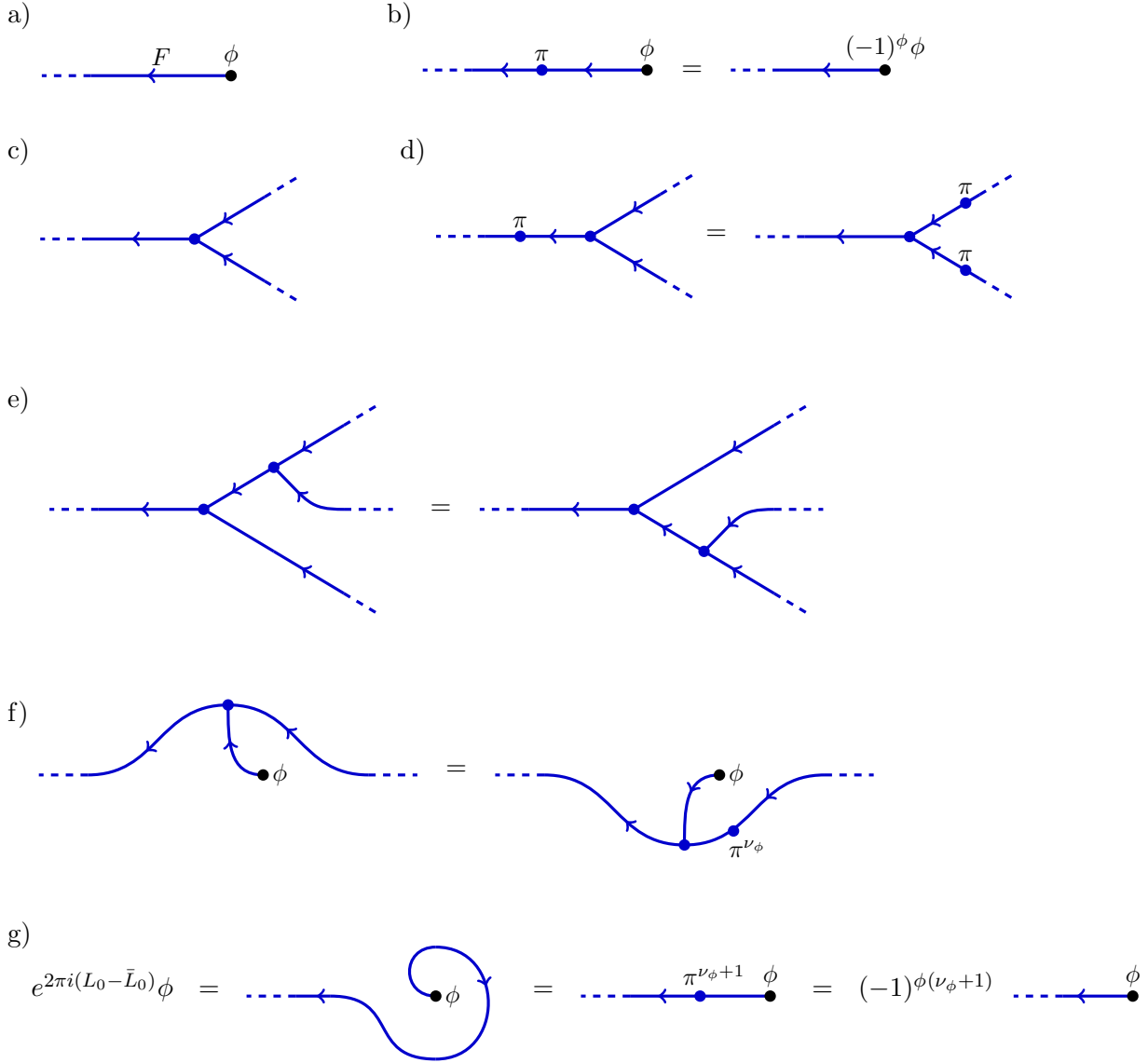

We denote by $\cH_F$ the space of disorder fields that sit at the start of the topological defect $F$ (Figure~\ref{fig:F-properties}\,(a)). 
Since $F$ is topological, $\cH_F$ carries a representation of the
holomorphic and anti-holomorphic copy of the Virasoro algebra.

The space $\cH_F$ is a super-vector space, that is, it is $\mathbb{Z}_2$-graded into an even and an odd component,
\be
	\cH_F  = \cH_F^\mathrm{ev} \oplus \cH_F^\mathrm{odd} \ .
\ee
We refer to this grading as {\em parity}, and for a homogeneous element 
$\phi \in \cH_F$ we write $|\phi| \in \{ 0 , 1\}$ -- or just $\phi$ if no confusion can arise -- for its parity. 
There is a second $\mathbb{Z}_2$-grading on $\cH_F$ whose components are called the Neveu-Schwarz and the Ramond sector,
\be
\cH_F  = \cH_F^\mathrm{NS} \oplus \cH_F^\mathrm{R} \ .
\ee
We will refer to this as the {\em spin grading}. 
It will be convenient to abbreviate $\cH_F^0 = \cH_F^\mathrm{NS}$ and $\cH_F^1 = \cH_F^\mathrm{R}$. Each $\cH_F^\nu$ is still parity graded, 
$\cH_F^\nu =\cH_F^\mathrm{\nu,ev} \oplus \cH_F^\mathrm{\nu,odd}$, so that altogether, $\cH_F$ is $\mathbb{Z}_2 \times \mathbb{Z}_2$-graded.
For a field $\phi \in \cH_F$ that is homogeneous with respect to the spin grading we write $\nu_\phi \in \{0,1\}$ for its degree.

We will refer to fields in $\cH_F$ as {\em bulk fields} of the fermionic CFT.

The defect $F$ has the following properties:
\begin{enumerate}
\item
There is a parity even weight zero\footnote{
	We use ``weight zero'' to mean that it behaves like a vacuum field, i.e.\ that it is annihilated by the translation operators $L_{-1}$ and $\bar L_{-1}$.
}
defect field $\pi$ on $F$ which implements parity on $F$ in the sense that the OPE of $\pi$ with a bulk field $\phi$ is $(-1)^{\pb{}} \phi$ (Figure~\ref{fig:F-properties}\,(b)). It will be convenient to write $\pi^0 = 1$ for the identity defect field and $\pi^1 = \pi$.

\item
There is a parity even weight zero defect junction joining two in-coming $F$-defects into an out-going $F$-defect (Figure~\ref{fig:F-properties}\,(c)). This junction commutes with $\pi$ in the sense shown in Figure~\ref{fig:F-properties}\,(d), and it is associative as shown in Figure~\ref{fig:F-properties}\,(e).

\item
Taking a bulk field $\phi$ of spin grade $\nu_\phi$ past an $F$-defect results in the insertion of $\pi^{\nu_\phi}$ as shown in Figure~\ref{fig:F-properties}\,(f).
\end{enumerate}
One can verify that the effect of a $2\pi$-rotation of a bulk field $\phi$ can be replaced by an insertion of $\pi^{\nu_\phi + 1}$ on the $F$-defect (Figure~\ref{fig:F-properties}\,(g)), see \cite[Lem.\,4.7]{Novak:2015ela}. Thus we get
\be\label{eq:2pi-rotation}
	e^{2\pi i(L_0 - \bar L_0)} \phi
		~=~
		(-1)^{\pb{} (\nu_\phi + 1) } \, \phi \ .
\ee
Denote the conformal spin of a field by $S_\phi := h_\phi - \bar h_\phi$. The above relation implies that 
\be\label{eq:spin-vs-gradings}
	S_\phi \in \begin{cases}
		\mathbb{Z} + \frac12 &; \text{ $\phi$ is a parity-odd NS-field ,}  \\
		\mathbb{Z} &; \text{ otherwise .}
		\end{cases}
\ee

\subsection{Bulk structure constants}

To define the OPE of bulk fields, we need to fix a convention for the spin structure in a neighbourhood of the fields. In the present formalism, this is done by requiring a particular pattern of defect lines. The convention we will use is that for $x>y$ real,
\be\label{eq:bulk-OPE}
\phi_i(x)\phi_j(y)
~=~
\begin{tikzpicture}[baseline=0]
\coordinate (v1) at (0,0);
\coordinate (v2) at (-1,0);
\coordinate (v3) at (-2,0);
\coordinate (v4) at (-3,0);
\coordinate (b) at (-1,+0.7);

\begin{scope}[very thick,blue!80!black,decoration={markings,mark=at position 0.55 with {\arrow{>}}}]
\draw[postaction={decorate}] (v1) .. controls ++(-0.5,0) and ++(0.5,0) .. (b) .. controls ++(-0.5,0) and ++(0.5,0.5) .. (v3);
\draw[postaction={decorate}] (v2) -- (v3);
\draw[postaction={decorate}] (v3) -- (v4);
\draw[dashed] (v4) -- ++(-0.7,0);
\end{scope}
\draw[very thick,black,fill=black] (v1) circle (0.060);
\draw[very thick,black,fill=black] (v2) circle (0.060);
\draw[very thick,blue!80!black,fill=blue!80!black] (v3) circle (0.060);
\node[below] at (v2) {\small $\phi_j(y)$};
\node[below] at (v1) {\small $\phi_i(x)$};
\end{tikzpicture}
=~
\sum_{k} \SC_{ij}^{~k} \, (x-y)^{\Delta_k -\Delta_i - \Delta_j} \, \phi_k(y) 
+ (\text{descendants})\ ,
\ee
where the $\phi_i$ are a basis of primary fields in $\cH_F$, and $\Delta_i = h_i +\bar h_i$ denotes the scaling dimension of $\phi_i$.

One could alternatively have chosen the convention that the $F$-defect
starting at $\phi_i$ passes below $\phi_j$. 
According to Figure~\ref{fig:F-properties}\,(f), the two choices are related by
\be\label{eq:above-below-relation}
\begin{tikzpicture}[baseline=0]
\coordinate (v1) at (0,0);
\coordinate (v2) at (-1,0);
\coordinate (v3) at (-2,0);
\coordinate (v4) at (-3,0);
\coordinate (b) at (-1,-0.7);

\begin{scope}[very thick,blue!80!black,decoration={markings,mark=at position 0.55 with {\arrow{>}}}]
\draw[postaction={decorate}] (v1) .. controls ++(-0.5,0) and ++(+0.5,0) .. (b) .. controls ++(-0.5,0) and ++(+0.5,-0.5) .. (v3);
\draw[postaction={decorate}] (v2) -- (v3);
\draw[postaction={decorate}] (v3) -- (v4);
\draw[dashed] (v4) -- ++(-0.7,0);
\end{scope}
\draw[very thick,black,fill=black] (v1) circle (0.060);
\draw[very thick,black,fill=black] (v2) circle (0.060);
\draw[very thick,blue!80!black,fill=blue!80!black] (v3) circle (0.060);
\node[above] at (v2) {\small $\phi_j(y)$};
\node[above] at (v1) {\small $\phi_i(x)$};
\end{tikzpicture}
~~=~~
\begin{tikzpicture}[baseline=0]
\coordinate (v1) at (0,0);
\coordinate (v2) at (-1,0);
\coordinate (v3) at (-2,0);
\coordinate (v4) at (-3,0);
\coordinate (b) at (-1,+0.7);
\coordinate (pi) at (-1.4,0.65);

\begin{scope}[very thick,blue!80!black,decoration={markings,mark=at position 0.55 with {\arrow{>}}}]
\draw[postaction={decorate}] (v1) .. controls ++(-0.5,0) and ++(+0.5,0) .. (b) .. controls ++(-0.5,0) and ++(+0.5,+0.5) .. (v3);
\draw[postaction={decorate}] (v2) -- (v3);
\draw[postaction={decorate}] (v3) -- (v4);
\draw[dashed] (v4) -- ++(-0.7,0);
\end{scope}
\draw[very thick,black,fill=black] (v1) circle (0.060);
\draw[very thick,black,fill=black] (v2) circle (0.060);
\draw[very thick,blue!80!black,fill=blue!80!black] (v3) circle (0.060);
\draw[very thick,blue!80!black,fill=blue!80!black] (pi) circle (0.060);
\node[below] at (v2) {\small $\phi_j(y)$};
\node[below] at (v1) {\small $\phi_i(x)$};
\node[above] at (pi) {\small $\pi^{\nu_j}$};
\end{tikzpicture}
\quad .
\ee
If we denote the structure constants computed in the $F$-defect-passes-below convention by $\SCvar_{ij}^{~k}$, then the resulting relation is
\be
	\SCvar_{ij}^{~k} = (-1)^{\pb{i} \nu_j} \, \SC_{ij}^{~k} \ .
\ee
In the following we will stick to the convention in \eqref{eq:bulk-OPE}.

\subsection{Crossing symmetry constraint}

\begin{figure}

a)~~
\begin{tikzpicture}[baseline=2em]
\coordinate (v1) at (0,-2.5);
\coordinate (v2) at (-2.0,-2.5);
\coordinate (v3) at (-{2.0-1},-2.5);
\coordinate (b) at (-{2.0+0.7},-{2.5+0.7});
\coordinate (w1) at (0,0);
\coordinate (w2) at (-2.0,0);
\coordinate (w3) at (-{2.0-1},0);
\coordinate (p1) at (-{2.0-2},-1.25);
\coordinate (p2) at (-{2.0-3},-1.25);
\coordinate (bb) at (-{2.0+0.5},+0.7);
\begin{scope}[very thick,blue!80!black,decoration={markings,mark=at position 0.55 with {\arrow{>}}}]
\draw[postaction={decorate}] (v1) .. controls ++(-0.5,0) and ++(+0.5,0) .. (b) .. controls ++(-0.5,0) and ++(+0.5,+0.5) .. (v3);
\draw[postaction={decorate}] (v2) -- (v3);
\draw[postaction={decorate}] (v3) .. controls ++(-0.5,0) and ++(+0.5,-0.5) .. (p1);

\draw[postaction={decorate}] (w1) .. controls ++(-0.5,0) and ++(+0.5,0) .. (bb) .. controls ++(-0.5,0) and ++(+0.5,+0.5) .. (w3);
\draw[postaction={decorate}] (w2) -- (w3);
\draw[postaction={decorate}] (w3) -- (p1);
\draw[postaction={decorate}] (p1) -- (p2);
\draw[dashed] (p2) -- ++(-0.7,0);
\end{scope}
\draw[very thick,black,fill=black] (v1) circle (0.060);
\draw[very thick,black,fill=black] (v2) circle (0.060);
\draw[very thick,black,fill=black] (w1) circle (0.060);
\draw[very thick,black,fill=black] (w2) circle (0.060);
\draw[very thick,blue!80!black,fill=blue!80!black] (v3) circle (0.060);
\draw[very thick,blue!80!black,fill=blue!80!black] (w3) circle (0.060);
\draw[very thick,blue!80!black,fill=blue!80!black] (p1) circle (0.060);

\draw[<->] (-0.5,0) --  node[below] {\small $a$}(-1.5,0);
\draw[<->] (0,-2.0) --  node[left] {\small $b$}(0,-0.5);

\node[above] at (v1) {\small $k$};
\node[above] at (v2) {\small $\ell$};
\node[above] at (w1) {\small $i$};
\node[above] at (w2) {\small $j$};
\end{tikzpicture}
\hspace{3em}
b)~~
\begin{tikzpicture}[baseline=2em]
\coordinate (v1) at (0,-2.5);
\coordinate (v2) at (-2.0,-2.5);
\coordinate (b) at (-{2.0+0.7},-{2.5+0.7});
\coordinate (w1) at (0,0);
\coordinate (w2) at (-2.0,0);
\coordinate (p1) at (-{2.0-2},-1.25);
\coordinate (p2) at (-{2.0-3},-1.25);
\coordinate (bb) at (-2.0,+0.7);
\coordinate (q1) at (-1,-1.25);
\coordinate (q2) at (-{2.0-1},-1.25);
\begin{scope}[very thick,blue!80!black,decoration={markings,mark=at position 0.55 with {\arrow{>}}}]
\draw[postaction={decorate}] (v1) .. controls ++(-0.5,0) and ++(+0.5,-0.5) .. (q1);
\draw[postaction={decorate}] (w1) .. controls ++(-0.5,0) and ++(+0.5,+0.5) .. (q1);
\draw[postaction={decorate}] (v2) .. controls ++(-0.5,0) and ++(+0.5,-0.5) .. (q2);
\draw[postaction={decorate}] (w2) .. controls ++(-0.5,0) and ++(+0.5,+0.5) .. (q2);
\draw[postaction={decorate}] (p1) -- (p2);
\draw[postaction={decorate}] (q2) -- (p1);
\draw[postaction={decorate}] (q1) .. controls ++(-0.5,0) and ++(+0.8,0) .. (bb) .. controls ++(-0.8,0) and ++(+0.5,+0.5) .. (p1);
\draw[dashed] (p2) -- ++(-0.7,0);
\end{scope}
\draw[very thick,black,fill=black] (v1) circle (0.060);
\draw[very thick,black,fill=black] (v2) circle (0.060);
\draw[very thick,black,fill=black] (w1) circle (0.060);
\draw[very thick,black,fill=black] (w2) circle (0.060);
\draw[very thick,blue!80!black,fill=blue!80!black] (q1) circle (0.060);
\draw[very thick,blue!80!black,fill=blue!80!black] (q2) circle (0.060);
\draw[very thick,blue!80!black,fill=blue!80!black] (p1) circle (0.060);
\node[above] at (v1) {\small $k$};
\node[above] at (v2) {\small $\ell$};
\node[above] at (w1) {\small $i$};
\node[above] at (w2) {\small $j$};
\end{tikzpicture}

\caption{a) The arrangement of $F$-defects needed to substitute the OPE in the $a\to 0$ limit. b) The corresponding configuration required in the $b \to 0$ limit.}
\label{fig:two-limits}
\end{figure}
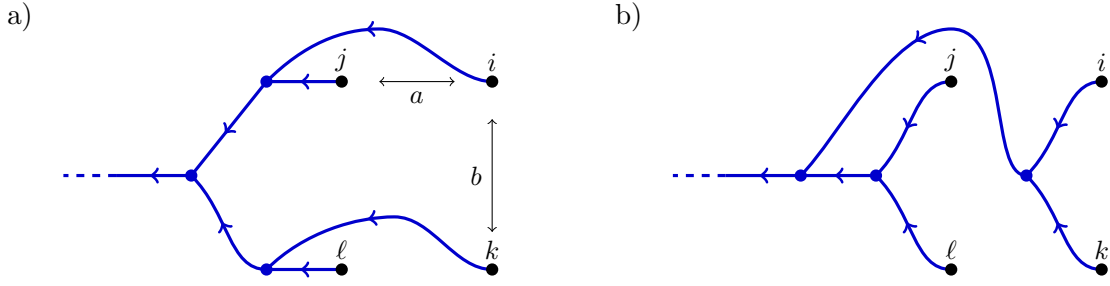

Consider a correlator $f(a,b)$ of four primary bulk fields $\phi_i$,
$\phi_j$, $\phi_k$, $\phi_\ell$, inserted at positions $a+ib$, $ib$, 
$a$, $0$ for some $a,b>0$,
\be
	f(a,b) = \big\langle  \phi_i(a+ib) \, \phi_j(ib) \, \phi_k(a) \, \phi_\ell(0) \big\rangle \ .
\ee
Here it is understood that the $F$-defects are placed as in
Figure~\ref{fig:two-limits}\,(a), and that the ordering relevant for
the parity signs is radial ordering. 

For simplicity, in the derivation of the crossing constraint
(and in that of similar constraints below) we restrict ourselves to
theories which are rational with respect to the Virasoro
symmetry. Extended chiral algebras can be treated in the same way, but
one has to account for two complications. Firstly, the fusing matrices
will in general carry multiplicity labels. Secondly, the leading
contribution in the OPE of two primary fields (with respect to the
extended symmetry) may be a descendent field, which makes the
definition of the OPE coefficients more involved. 

The defect arrangement in Figure~\ref{fig:two-limits}\,(a) is such
that in the $a \to 0$ limit, we can substitute the bulk OPE right
away. Hence in this limit we simply have 
\be
	f(a,b) ~\underset{a\to0}\sim~ 
	\sum_{p} \SC_{ij}^{~p} \SC_{k\ell}^{~p} \SC_{pp}^{~1}
	\,e^{-\pi i S_p} \,b^{-2 \Delta_p} 
	\cdot a^{2 \Delta_p - \Delta_i - \Delta_j - \Delta_k -
          \Delta_\ell} 
\;.
\label{eq:2.9}
\ee 
The phase $e^{-\pi i S_p} = e^{-\pi i (h_p - \bar h_p)}$ in
\eqref{eq:2.9} is determined by our convention on how to continue the
OPE \eqref{eq:bulk-OPE} to the configuration in figure
\ref{fig:two-limits} where the result of the OPE of fields $\phi_i$
with $\phi_j$ and $\phi_l$ with $\phi_k$ results in fields at
positions $0$ and $ib$. We choose to continue the OPE so that the
defect lines stay in the same topological arrangement and do not cross
through the field insertion points which determines the phase
uniquely.

In the $b \to 0$ limit we have to rearrange the $F$-defect before we are allowed to substitute the OPE. The relevant configuration is shown in Figure~\ref{fig:two-limits}\,(b). We have
\be
\begin{tikzpicture}[baseline=-2.8em]
\coordinate (v1) at (0,-2);
\coordinate (v2) at (-2.5,-2);
\coordinate (v3) at (-{2.5-1},-2);
\coordinate (b) at (-{2.5+0.7},-{2+0.7});
\coordinate (w1) at (0,0);
\coordinate (w2) at (-2.5,0);
\coordinate (w3) at (-{2.5-1},0);
\coordinate (p1) at (-{2.5-2},-1);
\coordinate (p2) at (-{2.5-3},-1);
\coordinate (bb) at (-{2.5+0.7},+0.7);
\begin{scope}[very thick,blue!80!black,decoration={markings,mark=at position 0.55 with {\arrow{>}}}]
\draw[postaction={decorate}] (v1) .. controls ++(-0.5,0) and ++(+0.5,0) .. (b) .. controls ++(-0.5,0) and ++(+0.5,+0.5) .. (v3);
\draw[postaction={decorate}] (v2) -- (v3);
\draw[postaction={decorate}] (v3) .. controls ++(-0.5,0) and ++(+0.5,-0.5) .. (p1);

\draw[postaction={decorate}] (w1) .. controls ++(-0.5,0) and ++(+0.5,0) .. (bb) .. controls ++(-0.5,0) and ++(+0.5,+0.5) .. (w3);
\draw[postaction={decorate}] (w2) -- (w3);
\draw[postaction={decorate}] (w3) -- (p1);
\draw[postaction={decorate}] (p1) -- (p2);
\draw[dashed] (p2) -- ++(-0.7,0);
\end{scope}
\draw[very thick,black,fill=black] (v1) circle (0.060);
\draw[very thick,black,fill=black] (v2) circle (0.060);
\draw[very thick,black,fill=black] (w1) circle (0.060);
\draw[very thick,black,fill=black] (w2) circle (0.060);
\draw[very thick,blue!80!black,fill=blue!80!black] (v3) circle (0.060);
\draw[very thick,blue!80!black,fill=blue!80!black] (w3) circle (0.060);
\draw[very thick,blue!80!black,fill=blue!80!black] (p1) circle (0.060);

\node[above] at (v1) {\small $k$};
\node[above] at (v2) {\small $\ell$};
\node[above] at (w1) {\small $i$};
\node[above] at (w2) {\small $j$};
\end{tikzpicture}
~~=~~
\begin{tikzpicture}[baseline=-2.8em]
\coordinate (v1) at (0,-2);
\coordinate (v2) at (-2.5,-2);
\coordinate (v3) at (-{2.5-1},-2);
\coordinate (b) at (-{2.5+0.7},-{2+0.7});
\coordinate (w1) at (0,0);
\coordinate (w2) at (-2.5,0);
\coordinate (w3) at (-{2.5-1},0);
\coordinate (p1) at (-{2.5-2},-1);
\coordinate (p2) at (-{2.5-3},-1);
\coordinate (bb) at (-{2.5+0.7},+0.7);
\coordinate (pi) at (-1.65,-0.2);

\begin{scope}[very thick,blue!80!black,decoration={markings,mark=at position 0.55 with {\arrow{>}}}]
\draw[postaction={decorate}] (v1) .. controls ++(-1,0) and ++(0,-1) .. (bb);
\draw[postaction={decorate}] (v2) -- (v3);
\draw[postaction={decorate}] (v3) .. controls ++(-0.5,0) and ++(+0.5,-0.5) .. (p1);
\draw[postaction={decorate}] (w1) .. controls ++(-0.5,0) and ++(+0.5,0) .. (bb);
\draw[postaction={decorate}] (bb) .. controls ++(-0.5,0) and ++(+0.5,+0.5) .. (w3);
\draw[postaction={decorate}] (w2) -- (w3);
\draw[postaction={decorate}] (w3) -- (p1);
\draw[postaction={decorate}] (p1) -- (p2);
\draw[dashed] (p2) -- ++(-0.7,0);
\end{scope}
\draw[very thick,black,fill=black] (v1) circle (0.060);
\draw[very thick,black,fill=black] (v2) circle (0.060);
\draw[very thick,black,fill=black] (w1) circle (0.060);
\draw[very thick,black,fill=black] (w2) circle (0.060);
\draw[very thick,blue!80!black,fill=blue!80!black] (bb) circle (0.060);
\draw[very thick,blue!80!black,fill=blue!80!black] (w3) circle (0.060);
\draw[very thick,blue!80!black,fill=blue!80!black] (p1) circle (0.060);
\draw[very thick,blue!80!black,fill=blue!80!black] (pi) circle (0.060);

\node[above] at (v1) {\small $k$};
\node[above] at (v2) {\small $\ell$};
\node[above] at (w1) {\small $i$};
\node[above] at (w2) {\small $j$};

\node[right] at (pi) {\small $\pi^{\nu_j}$};
\end{tikzpicture}
\quad .
\ee
Here we used the associativity of the junction field and the effect of dragging an $F$-defect through a field, see Figure~\ref{fig:F-properties}\,(e),\,(f). The $\pi$-insertion contributes the sign factor $(-1)^{\nu_j \pb{k}}$. The resulting graph of $F$-defects can be brought to the form in Figure~\ref{fig:two-limits}\,(b) by using the associativity relation once more.

We see that bringing the spin structure to the form required for the OPE contributes a sign factor $(-1)^{\nu_j \pb{k}}$. Another sign arises from parity as the order of $\phi_j$ and $\phi_k$ changes. Altogether we get
\be
	f(a,b) ~\underset{b\to0}\sim~ 
(-1)^{(\pb{j} + \nu_j)\,\pb{k}}	
\sum_{q} \SC_{ik}^{~q} \, \SC_{j\ell}^{~q} \, \SC_{qq}^{~1} \,
e^{\frac{\pi i }2(2 S_q - S_i - S_j - S_k - S_\ell)}  \,
a^{-2 \Delta_q} \cdot b^{2 \Delta_q - \Delta_i - \Delta_j - \Delta_k - \Delta_\ell}
 \ .
\ee
The rest of the computation is a standard manipulation of conformal blocks. The overall result is
\be\label{eq:bulk-crossing}
\SC_{ik}^{~q} \, \SC_{j\ell}^{~q} \, \SC_{qq}^{~1} 
~=~
(-1)^{(\pb{j} + \nu_j)\,\pb{k}}	
\,
\sum_p e^{\pi i(S_i+S_\ell-S_p-S_q)} 
\,
F_{pq}\big[ \begin{smallmatrix} j&\ell \\ i & k \end{smallmatrix} \big]
\,
F_{\bar p \bar q}\big[ \begin{smallmatrix} \bar \jmath & \bar \ell \\ \bar \imath & \bar k \end{smallmatrix} \big]
\,
\SC_{ij}^{~p}  \, \SC_{k\ell}^{~p} \, \SC_{pp}^{~1}
\ .
\ee
Here, the entries ``\,$i$\,'' and ``\,$\bar \imath$\,'' in the two F-matrices refer to the holomorphic and antiholomorphic conformal weight $(h_i,\bar h_i)$ of $\phi_i$, etc.

\subsection{Symmetry properties of structure constants}\label{sec:bulksym}

The four-point crossing relation determines how the structure constants behave under permutation of indices. Define
\be\label{eq:lower-index}
	\SC_{ijk} = \SC_{ij}^{~k} \, \SC_{kk}^{~1} \ .
\ee
We obtain the following two relations by setting
$\phi_j=1$ and $\phi_\ell=1$ in \eqref{eq:bulk-crossing}, respectively:
\be
	\SC_{k\ell i} = \SC_{ik\ell}
	\qquad , \qquad
	\SC_{ijk}
		= (-1)^{(\pb{j} + \nu_j)\,\pb{k}} \, e^{\pi i (S_k+S_j-S_i)} \, \SC_{ikj} \ .
\label{eq:cyclic}
\ee
In particular, the $\SC_{ijk}$ are cyclically symmetric. Using this we
can rewrite the second equality above as a relation between
$\SC_{jki}$ and $\SC_{kji}$. After relabelling and dividing by
$\SC_{kk}^{~1}$ this implies 
\be
\SC_{ji}^{~k} 
~=~
(-1)^{\pb{i}\,(\pb{j} + \nu_j)} \,
e^{\pi i (S_i+S_j-S_k)} \,
\SC_{ij}^{~k} 
~\overset{(*)}=~
(-1)^{(\pb{i} + \nu_i)\,\pb{j}} \,
e^{\pi i (S_k-S_i-S_j)} \,
\SC_{ij}^{~k} \ .
\label{eq:sym1}
\ee
In step $(*)$ we used that $e^{2 \pi i S_i} = (-1)^{\pb{i}\,(\nu_i+1)}$, etc., from \eqref{eq:2pi-rotation} and the fact that the OPE preserves the parity and spin gradings.

For the description of the classifying algebra below it will be important that structure constants involving only spinless fields are symmetric:
\be\label{eq:spinless-SC}
\text{If}
\quad 
S_i=S_j=S_k = 0
\quad
\text{then}
\quad
\SC_{ij}^{~k}  =  \SC_{ji}^{~k}  \ .
\ee
To see this, note that the only situation in which both expressions in \eqref{eq:sym1} produce a sign is $\phi_i=\phi_j=1$ and $\nu_i=\nu_j=0$.
But then \eqref{eq:spin-vs-gradings} imposes $S_i, S_j \in \mathbb{Z} + \frac12$, which we excluded.

\subsection{Parity shift of the Ramond sector}\label{sec:shiftRamond}

As we have seen in section~\ref{sec:topdefF}, the state space $\cH_F$ of a fermionic CFT carries two $\mathbb{Z}_2$-gradings, namely even/odd and NS/R. Accordingly it splits into four direct summands,
\begin{align}
	&\cH_F^\mathrm{NS,ev}
	&\cH_F^\mathrm{R,ev}
\nonumber \\	
	&\cH_F^\mathrm{NS,odd}
	&\cH_F^\mathrm{R,odd}
\label{eq:HF-foursector}
\end{align}
As before, for a primary bulk field $\phi_i$, its parity is $|\phi_i| \in \{0,1\}$ (with $0$ being even) and its spin grade is $\nu_i \in \{0,1\}$ (with $0$ being NS).
Let $\SC_{ij}^{~k}$ be a solution of the bulk crossing relation \eqref{eq:bulk-crossing}.

Given such a solution, we can construct a new fermionic CFT as follows. The new state space $\widetilde\cH_F$ agrees with the old one, except for a parity shift in the Ramond sector:
\begin{align}
\widetilde\cH_F^\mathrm{NS,ev} &:= \cH_F^\mathrm{NS,ev}
& \widetilde\cH_F^\mathrm{R,ev}  &:= \cH_F^\mathrm{R,odd}
\nonumber \\	
\widetilde\cH_F^\mathrm{NS,odd} &:= \cH_F^\mathrm{NS,odd}
&\widetilde\cH_F^\mathrm{R,odd} &:= \cH_F^\mathrm{R,ev}
\end{align}
That is, the old and new gradings are related by
\be
|\widetilde\phi_i| = |\phi_i| + \nu_i
\qquad , \qquad
\widetilde\nu_i = \nu_i \ .
\ee
The new structure constants are related to the old ones by signs,
\be\label{eq:SC-R-parity-shift}
	\widetilde C_{ij}^{~k}
	:=
	(-1)^{\nu_i \phi_j}\,
	\SC_{ij}^{~k} \ .
\ee
It is a straightforward computation to see that $\widetilde C_{ij}^{~k}$ again solves the crossing relation \eqref{eq:bulk-crossing} (and hence also has the symmetry properties stated in section~\ref{sec:bulksym}).

Applying the shift operation twice produces the structure constants $(-1)^{\nu_i \nu_j}\,\SC_{ij}^{~k}$. The extra sign can be absorbed into the normalisation of the fields (e.g.\ multiply all Ramond fields by $i$), so that this reproduces the theory one started from.\footnote{
	It is possible to modify \eqref{eq:SC-R-parity-shift} so that applying the parity shift twice gives back precisely the structure constants one started from. For example, with $\gamma_a = e^{-\pi i (S_a + \phi_a/2)}$ one can set $\widetilde C_{ij}^{~k}
	=
	(-1)^{\nu_i \phi_j} \, \gamma_i \gamma_j / \gamma_k \, \SC_{ij}^{~k}
	=
	(-1)^{(\nu_i+\phi_i) \phi_j} \, e^{\pi i (S_k-S_i-S_j)} \, \SC_{ij}^{~k}
	= 
	\SC_{ji}^{~k}
	$, where in the last step we used \eqref{eq:sym1}.
}

The parity shifted theory may or may not be isomorphic to the
unshifted theory. For example, parity shifting the fermionic Ising
model produces an equivalent theory, while parity shifting the
fermionic minimal model $\FM(3,8)$ produces an inequivalent theory,
see sections~\ref{sec:IsingBulk} and~\ref{sec:susyly}.

\section{Classifying algebras}
\label{sec:ca}

In this section we use the bulk structure constants of a fermionic CFT
to define several types of fermionic classifying algebras. These are
semisimple super-algebras, graded by parity, whose direct summands are
in 1-1 correspondence to boundary conditions
(section~\ref{sec:bndOPE+bnd-class-alg}), defects, or interfaces
(section~\ref{sec:def-class-alg}), depending on the algebra under
consideration. In section~\ref{sec:class-parityshift} we observe that
either one of these classifying algebras in a fermionic CFT is isomorphic
to the corresponding algebra in the parity shifted CFT as an ungraded
algebra, but typically not as a super-algebra. 

\subsection{Bulk-boundary OPE and boundary classifying algebra}\label{sec:bndOPE+bnd-class-alg}

\subsubsection{Conventions for bulk-boundary OPE and boundary OPE}

We now consider the theory on the upper half plane with some
conformally invariant boundary condition $\alpha$  
placed on the real line. As for bulk fields, boundary fields will also serve as starting point for an $F$-defect. We denote the space of boundary fields by $\cH^{(\alpha)}_F$. It is again $\mathbb{Z}_2$-graded by parity,\footnote{
	If there are holomorphic and antiholomorphic fermions, one can define a monodromy for fermion fields around boundary (changing) fields. In this sense the space of boundary fields may also split into NS- and R-sectors. But this split does not have a geometric counterpart in the present setting: we consider a surface with a single spin structure, and near a boundary point a spin structure is unique up to isomorphism.
}
\be
\cH^{(\alpha)}_F ~=~ \cH^{(\alpha),\mathrm{ev}}_F \, \oplus \,
\cH^{(\alpha),\mathrm{odd}}_F \ . 
\ee
Our convention for the defect arrangement near an insertion of a
boundary field $\psi \in \cH^{(\alpha)}_F$ is: 
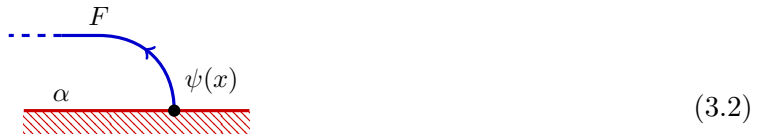
\be
\begin{tikzpicture}[baseline=-0.2em]
\coordinate (phi) at (0,1);
\coordinate (psi) at (0,0);
\coordinate (end) at (-1,1);
\coordinate (b1) at (-2,0);
\coordinate (b2) at (1,0);

\begin{scope}[very thick,blue!80!black,decoration={markings,mark=at position 0.6 with {\arrow{>}}}]
\draw[postaction={decorate}] (psi) .. controls ++(0,0.6) and ++(0.6,0) .. (end);
\draw (end) -- ++(-0.5,0);
\draw[dashed] (end)++(-0.5,0) -- ++(-0.7,0);
\end{scope}
\draw[very thick,red!80!black] (b1) -- (b2);
\fill[pattern=north west lines, pattern color=red!80!black] ([shift={(0,-0.3)}]b1)  rectangle (b2);
\draw[very thick,black,fill=black] (psi) circle (0.060);

\node[above] at (end) {\small $F$};
\node at ([shift={(0.5,0.4)}]psi) {\small $\psi(x)$};
\node at ([shift={(0.5,0.2)}]b1) {\small $\alpha$};
\end{tikzpicture}
\ee

The bulk-boundary structure constants $B^\phi_\psi$ are defined as, for a primary bulk field $\phi \in \cH_F$,
\be
\phi(x+iy)
= \sum_{\psi} B^\phi_\psi \, (2y)^{h_\psi - \Delta_\phi} \, \psi(x) + (\text{desc.}) \ ,
\ee
where the sum runs over a basis of primary fields $\psi$ in $\cH^{(\alpha)}_F$. 
In pictures the above relation looks as follows,
\be
\begin{tikzpicture}[baseline=-0.2em]
\coordinate (phi) at (0,1);
\coordinate (psi) at (0,0);
\coordinate (end) at (-1,1);
\coordinate (b1) at (-2,0);
\coordinate (b2) at (1,0);

\begin{scope}[very thick,blue!80!black,decoration={markings,mark=at position 0.6 with {\arrow{>}}}]
\draw[postaction={decorate}] (phi) -- (end);
\draw (end) -- ++(-0.5,0);
\draw[dashed] (end)++(-0.5,0) -- ++(-0.7,0);
\end{scope}
\draw[very thick,black,fill=black] (phi) circle (0.060);
\draw[very thick,red!80!black] (b1) -- (b2);
\fill[pattern=north west lines, pattern color=red!80!black] ([shift={(0,-0.3)}]b1)  rectangle (b2);

\node[above] at (end) {\small $F$};
\node at ([shift={(0.2,-0.4)}]phi) {\small $\phi(x+iy)$};
\node at ([shift={(0.5,0.2)}]b1) {\small $\alpha$};
\end{tikzpicture}
~~\sim~~
\sum_\psi
B^\phi_\psi \, (2y)^{h_\psi - \Delta_\phi} ~ 
\begin{tikzpicture}[baseline=-0.2em]
\coordinate (phi) at (0,1);
\coordinate (psi) at (0,0);
\coordinate (end) at (-1,1);
\coordinate (b1) at (-2,0);
\coordinate (b2) at (1,0);

\begin{scope}[very thick,blue!80!black,decoration={markings,mark=at position 0.6 with {\arrow{>}}}]
\draw[postaction={decorate}] (psi) .. controls ++(0,0.6) and ++(0.6,0) .. (end);
\draw (end) -- ++(-0.5,0);
\draw[dashed] (end)++(-0.5,0) -- ++(-0.7,0);
\end{scope}
\draw[very thick,red!80!black] (b1) -- (b2);
\fill[pattern=north west lines, pattern color=red!80!black] ([shift={(0,-0.3)}]b1)  rectangle (b2);
\draw[very thick,black,fill=black] (psi) circle (0.060);

\node[above] at (end) {\small $F$};
\node at ([shift={(0.5,0.4)}]psi) {\small $\psi(x)$};
\node at ([shift={(0.5,0.2)}]b1) {\small $\alpha$};
\end{tikzpicture}
\qquad .
\ee
We will also use the OPE of boundary fields, for which our convention is, for $x>y$,
\be
\begin{tikzpicture}[baseline=-0.2em]
\coordinate (psi1) at (1.5,0);
\coordinate (end2) at (-1,1);
\coordinate (psi2) at (0,0);
\coordinate (end1) at (-1,2);
\coordinate (b1) at (-3,0);
\coordinate (b2) at (3,0);
\coordinate (m) at (-1.8,1.5);

\begin{scope}[very thick,blue!80!black,decoration={markings,mark=at position 0.6 with {\arrow{>}}}]
\draw[postaction={decorate}] (psi1) .. controls ++(0,1.2) and ++(1.2,0) .. (end1);
\draw[postaction={decorate}] (psi2) .. controls ++(0,0.6) and ++(0.6,0) .. (end2);
\draw[postaction={decorate}] (m) -- ++(-1,0);
\draw[dashed] (m)++(-1,0) -- ++(-0.7,0);
\draw (end1) .. controls ++(-0.4,0) and ++(0.2,0.2) .. (m);
\draw (end2) .. controls ++(-0.4,0) and ++(0.2,-0.2) .. (m);
\end{scope}
\draw[very thick,red!80!black] (b1) -- (b2);
\fill[pattern=north west lines, pattern color=red!80!black] ([shift={(0,-0.3)}]b1)  rectangle (b2);
\draw[very thick,black,fill=black] (psi1) circle (0.060);
\draw[very thick,black,fill=black] (psi2) circle (0.060);
\draw[very thick,blue!80!black,fill=blue!80!black] (m) circle (0.060);

\node[above] at ([shift={(-0.7,0)}]m) {\small $F$};
\node at ([shift={(0.55,0.35)}]psi1) {\small $\psi_1(x)$};
\node at ([shift={(0.55,0.35)}]psi2) {\small $\psi_2(y)$};
\node at ([shift={(0.5,0.2)}]b1) {\small $\alpha$};
\end{tikzpicture}
\quad \sim \quad
\sum_r c^{~~r}_{12} 
~ (x-y)^{h_r - h_1 - h_2} \quad
\begin{tikzpicture}[baseline=-0.2em]
\coordinate (phi) at (0,1);
\coordinate (psi) at (0,0);
\coordinate (end) at (-1,1);
\coordinate (b1) at (-2,0);
\coordinate (b2) at (1,0);

\begin{scope}[very thick,blue!80!black,decoration={markings,mark=at position 0.6 with {\arrow{>}}}]
\draw[postaction={decorate}] (psi) .. controls ++(0,0.6) and ++(0.6,0) .. (end);
\draw (end) -- ++(-0.5,0);
\draw[dashed] (end)++(-0.5,0) -- ++(-0.7,0);
\end{scope}
\draw[very thick,red!80!black] (b1) -- (b2);
\fill[pattern=north west lines, pattern color=red!80!black] ([shift={(0,-0.3)}]b1)  rectangle (b2);
\draw[very thick,black,fill=black] (psi) circle (0.060);

\node[above] at (end) {\small $F$};
\node at ([shift={(0.5,0.4)}]psi) {\small $\psi_r(y)$};
\node at ([shift={(0.5,0.2)}]b1) {\small $\alpha$};
\end{tikzpicture}
\ee
where the sum runs over a basis of primary fields $\psi_r$ in $\cH^{(\alpha)}_F$.

We will only consider boundary conditions for which the bulk-boundary OPE and the boundary OPE preserve parity.

\subsubsection{Bulk-boundary crossing relation}

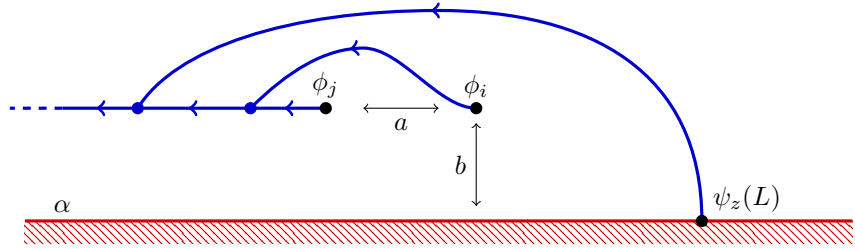
\begin{figure}[t]
	\begin{center}
		\begin{tikzpicture}[baseline=0]
		\coordinate (v1) at (0,1.5);
		\coordinate (v2) at (-2,1.5);
		\coordinate (v3) at (-3,1.5);
		\coordinate (v4) at (-4.5,1.5);
		\coordinate (v5) at (-5.5,1.5);
		\coordinate (b) at (-1.5,+2.3);
		\coordinate (psi) at (3,0);
		
		\coordinate (b1) at (-6,0);
		\coordinate (b2) at (5,0);
		
		\draw[very thick,red!80!black] (b1) -- (b2);
		\fill[pattern=north west lines, pattern color=red!80!black] ([shift={(0,-0.3)}]b1)  rectangle (b2);
		
		\begin{scope}[very thick,blue!80!black,decoration={markings,mark=at position 0.55 with {\arrow{>}}}]
		\draw[postaction={decorate}] (v1) .. controls ++(-0.5,0) and ++(0.5,0) .. (b) .. controls ++(-0.5,0) and ++(0.5,0.5) .. (v3);
		\draw[postaction={decorate}] (psi) .. controls ++(0,2) and ++(2,0) .. ([shift={(1,0.5)}]b) .. controls ++(-2,0) and ++(0.5,0.8) .. (v4);
		\draw[postaction={decorate}] (v2) -- (v3);
		\draw[postaction={decorate}] (v3) -- (v4);
		\draw[postaction={decorate}] (v4) -- (v5);
		\draw[dashed] (v5) -- ++(-0.7,0);
		\end{scope}
		\draw[very thick,black,fill=black] (v1) circle (0.060);
		\draw[very thick,black,fill=black] (v2) circle (0.060);
		\draw[very thick,black,fill=black] (psi) circle (0.060);
		\draw[very thick,blue!80!black,fill=blue!80!black] (v3) circle (0.060);
		\draw[very thick,blue!80!black,fill=blue!80!black] (v4) circle (0.060);
		
		\draw[<->] ([shift={(-0.5,0)}]v1) --  node[below] {\small $a$} ([shift={(0.5,0)}]v2);
		\draw[<->] ([shift={(0,-0.2)}]v1) --  node[left] {\small $b$}(0,0.2);
		
		\node[above] at (v2) {\small $\phi_j$};
		\node[above] at (v1) {\small $\phi_i$};
		\node at ([shift={(0.5,0.2)}]b1) {\small $\alpha$};
		\node at ([shift={(0.6,0.3)}]psi) {\small $\psi_z(L)$};
		\end{tikzpicture}
	\end{center}
	\caption{Two bulk fields and one boundary field together with their defect lines as used in computing the crossing constraint linking 
		bulk-boundary couplings and bulk structure constants. In the correlator $g(a,b)$, 
		the boundary field $\psi_z$ inserted at $L$ is moved off to $\infty$.
	}
	\label{fig:BBb-corr}
\end{figure}

Let $g(a,b)$ stand for the correlator of two bulk fields and one boundary field shown in Figure~\ref{fig:BBb-corr}. The crossing relation is obtained by comparing the $a\to 0$ and $b \to 0$ limit. It turns out that this relation does not involve parity signs or spin structure signs. The computation is thus standard and we will be very brief. The asymptotic behaviour of $g(a,b)$ in the two limits is
\begin{align}
g(a,b) &\underset{a\to 0}\sim
\sum_k \SC_{ij}^{~k} \, B^k_z \, c_{zz}^{~~1}
\, (2b)^{h_z-\Delta_k} 
\cdot
a^{\Delta_k - \Delta_i - \Delta_j} \ ,
\nonumber \\
g(a,b) &\underset{b\to 0}\sim
\sum_{x,y} B^i_x \, B^j_y 
\, c_{xy}^{~~z} \, c_{zz}^{~~1} \, 
a^{h_z-h_x-h_y} \cdot
(2b)^{h_x+h_y-\Delta_i - \Delta_j} \ .
\end{align}
The crossing relation now follows from a computation with five-point Virasoro conformal blocks similar to those in \cite{Lewellen:1991tb},
and the result is
\begin{align}
B^i_x 
~
B^j_y 
~
c^{~~z}_{xy}
~=~
\sum_{k,r}
\SC_{ij}^{~k} 
~
B^k_z 
~
&e^{\frac{i \pi}2(S_k - S_i - S_j)} 
\,
e^{\frac{i \pi}2(h_x-h_y-h_z)}
\notag\\
& \times ~
e^{i \pi(h_r- \overline h_i)}
F_{k,r}\big[ \begin{smallmatrix} \bar k&j\\z&i \end{smallmatrix} \big]
F_{\bar k,y}\big[ \begin{smallmatrix} j&\bar\jmath\\r&\bar\imath \end{smallmatrix} \big]
F_{r,x}\big[ \begin{smallmatrix} i&\bar\imath\\z&y \end{smallmatrix}
\big] \ .
\label{eq:BB2}
\end{align}
If the boundary field $\psi_z$ has weight zero (but could still be of either parity, see next subsection), this relation simplifies to
\be
B^i_x 
~
B^j_y 
~
c^{~~z}_{xy}
~=~
\sum_k
\Big(
\SC_{ij}^{~k} 
\,
e^{\frac{i\pi}{2}(S_i - S_j)}
\,
F_{k,y}\big[  
       \begin{smallmatrix}  j&\bar\jmath\\i& \bar\imath \end{smallmatrix} 
       \big]
\Big)
~
B^k_z 
\ ,
\label{eq:BB1}
\ee
where we used that $B^k_z \neq 0$ requires $h_k = \overline h_k$.
If the boundary fields $\psi_x$, $\psi_y$, $\psi_z$ all have weight
zero (but could again be of either parity), this relation simplifies
further to
\be
B^i_x 
~
B^j_y 
~
c^{~~z}_{xy}
~=~
\sum_k
\Big(
\SC_{ij}^{~k} 
\,
F_{k,0}\big[ \begin{smallmatrix}  j&j\\i& i \end{smallmatrix} \big]
\Big)
~
B^k_z 
\ .
\label{eq:BB}
\ee
The corresponding relation for bosonic theories was derived in \cite{Cardy:1991tv}.

\subsubsection{Boundary classifying algebra:  Bosonic case}

Let us first recall the situation when there are just bosonic (i.e.\ parity even)
fields and consider the space of weight zero boundary fields which can arise
in the bulk-boundary OPE. 
Looking at the right hand side of the crossing constraint \eqref{eq:BB},
this is symmetric under $i \leftrightarrow j$ and as a consequence
$c_{xy}^{~~z}$ must be symmetric under $x \leftrightarrow y$.
This means that the  weight
zero boundary fields which arise in the bulk-boundary OPE form a commutative, associative 
algebra. If one restricts one's attention to situations where this algebra is semisimple, by the Wedderburn theorem it
must be equal to a direct
sum of copies of $\mathbb C$, each of which can be considered as the
identity operator on an elementary boundary condition.

Due to this observation, it is enough to consider elementary boundary
conditions, i.e.\ boundary conditions on which there is a single boundary field of weight 0 (the identity field). The bosonic version of
\eqref{eq:BB} reduces to  
\be
B^i
~
B^j 
~=~
\sum_k
\Big(
C_{ij}^{~k} 
\,
F_{k,0}\big[ \begin{smallmatrix}  j&j\\i& i \end{smallmatrix} \big]
\Big)
~
B^k
\;.
\label{eq:BBbosonix}
\ee
This equation can be simply seen as a set of equations for the
constants $B^i$, but as shown in \cite{acafbc,Fuchs:2009zr}
this is also the
defining relation for a commutative associative algebra with
generators $B^k$, the \textit{(bosonic) boundary classifying algebra}.
From this point of view, the elementary boundary
conditions correspond to the one-dimensional representations of this
algebra,
or, equivalently, to the number of summands $\mathbb{C}$ in the direct sum decomposition of \eqref{eq:BBbosonix}.

\subsubsection{Boundary classifying algebra:  Fermionic case}
\label{sec:ferm-class-alg}

In the fermionic case, we equally take the coefficients 
$\SC_{ij}^{~k} F_{k,0}\big[ \begin{smallmatrix}  j&j\\i& i \end{smallmatrix} \big]$  
in \eqref{eq:BB} as the structure constants of an algebra with generators $B^i$,
\be
B^i
~
B^j 
~=~
\sum_k
\Big(
\SC_{ij}^{~k} 
\,
F_{k,0}\big[ \begin{smallmatrix}  j&j\\i& i \end{smallmatrix} \big]
\Big)
~
B^k
\ ,
\label{eq:BB'}
\ee
where the indices $i,j,k$ now run over all spinless fields of either
parity. 
We denote this algebra by $\hat \cB$ and call it the {\em fermionic boundary classifying algebra}.
In the examples we consider in sections~\ref{sec:Ising} and \ref{sec:furtherVir}, this is indeed an associative algebra which is in addition semisimple. In \cite{RSW-prep} we will show that \eqref{eq:BB'} defines a semisimple associative algebra for fermionic rational CFTs in general.

While the
$\SC_{ij}{}^k$ are not necessarily symmetric under
$i\leftrightarrow j$, 
we saw in \eqref{eq:spinless-SC} that
they are symmetric for all $i,j$ and $k$ which have
spin 0 and which can therefore couple to boundary fields of weight
0. 
It follows that the algebra $\hat\cB$ is commutative, $B^iB^j = B^jB^i$.

Since the bulk fields are graded by their parity, 
we can also view $\hat\cB$ as a super-algebra with the 
parity of the generator $B^i$ being the same as that of the field $\phi_i$. As the above commutativity relation does not involve parity signs, $\hat\cB$ is in general not super-commutative.
Note that $\hat\cB$ is in fact bi-graded, by parity and by spin, since
the bulk structure constants $\SC_{ij}{}^k$ preserve both these
gradings. 

Let us assume that $\hat\cB$ is semisimple. Since $\hat\cB$ is
commutative (rather than super-commutative), the 
super-Wedderburn theorem~\cite[Cor.\,2.12]{Jozefiak:1988}
states that the algebra splits into
a sum of copies of $\mathbb C^{1|0} = \mathbb C$ and the Clifford algebra
$Q(2)\equiv C\ell_1$ which is two-dimensional with one odd generator, $a$,
satisfying $a^2=1$. In other words, $\hat\cB$ is equal to a sum of $m$ copies of $\mathbb C$ with
generators $e_\alpha$ (where $\alpha$ labels the copy) and $n$ copies of $C\ell_1$ with even generator
$f_\beta$ and odd generator $a_\beta$, 
\be
\hat \cB 
= \left( \bigoplus^m \mathbb C e_\alpha \right)
\oplus 
\left( \bigoplus^n (\mathbb C f_\beta \oplus \mathbb C a_\beta) \right)
\ .
\label{eq:hatB-sum-decomp}
\ee
These generators satisfy the relations
\begin{align}
e_\alpha^2 &= e_\alpha
~,&
f_\beta^2 &= f_\beta
~,&
f_\beta a_\beta &= a_\beta
~,&
a_\beta^2 &= f_\beta
~,&
x_\gamma y_\delta &= 0\;,\; \gamma \neq \delta
\ .
\label{eq:efa}
\end{align}

After analysing $\hat\cB$ in some detail, let us return to the bulk-boundary crossing constraint \eqref{eq:BB}.
Analogous to the bosonic case, we conclude that
the weight zero boundary fields which arise in the bulk-boundary OPE form an
associative  
super-algebra which is commutative but not necessarily super-commutative, and which is in addition semisimple.\footnote{
Semisimplicity follows from that of $\hat\cB$. Indeed, the coefficients $B^i_x$ in \eqref{eq:BB} can be understood as the coefficients of a homomorphism of super-algebras from $\hat\cB$ to the algebra of weight zero boundary fields. Since $\hat\cB$ is by assumption semisimple, so is its image under a homomorphism.	
} 
Again by the super-Wedderburn theorem, this algebra of boundary fields decomposes into a direct sum of $\mathbb{C}$'s and $C\ell_1$'s. We will call a boundary condition of the fermionic CFT {\em elementary} if its weight zero boundary fields consist of either exactly $\mathbb{C}$ or exactly $C\ell_1$. Note that in the standard basis of $\mathbb{C}$ and $C\ell_1$, the boundary structure constants $c_{xy}^{~~z}$ in \eqref{eq:BB} are either $0$ or $1$.

In the case of $C\ell_1$ there is a weight zero boundary fermion $a$
which satisfies $a^2=1$. Such weight-zero fermions are well-known
and appear in treatments of the boundary Ising model, for
example in~\cite{GZ,Chatterjee:1993ca,Toth:2006tj,Konechny:2018ujl}. 

We note that as an ungraded algebra, $C\ell_1$ is isomorphic to
$\mathbb{C} \oplus \mathbb{C}$, but the corresponding generators would
not have a fixed parity. Consequently, it is not possible to split a
boundary condition with weight zero field content $C\ell_1$ into two
more elementary boundaries without breaking parity-preservation of the
bulk-boundary OPE. 

We conclude that the indices $\alpha$ in \eqref{eq:hatB-sum-decomp}
label the different elementary boundary conditions: 
Each pair $\{ f_\beta, a_\beta\}$ corresponds to a boundary
condition which supports an odd weight 0 field, while  
each generator $e_\alpha$ corresponds to a boundary
condition which does not.
The bulk-boundary structure constants are given by the action of the
algebra elements $B^i$ on the generators,
\be\label{eq:class-alg-and-bb-couplings}
B^i e_\alpha = B^{(\alpha)\,i}_1\, e_\alpha
\ , \quad
B^i f_\beta
= B^{(\beta)\,i}_1\, f_\beta + B^{(\beta)\,i}_a\, a_\beta
\ , \quad
B^i a_\beta = B^{(\beta)\,i}_1\, a_\beta + B^{(\beta)\,i}_a\,
f_\beta
\ .
\ee
Here, we added the superscripts $(\alpha)$, $(\beta)$ to distinguish different solutions to \eqref{eq:BB}.

Analogous to the bosonic case, there is a relation between elementary boundary conditions and representations of the fermionic boundary classifying algebra. Namely, 
\begin{align}
&\{ \text{ elementary boundary conditions } \}
\nonumber \\
&\qquad\cong
\{ \text{ irreducible $\mathbb{Z}_2$-graded $\hat\cB$-modules that are submodules of $\hat\cB$ } \}
\nonumber \\
&\qquad\cong
\{ \text{ irreducible $\mathbb{Z}_2$-graded $\hat\cB$-modules up to even or odd isomorphisms } \} \ .
\label{eq:el-bc-vs-irrep}
\end{align}
Both equivalences follow from the description of elementary boundary conditions in terms of direct summands in \eqref{eq:hatB-sum-decomp}. Note, however, that elementary boundary conditions {\em do not} correspond to irreducible $\mathbb{Z}_2$-graded $\hat\cB$-modules up to only even isomorphism (this would count summands $\mathbb C$ with a factor of two and summands $C\ell_1$ with a factor of one)\footnote{Nonetheless, we will see in \cite{RSW-prep} that it is in fact natural to say that each elementary boundary condition comes in two varieties, which differ by an overall sign in the boundary state (this then counts {\em each} elementary boundary condition with a factor of two).}.

As can be seen from the structure of the boundary classifying algebra
of fermionic theories, restricting to the bosonic generators and
removing the fermionic generators reduces each copy of $C\ell_1=\mathbb C^{1|1}$ to just
$\mathbb C$. In this way we see that the elementary boundary conditions of a
fermionic theory are in 1-1 correspondence with the elementary
boundary conditions of its bosonic projection (the related ``spin
theory''); consideration of the fermionic generators allows one to see
which boundary conditions support a fermionic weight zero field and
are hence ``supersymmetric'' as we discuss below.

\subsection{Bulk-defect OPE and defect classifying algebra}\label{sec:def-class-alg}

Topological defects in a conformal field theory $\cal C$ have been long known
to be equivalent to a particular class of conformal
boundary conditions (a special case of so-called permutation boundary conditions \cite{Recknagel:2002qq}) 
on the doubled model 
$\cal C \times \overline{\cal C}$, 
and so can be studied from that perspective. It can, however,
also be useful to consider them in their own right and that is what we
do here. 

A topological defect will always have a non-trivial set of defect
fields, as it will always support the full set of bulk fields; in the
case of the trivial identity defect this is exactly the space of
defect fields, but in general it will be larger. 

It does not add any difficulty to generalise the situation slightly to topological interfaces with a CFT $\cal C$ above the interface and a CFT $\cal C'$ below the interface. Below we will sometimes
use the term defect to include both situations, with $\cal C$ and $\cal C'$ being equal or different.

The sewing constraint for topological defects is almost identical to
the bulk crossing symmetry constraint \eqref{eq:bulk-crossing}, with the replacement of some bulk
structure constants by bulk-defect structure constants and
defect-defect structure constants.
We shall here focus only on weight zero fields on the defect as this
is sufficient to derive the defect classifying algebra.
The bulk-defect structure constants in the OPE of two bulk fields
$\phi_i$ (above the defect) and $\phi_k'$ (below the defect)
to a weight zero defect field
$\vartheta_z$
will be denoted by $D^{ik}_z$
and the OPE of weight-zero defect fields
will be taken to 
have structure constants $c_{xy}{}^z$:
\begin{align}
&\phi_i(a + ib)\, \phi'_k(a - ib) ~=~ \delta_{h_i, h_k}
\delta_{\overline h_i,\overline h_k}\,\frac{D^{ik}_z}{|b|^{2\Delta_i}} \,\vartheta_z(a) + \ldots \quad ,
\nonumber \\	
&\vartheta_x \vartheta_y \,=\, c_{xy}{}^z \vartheta_z
\ .
\end{align}
The result is the defect sewing constraint
\be\label{eq:defect}
D^{ik}_{~x} \, D^{j\ell}_{~y} \, c_{xy}{}^z \,
=
(-1)^{(\pb{j} + \nu_j)\,\pb{k}}	
\,
\sum_{p,q} e^{\pi i(S_i+S_\ell-S_p)} 
\,
F_{p,0}\big[ \begin{smallmatrix} j&\ell \\ i & k \end{smallmatrix} \big]
\,
F_{\bar p,0}\big[ \begin{smallmatrix} \bar \jmath & \bar \ell \\ \bar \imath & \bar k \end{smallmatrix} \big]
\,
\SC_{ij}^{~p}  \, 
\SC'\!\!{}_{k\ell}^{~\,q} 
\, D^{pq}_{~z} \ .
\ee
As in the boundary case, we can consider the couplings to the weight
zero fields as generators $D^{ik}$ of an algebra $\hat\cD$ -- the {\em fermionic defect classifying algebra} -- defined in terms of the constants on the right hand side of \eqref{eq:defect},
\be\label{eq:defect-algebra}
 D^{ik} \,  D^{j\ell}
=
(-1)^{(\pb{j} + \nu_j)\,\pb{k}}	
\,
\sum_{p,q} e^{\pi i(S_i+S_\ell-S_p)} 
\,
F_{p,0}\big[ \begin{smallmatrix} j&\ell \\ i & k \end{smallmatrix} \big]
\,
F_{\bar p,0}\big[ \begin{smallmatrix} \bar \jmath & \bar \ell \\ \bar \imath & \bar k \end{smallmatrix} \big]
\,
\SC_{ij}^{~p}  \, \SC'\!\!{}_{k\ell}^{~\,q}  \, D^{pq}
\ .
\ee
Here, the pairs $ik$, $j\ell$ and $pq$ have to satisfy the condition that for each pair $xy$ the bulk fields $\phi_x$ and $\phi'_y$ have the same left conformal weight and the same right conformal weight. This implies, for example, that $F_{p,0}\big[ \begin{smallmatrix} j&\ell \\ i & k \end{smallmatrix} \big] = F_{p,0}\big[ \begin{smallmatrix} j& j \\ i & i \end{smallmatrix} \big]$.

The algebra $\hat\cD$ is commutative and associative (at least in fermionic rational CFTs \cite{RSW-prep}). We take the generators
$D^{ij}$ to inherit the parity of the product of fields
$\phi_i\phi_j'$, so that $\hat\cD$ is also a super-algebra (not necessarily super-commutative).
As in the boundary situation, we assume that $\hat\cD$ is semisimple, in which case it decomposes into a direct sum of copies of $\mathbb C$ and $C\ell_1$ which are in 1-1 correspondence with the solutions of the defect sewing constraints and
hence in correspondence with the defect
operators (up to an overall sign) and elementary defect conditions.

In the case $\cal C = \cal C'$, one generic solution for the
$D^{ik}_{~z}$ is provided by the trivial defect. The only weight zero
field on the trivial defect is the identity bulk field $1$, and we can
set $D^{ik}_{~1} = \SC_{ik}^{~1}$. The bulk-defect crossing relation
\eqref{eq:defect} then turns into a special case of the bulk crossing
relation \eqref{eq:bulk-crossing}. 

Interfaces between fermionic and bosonic theories can also be treated
by the above classifying algebra by simply choosing one of the two
CFTs to be purely even. 

In the purely bosonic case, classifying algebras for topological defects were studied in \cite{FSS07}. 

As opposed to case of boundary conditions, the number of defects of a
fermionic theory is larger than the number of defects of its bosonic
projection since there is a bosonic generator $D^{\phi\phi}$ of ${\cal D}$
for each field $\phi$, whether it is bosonic or fermionic. Each defect
of the bosonic theory can be associated with one or more defects in
the fermionic theory (as it happens, in the examples we have looked
at, each bosonic defect is associated to two defects in the fermionic
theory).

\subsection{Classifying algebras in the parity shifted theory}\label{sec:class-parityshift}

When comparing the various classifying algebras between a fermionic
theory and its parity-shifted version, one finds that they are
isomorphic as ungraded algebras, and that one can give an isomorphism
by a simple rescaling of the generators. 

However, we stress that the fermionic classifying algebras will in
general {\em not} be isomorphic as super-algebras (with
$\mathbb{Z}_2$-grading given by parity). We will see this explicitly
in the example of $\FM(3,8)$ in section~\ref{sec:susyly}. 

\subsubsection{Fermionic boundary classifying algebra}

Let us write $B^i$ for the generators in the unshifted theory as in
section~\ref{sec:ferm-class-alg} and denote the structure constants of
the fermionic boundary classifying algebra by $\beta_{ij}^{~k}$, such
that \eqref{eq:BB'} becomes $B^i B^j = \sum_k \beta_{ij}^{~k} B^k$.  
The bulk structure constants in the parity-shifted theory are given in
\eqref{eq:SC-R-parity-shift}. If we denote the generators and
structure constants of the classifying algebra of the parity shifted
theory by $\widetilde B^i$ and $\widetilde\beta_{ij}^{~k}$, we get 
\be
\widetilde B^i \, \widetilde B^j ~=~ \sum_k \widetilde\beta_{ij}^{~k} \, \widetilde B^k
\qquad
\text{with}
\quad
\widetilde\beta_{ij}^{~k} = (-1)^{\nu_i \phi_j} \beta_{ij}^{~k} \ .
\ee
In general, the factor $(-1)^{\nu_i \phi_j}$ cannot be absorbed into a rescaling,\footnote{ 
	The reason for this is that $c_{i,j} = (-1)^{\nu_i \phi_j}$
	is a non-exact two-cocycle on $\mathbb{Z}_2 \times \mathbb{Z}_2$ as witnessed by the fact that it is not symmetric in $i,j$.}
but since certain of the $\beta_{ij}^{~k}$ are zero, here this will be possible. Explicitly, we may identify $\widetilde B^i = \lambda_i B^i$ to get
\be
	\widetilde\beta_{ij}^{~k}
	~=~
	\frac{\lambda_i \,\lambda_j}{\lambda_k} \,\beta_{ij}^{~k}
\qquad
\text{with}
\quad
	\lambda_x = e^{\frac{\pi i}2 \phi_x} \ .	
\label{eq:bnd-class-alg-shift-ident}
\ee
The scalar coefficients $\lambda$ are not unique, and we just exhibit one possible solution.
To verify the above equality, first note that $\lambda_i \lambda_j/\lambda_k = (-1)^{\phi_i\phi_j}$. One thus
needs to check that $(-1)^{\nu_i \phi_j}=(-1)^{\phi_i \phi_j}$ whenever $\beta_{ij}^{~k} \neq 0$. Since all NS-sector generators in the boundary classifying algebra are parity-even, we only need to consider the case that $B^i$ and $B^j$ are from the R-sector, and that $B^i$ is even and $B^j$ is odd. But then $\beta_{ij}^{~k} = 0$, as $B^k$ would then need to be odd and in the NS-sector.

\subsubsection{Fermionic defect classifying algebra}

Let us use the letter ``$U$'' for the unshifted theory and ``$P$'' for the parity shifted theory. Here we will give the relation between the defect classifying algebras of types $U$-$U$, $U$-$P$, $P$-$U$ and $P$-$P$, 
where the first letter refers to the theory in the upper half plane.
The $U$-$U$-case is as given in section~\ref{sec:def-class-alg} and will be our reference case:
\be
D^{ik} \,  D^{j\ell}
=
\sum_{p,q} 
\beta^{(UU)\,pq}_{ik,j\ell}
\, D^{pq} \ ,
\ee
with $\beta^{(UU)\,pq}_{ik,j\ell}$ determined by \eqref{eq:defect-algebra}.
A computation similar to the boundary classifying algebra gives
(all parities are stated with respect to the unshifted theory)
\begin{align}
\beta^{(UP)\,pq}_{ik,j\ell}
&~=~ (-1)^{\nu_k(\phi_j+\phi_\ell + \nu_\ell)} \beta^{(UU)\,pq}_{ik,j\ell}
~=~ \frac{\lambda_{ik}\,\lambda_{j\ell}}{\lambda_{pq}} \, \beta^{(UU)\,pq}_{ik,j\ell}
~,
&&\lambda_{xy} = e^{\frac{\pi i}2 (\phi_x+\phi_y-2\phi_x\phi_y+\nu_y)}
~,
\nonumber\\
\beta^{(PU)\,pq}_{ik,j\ell}
&~=~ (-1)^{\nu_i \phi_j + \nu_\ell\phi_k} \beta^{(UU)\,pq}_{ik,j\ell}
~=~ \frac{\lambda'_{ik}\,\lambda'_{j\ell}}{\lambda'_{pq}} \, \beta^{(UU)\,pq}_{ik,j\ell}
~,
&&\lambda'_{xy} = e^{\frac{\pi i}2 (\phi_x+\phi_y-2\phi_x\phi_y)} \, e^{2 \pi i S_y}
~,
\nonumber\\
\beta^{(PP)\,pq}_{ik,j\ell}
&~=~ (-1)^{\nu_\ell \phi_k + \nu_k \phi_\ell} \beta^{(UU)\,pq}_{ik,j\ell}
~=~ \frac{\lambda''_{ik}\,\lambda''_{j\ell}}{\lambda''_{pq}} \, \beta^{(UU)\,pq}_{ik,j\ell}
~,
&&\lambda''_{xy} = e^{2 \pi i S_y} \ .
\label{eq:defect-class-ident}
\end{align}
To verify these equations one needs to use that for a generator $D^{xy}$ one always has $\nu_x = \nu_y$ (but not necessarily $\phi_x=\phi_y$) as otherwise the bulk-defect OPE does not contain $1$ or $a$.

\section{Fermionic Virasoro minimal models}
\label{sec:fermVir}

The examples we shall consider are all Virasoro minimal, i.e.\ the
Hilbert space is formed from a finite set of Virasoro
representations. This might seem odd, as it might be more natural to
start with a super-algebra for which there is naturally a fermionic
interpretation, but as we shall see it provides ample examples, not
only infinite series which extend the bosonic Virasoro minimal models
but also the simplest example, a free fermion, as well as examples
with super-Virasoro symmetry and extended (W-algebra type) fermionic
algebras.

In this section we give the bulk structure constants for fermionic A-
and D-type minimal models, in section~\ref{sec:Ising} we treat the
free fermion in detail, and in section~\ref{sec:furtherVir} we give
further Virasoro examples. 

Consider the Virasoro algebra at the minimal model central charge
$c(p,q)=1-6(p-q)^2/(pq)$ with
$p,q$ coprime integers greater than 1. 
We recall that the possible Virasoro representations are labelled by
two integers $(r,s)$ where $1\leq r <p$  and $1\leq s < q$ with the
identification
$(r,s)\simeq(p-r,q-s)$; for more details see \cite{YBk}.

The fermionic generator $G$ we consider has Kac-labels $(1,q-1) \sim
(p-1,1)$ and is the unique non-trivial simple current at that central
charge.  Note that for $G$ to indeed be different from the vacuum
representation we actually need $p,q>2$. The conformal weight of $G$
is $h_G = \frac14(p-2)(q-2)$. The condition that $h_G \in
\mathbb{Z}+\frac12$ amounts to
\be\label{eq:pq-FM}
	p=2n+1~,~q=4k \ , \qquad \text{or equivalently} ~~ p=4k~,~q=2n+1 \ ,
\ee
$k,n \ge 1$,
in which case $h_G = \frac 12 (2n-1)(2k-1)$. 
The first few values of $h_G$ are realised in the following models:
\be\label{eq:models-by-weight}
{\renewcommand{\arraystretch}{1.4}
\begin{array}{c|c|c|c|c|c|c}
h_G & \frac12  & \frac32  & \frac52  & \frac72  & \frac92 & \dots
\\
\hline
(p,q) & (3,4) & (3,8),\,(4,5) & \raisebox{.35em}{\rule{3em}{0.7pt}}\hspace{-2.8em}(3,12),\,(4,7) & (3,16),\,(4,9) & (3,20),\,(4,11),\,(5,8)
&\dots
\\
\hline
c(p,q) & \frac12 
& -\frac{21}{4} ~~,~~ \frac{7}{10} 
& \phantom{-\frac{21}{4}} ~~~ -\frac{13}{14}
& -\frac{161}{8} ~,~ -\frac{19}{6} 
& -\frac{279}{10} \,,~ -\frac{125}{22}  \,,~ -\frac{7}{20}
& \dots
\end{array}
} 
\ee The next value of $h_G$ realised in a unitary model is $h_G =
\frac{15}2$ at central charge $c(7,8) = \frac{25}{28}$.

Recall that the possible different bosonic field theories with this
central charge are labelled by a pair of simply-laced Lie algebras
with Coxeter numbers $p,q$ \cite{Cappelli:1987xt}.  We will give two
explicit solutions to the fermionic crossing relation
\eqref{eq:bulk-crossing}.  One is an extension of the minimal model of
type $M(A_{p-1},A_{q-1})$ which we will call the fermionic Virasoro
minimal model $\FM(A_{p-1},A_{q-1})$ or $\FM(p,q)$ for short. The
other is an extension of the minimal model $M(A_{p-1},D_{q/2+1})$
which we will call $\FM(A_{p-1},D_{q/2+1})$ or $\widetilde\FM(p,q)$
for short.  By an extension we mean that the fermionic model contains
the full bosonic field theory as a sub-theory. The two models
$\FM(p,q)$ and $\widetilde\FM(p,q)$ are obtained from each other by
shifting the Ramond sector parity as in section~\ref{sec:shiftRamond}.

\subsection{A-type fermionic models}
\label{ssec:fvmm}

Let $\mathcal{I}$ be an indexing set for the Kac-table modulo its $\mathbb{Z}_2$-identification $(r,s) \sim (p-r,q-s)$,
and let $M_a$, $a \in \mathcal{I}$ be the corresponding irreducible Virasoro representation.
The splitting into NS- and R-sector depends on the sign of the ratio of $S$-matrices $S_{G,a}/S_{G,0} \in \{ \pm 1 \}$, with $+1$ being the NS-sector and $-1$ the R-sector. Explicitly, for $a=(r,s)$ we have $S_{G,a}/S_{G,0}=(-1)^{qr+ps+1}$ so that the set $\mathcal{I}$ splits as
\be
\mathcal{I}_{\mathrm{NS}} = \big\{\, (r,s) \in \mathcal{I} \,\big|\, qr+ps \text{ odd} \, \big\}
\quad ,
 \qquad
\mathcal{I}_{\mathrm{R}} = \big\{\, (r,s) \in \mathcal{I} \,\big|\, qr+ps \text{ even} \, \big\} \ .
\ee
The state space $\cH_F$ of $\FM(A_{p-1},A_{q-1})$ splits into the four sectors in \eqref{eq:HF-foursector} as follows:
\begin{align}
	\cH_F^\mathrm{NS,ev} &:=  \bigoplus_{a \in \mathcal{I}_{\mathrm{NS}}} M_a \otimes \overline M_a \ ,
	& \cH_F^\mathrm{R,ev}  &:= \bigoplus_{a \in \mathcal{I}_{\mathrm{R}}} M_a \otimes \overline M_a \ ,
	\nonumber \\	
	\cH_F^\mathrm{NS,odd} &:= \bigoplus_{a \in \mathcal{I}_{\mathrm{NS}}} M_a \otimes \overline M_{Ga} \ ,
	&\cH_F^\mathrm{R,odd} &:= \bigoplus_{a \in \mathcal{I}_{\mathrm{R}}} M_a \otimes \overline M_{Ga} \ ,
\end{align}
where $Ga$ is the result of the fusion product of $G$ and $a$. That is, for $a=(r,s)$ we have $Ga = (r,q-s) \sim (p-r,s)$. 
Note that $\cH_F^\mathrm{NS,odd}$ contains a holomorphic field of weight $(h_G,0)$ and an anti-holomorphic one of weight $(0,h_G)$. 

We will use the notation $\phi_a^e$ for the primary field in $\cH_F^\mathrm{ev}$ of conformal weights $(h_a,h_a)$, and $\phi_a^o$ for the primary field in $\cH_F^\mathrm{odd}$ of conformal weights $(h_a,h_{Ga})$. This notation is slightly asymmetric in that $\phi_G^o$ is the field of weight $(h_G,0)$ while $\phi^o_1$ is that of weight $(0,h_G)$.
Suppose we take $q=4k$ in \eqref{eq:pq-FM}. Then $Ga=a$ if and only if $a = (r,\frac q2) = (r,2k)$, so that for Kac labels of this form we have $a \in \cI_\mathrm{R}$ and fields with conformal weights $(h_a,h_a)$ occur with multiplicity two, once as $\phi^e_a \in  \cH_F^\mathrm{R,ev}$ and once as $\phi^o_a \in  \cH_F^\mathrm{R,odd}$.

For the structure constants we use the notation
\be
	\phi_a^\alpha(x) \, \phi_b^\beta(y)
	~\sim~ 
	\sum_{c \in \mathcal{I}} \SC_{ab}^{(\alpha\beta)\,c}
	\, 
	(x-y)^{\Delta_c-\Delta_a-\Delta_b} 
	\,
	\phi_c^{\alpha+\beta} \ ,
\ee
where $\alpha,\beta \in \{e,o\}$ and $\alpha+\beta$ stands for the
parity of the product. One solution to the bulk crossing relation \eqref{eq:bulk-crossing} is given by 
\begin{align}
\SC_{ab}^{(ee)\,c} &= 
\frac{\lambda_a^e \, \lambda_b^e}{\lambda_c^e}\, 
F_{0,c}\big[ \begin{smallmatrix} a&b\\a&b \end{smallmatrix} \big] \ ,
&
\SC_{ab}^{(oo)\,c} &= 
\frac{\lambda_a^o \, \lambda_b^o}{\lambda_c^e}\,
e^{\pi i (h_{Ga} - h_a - h_{G})} F_{G,c}\big[ \begin{smallmatrix} a&b\\Ga&Gb \end{smallmatrix} \big]
\ ,
\nonumber \\
\SC_{ab}^{(oe)\,c} &= 
\frac{\lambda_a^o \, \lambda_b^e}{\lambda_c^o}\,
F_{0,c}\big[ \begin{smallmatrix} a&b\\a&b \end{smallmatrix} \big] \, F_{a,Gc}\big[ \begin{smallmatrix} Ga&b\\G&c \end{smallmatrix} \big] \ ,
&
\SC_{ab}^{(eo)\,c} &=
e^{\pi i ( (h_c-h_{Gc}) - (h_b-h_{Gb}))} \, \SC_{ba}^{(oe)\,c} \ .
\label{eq:cdefs}
\end{align}
The $\lambda^{e/o}_i \in \mathbb{C}^\times$ are normalisation constants which can be chosen at will. 
The proof that these constants indeed solve the crossing constraint will be given in \cite{RSW-prep}. There it will also be shown that 
one can find a topological defect $F$ with the required properties.\footnote{
	In the setting of \cite[Sec.\,5]{Novak:2015ela}, this amounts to $F$ being the superposition of the identity defect and a parity shifted version of the topological defect labelled by the representation $G$. (What we call $F$ here is called $A$ in \cite{Novak:2015ela}.)
}

One standard normalisation is to make a choice of square roots
\be
	\lambda^e_a = \Big( \SC_{aa}^{(ee)\,1}\big|_{\lambda=1} \Big)^{-\frac12}
	\quad , \quad
	\lambda^o_a = \Big( \SC_{aa}^{(oo)\,1}\big|_{\lambda=1} \Big)^{-\frac12} \ .
\ee
With this choice one has $\SC_{aa}^{(ee)\,1}=\SC_{aa}^{(oo)\,1}=1$.

This concludes the definition of the fermionic minimal model  $\FM(A_{p-1},A_{q-1})=\FM(p,q)$, where $p,q$ are as in \eqref{eq:pq-FM}.

\subsubsection*{Bosonic subtheory}

The bosonic subtheory of $\FM(p,q)$ is the restriction to the parity even subspace of $\cH_F$. This can alternatively be understood as the result of summing over spin structures. Explicitly we have
\be
   \cH_F^\mathrm{ev} =  \bigoplus_{a \in \mathcal{I}} M_a \otimes \overline M_a \ ,
\ee
which agrees with the state space of the A-type (bosonic) minimal model $M(A_{p-1},A_{q-1})$. Restricting the structure constants to
the even subsector and choosing the normalisation $\lambda^e_a = S_{0a} / S_{00}$, 
the solution \eqref{eq:cdefs} precisely recovers the bulk structure constants for the A-type models in terms of $F$-matrices as given in \cite{runkel},
\be
 C_{ab}{}^c = \frac{S_{0a} S_{0b} }{ S_{00} S_{0c} } 
 \,
 F_{0c}\big[\begin{smallmatrix} a & b \\ a &  b \end{smallmatrix}\big]
\;.
\ee
Note that with these conventions, the coupling to the identity field 
is not normalised to 1.

\subsection{D-type fermionic models}
\label{ssec:fvmm-D}

We now consider the theory $\widetilde\FM(p,q)$ obtained from $\FM(p,q)$ by shifting parity in the Ramond sector as in section~\ref{sec:shiftRamond}. We will assume that $p$ is odd and $q$ is even, which according to \eqref{eq:pq-FM} implies that $q \in 4 \mathbb{Z}$.
We will also denote $\widetilde\FM(p,q)$ as $\FM(A_{p-1},D_{q/2+1})$ which will be justified later by the restriction to the bosonic subtheory.
The state space $\widetilde\cH_F$ of $\widetilde\FM(p,q)$ splits as
\begin{align}
	\widetilde\cH_F^\mathrm{NS,ev} &:=  \bigoplus_{a \in \mathcal{I}_{\mathrm{NS}}} M_a \otimes \overline M_a \ ,
	& \widetilde\cH_F^\mathrm{R,ev}  &:= \bigoplus_{a \in \mathcal{I}_{\mathrm{R}}} M_a \otimes \overline M_{Ga} \ ,
	\nonumber \\	
	\widetilde\cH_F^\mathrm{NS,odd} &:= \bigoplus_{a \in \mathcal{I}_{\mathrm{NS}}} M_a \otimes \overline M_{Ga} \ ,
	&\widetilde\cH_F^\mathrm{R,odd} &:= \bigoplus_{a \in \mathcal{I}_{\mathrm{R}}} M_a \otimes \overline M_{a} \ .
\end{align}
As opposed to $FM(p,q)$, shifting the antiholomorphic label by $G$ or not no longer corresponds to the field being odd. We therefore label the fields as, for $a \in \mathcal{I}$,
\be
	\tilde\phi^u_a
	 \quad \text{: weight }
	(h_a,h_a)
	\qquad
	\tilde\phi^s_a \quad \text{: weight }(h_a,h_{Ga})	
\ee
where $u$ stands for ``unshifted'' and $s$ for ``shifted'' (referring to the shift by $G$). The spin grading of $\tilde\phi_a^{u/s}$ is $\nu_a$ as before, but the parity has changed,
\be
	|\tilde\phi^u_a| = \nu_a 
	\quad , \quad
	|\tilde\phi^s_a| = \nu_a + 1 \ .	
\ee
As in the A-type case, for $a=(r,\frac q2)$ we have $a \in \cI_\mathrm{R}$, $Ga=a$, and both $\tilde\phi^u_a \in \widetilde\cH_F^\mathrm{R,odd}$ and $\tilde\phi^s_a \in \widetilde\cH_F^\mathrm{R,ev}$ have conformal weights $(h_a,h_a)$, so that these weights occur with multiplicity two.

The structure constants of $\widetilde\FM(p,q)$ are obtained from \eqref{eq:cdefs} by the transformation \eqref{eq:SC-R-parity-shift}:
\begin{align}
	\widetilde C_{ab}^{(uu)\,c} &= 
	\frac{\lambda_a^u \, \lambda_b^u}{\lambda_c^u}\, 
	F_{0,c}\big[ \begin{smallmatrix} a&b\\a&b \end{smallmatrix} \big] \ ,
	&
	\widetilde C_{ab}^{(ss)\,c} &= 
\frac{\lambda_a^s \, \lambda_b^s}{\lambda_c^u}\,
(-1)^{\nu_a}\,
e^{\pi i (h_{Ga} - h_a - h_{G})} F_{G,c}\big[ \begin{smallmatrix} a&b\\Ga&Gb \end{smallmatrix} \big]
\ ,
\nonumber \\
	\widetilde C_{ab}^{(su)\,c} &= 
\frac{\lambda_a^s \, \lambda_b^u}{\lambda_c^s}\,
F_{0,c}\big[ \begin{smallmatrix} a&b\\a&b \end{smallmatrix} \big] \, F_{a,Gc}\big[ \begin{smallmatrix} Ga&b\\G&c \end{smallmatrix} \big] \ ,
	&
	\widetilde C_{ab}^{(us)\,c} &=
(-1)^{\nu_a}\,
	e^{\pi i ( (h_c-h_{Gc}) - (h_b-h_{Gb}))} \, \widetilde C_{ba}^{(su)\,c} \ .
	\label{eq:cdefs-D}
\end{align}

\subsubsection*{Bosonic subtheory}

The even subspace of $\widetilde\cH_F$ is
\be
\widetilde\cH_F^\mathrm{ev} ~=~  \bigoplus_{a \in \mathcal{I}_{\mathrm{NS}}} \hspace{-0.2em} M_a \otimes \overline M_a 
~\oplus~
\bigoplus_{a \in \mathcal{I}_{\mathrm{R}}} \hspace{-0.2em} M_a \otimes \overline M_{Ga}
\ .
\ee
which is the state space of the D-type minimal model $M(A_{p-1},D_{q/2+1})$.
Note that since $q \in 4\mathbb{Z}$, this model is always of $D_\mathrm{odd}$-type (a permutation modular invariant).
When restricting the structure constants to the even subsector, i.e.\ to $\phi^u_a$, $a \in \mathcal{I}_{\mathrm{NS}}$ and $\phi^s_a$, $a \in \mathcal{I}_{\mathrm{R}}$, the expression \eqref{eq:cdefs-D} reproduces the structure constants of the D-series bosonic Virasoro minimal model found in \cite{Runkel:1999dz}.

Since the comparison requires a bit of calculation, we give some details. The expression in \cite{Runkel:1999dz} is
\begin{align}
C_{i_\alpha\,j_\beta}^{~~~m_\gamma} ~=~ 
&\exp\big(i\tfrac\pi{2}(h_{\omega_\gamma}{+}h_{\omega_\alpha}{-}h_{\omega_\beta}{+}2(h_j{-}h_r)
{+}h_m{-}\bar h_m{-}h_i{+}\bar h_i{-}h_j{+}\bar h_j)\big)
\nonumber \\
&\times
\frac{{}^{(\omega)}B_{i_\alpha}^{~\omega_\alpha} \,
	{}^{(\omega)}B_{j_\beta}^{~\omega_\beta} }{ {}^{(\omega)}B_{m_\gamma}^{~\omega_\gamma} }
\,
F_{\mu_u,\omega_\gamma}\big[ \begin{smallmatrix} \mu_u & \mu_u \\ \omega_\alpha & \omega_\beta \end{smallmatrix} \big]
\,
F_{\omega_\beta,r}\big[ \begin{smallmatrix} \omega_\gamma & \bar j \\ \omega_\alpha & j \end{smallmatrix} \big] 
\,
F_{\omega_\alpha, m}\big[ \begin{smallmatrix} \bar i & r \\ i & j  \end{smallmatrix} \big] 
\,
F_{r, \bar m}\big[ \begin{smallmatrix} \bar i & \bar j \\ m & \omega_\gamma  \end{smallmatrix} \big] 
\ .
\label{eq:Dodd-bos-sc}
\end{align}
The notation is as follows. The indices $\alpha,\beta,\gamma$ take values in $\{e,o\}$. The index $i_e$ corresponds to the field $\phi_i^u$ and $i_o$ to $\phi_i^s$. 
Furthermore, $\omega_e=(1,1)$ and $\omega_o=G$. For $i_e$ we have $\bar h_i = h_i$ and for $i_o$ we have $\bar h_i = h_{Gi}$. The label $r$ is given by $r=j$ if $\alpha=e$ and $r=Gj$ if $\alpha{=}o$. 
The constants ${}^{(\omega)}B_{i_\alpha}^{~\omega_\alpha}$, etc., are certain bulk-boundary structure constants which will be absorbed into the normalisation coefficients (together with a phase for the $G$-shifted fields),
\be
	\lambda_a^u = {}^{(\omega)}B_{a_e}^{~\omega_e}
	\quad , \quad
	\lambda_a^s = {}^{(\omega)}B_{a_o}^{~\omega_o} \,\, 
e^{i\frac\pi2 (h_a - h_{Ga}-h_G)}
	\ .
\ee
The label $\mu_u$ is the Kac-label $(1,\frac q2)$, which is a fixed
point for $G$. The $F$-matrix entry
$F_{\mu_u,\omega_\gamma}\big[ \begin{smallmatrix} \mu_u & \mu_u
    \\ \omega_\alpha & \omega_\beta \end{smallmatrix} \big]$ is either
$0$ or $1$, depending on whether the $\mathbb{Z}_2$ fusion rules in
the OPE are obeyed. This grading rule holds by construction for
\eqref{eq:cdefs-D}, so that this $F$-matrix coefficient can be
dropped. One can now check sector by sector that the restriction of
\eqref{eq:cdefs-D} to the even subsector agrees with
\eqref{eq:Dodd-bos-sc}. In sector ``$su$'', i.e.\ for $\alpha=o$,
$\beta = e$, this requires an $F$-matrix identity: 
\begin{align}
&F_{0,Gb}\big[ \begin{smallmatrix}  G & b \\ G & b \end{smallmatrix} \big] 
\,
F_{G,c}\big[ \begin{smallmatrix} Ga & Gb \\ a & b \end{smallmatrix} \big] 
\,
F_{Gb,Gc}\big[ \begin{smallmatrix} Ga & b \\ c & G \end{smallmatrix} \big] 
\,
e^{\pi i (h_G+h_b-h_{Gb})}
\nonumber \\
&=~
F_{0,c}\big[ \begin{smallmatrix} a & b \\ a & b \end{smallmatrix} \big] 
\,
F_{a,Gc}\big[ \begin{smallmatrix} Ga & b \\ G & c \end{smallmatrix} \big] 
\,
e^{\pi i ((h_G+h_a-h_{Ga})-(h_G+h_c-h_{Gc}))}
\ .
\end{align}
Finally, note that since we restrict to the even subsector, for $\phi^u_a$ we have $\nu_a=0$ while for $\phi_a^s$ we have $\nu_a=1$. Accordingly, the factor $(-1)^{\nu_a}$ is equal to $1$ in $\widetilde C_{ab}^{(su)\,c}$ and equal to $-1$ in $\widetilde C_{ab}^{(ss)\,c}$. In the latter case, the minus sign cancels against $e^{-2\pi i h_G}$.

\medskip

Finally, let us note that fermionic minimal model $\FM(p,q)$ and its
parity shifted cousin $\widetilde\FM(p,q)$ are non-isomorphic (as
graded theories) whenever the corresponding $D_\text{odd}$-diagram
differs from an $A$-diagram.  
Since $q \in 4\mathbb{Z}$, this happens for $q \ge 8$. On the
other hand, the fermionic Ising model $\FM(3,4)$, which we will treat
in detail in the next section, is isomorphic to $\widetilde\FM(3,4)$. 

\section{The Ising model and the free fermion}
\label{sec:Ising}

The Ising model is the Virasoro minimal model $M(3,4)$ and the
relevant data is given in appendix \ref{app:ising}.  
The fermionic Ising model $\FM(3,4)$ is really the theory of the free
fermion - the field $\phi_\epsilon^o = \psi$ is a free holomorphic
fermion and $\phi_1^o = \bar\psi$ is a free anti-holomorphic fermion. 
Altogether, the Virasoro primary fields in the fermionic model are:
\begin{equation}
\begin{array}{c|cc|c|c}
 \text{spinless} & \multicolumn{3}{c|}{\text{even}} & \text{odd}\\ 
 \text{fields} & \multicolumn{2}{c|}{\text{NS}} & \text{R} & \text{R}\\ 
     & 1  & \eps & \sigma & \mu 
 \\[0.1em] \hline &&&& \\[-0.9em]
 h = \bar h & 0 & \tfrac12 & \tfrac1{16}  & \tfrac1{16} 
\end{array}
\qquad
\begin{array}{c|cc}
	\text{fields} & \multicolumn{2}{c}{\text{odd}\,\&\,\text{NS}} \\ 
	\text{w.\ spin} &  \psi & \bar\psi
 \\[0.1em] \hline && \\[-0.9em]
	(h,\bar h) & (\tfrac12,0) & (0,\tfrac12)  
\end{array}
\end{equation}

\subsection{Bulk structure constants}\label{sec:IsingBulk}

The structure constants of the free fermion can be read off from
\eqref{eq:cdefs}, but in this simple model it is easy to compute them
directly from the crossing constraint \eqref{eq:bulk-crossing}, and
this is what we will do. 

For all fields we set $C_{\phi\phi}^{~~1} = 1$, which fixes all normalisations up to signs. The structure constants of the even fields are those of the Ising
model and are of course well-known: up to the symmetry properties \eqref{eq:cyclic} the only remaining structure constant is
\be\label{eq:Ising-bulk-sse}
C_{\sigma\sigma}^{~~\epsilon} = \frac 12 ~,
\ee
which also fixes the sign-freedom in the normalisation of $\eps$.
The structure constants involving even and odd fields in the fermionic 
model have been considered before and some are given in
\cite{Chatterjee:1993ca} and \cite[Sec.\,12.3.3]{YBk}, 
but these are only partial results and it is not clear how the
various signs were chosen nor how the full consistency could be
checked. 
Using our formalism, we state and solve the sewing constraints, and
taking the same normalisation of the primaries as \cite{YBk} we agree
with the partial results stated there.

The odd fields in the fermionic model are $\psi, \bar\psi$ and $\mu$. 
Setting $i=j=\psi$, $k=l=\bar\psi$ in bulk crossing relation \eqref{eq:bulk-crossing} results in 
$(\SC_{\psi\bar\psi}^{~~~\eps})^2 = - \SC_{\psi\psi}^{~~1} \SC_{\bar\psi\bar\psi}^{~~1} = -1$.
We link the remaining free signs in the normalisation of
$\psi$ and $\bar\psi$ by setting $\eps = i \psi \bar\psi$, or, in other words,  
\be\label{eq:Ising-p-pb-e}
\SC_{\psi\bar\psi}^{~~~\eps} = -i \ .
\ee

The symmetry properties \eqref{eq:cyclic} relate any two permutations of the three primaries in a structure constants. Up to such permutations, the remaining structure constants involving two odd fields (and no identity field) are
\be\label{eq:FF-indep-odd-SC}
\SC_{\mu\mu\eps} 
\quad , \quad
\SC_{\mu\psi\sigma} 
\quad , \quad
\SC_{\mu\bar\psi\sigma}
\ .
\ee
To fix their value, we consider the crossing constraint
\eqref{eq:bulk-crossing} for the following four choices of parameters:
a) $i=\psi$, $k=\psib$, $j=\ell=\sigma$ and $q=\eps$; b) the same with
$j=\ell=\mu$; c) $i=j=\sigma$, $k=l=\mu$, $q=\psi$; and finally d) the
same with $q=\bar\psi$, 
\begin{subequations}
\begin{align}
\label{eq:FF-cross-a}
\SC_{\psi \bar\psi}^{~~\eps} \, \SC_{\sigma\sigma}^{~~\eps} 
&~=
- e^{\pi i / 2} 
\,
\SC_{\psi \sigma}^{~~\mu}  \, \SC_{\psib \sigma}^{~~\mu} 
\ ,
\\
\label{eq:FF-cross-b}
\SC_{\psi \bar\psi}^{~~\eps} \, \SC_{\mu\mu}^{~~\eps} 
&~=~
e^{\pi i / 2} 
\,
\SC_{\psi \mu}^{~~\sigma}  \, \SC_{\psib \mu}^{~~\sigma} 
\ ,
\\
\label{eq:FF-cross-c}
\big(\SC_{\sigma\mu}^{~~\psi} \big)^2 
&~=
- e^{-\pi i/2}
\left(
\SC_{\sigma\sigma}^{~~1} \, \SC_{\mu\mu}^{~~1} \, 
F_{0\epsilon} \big [ \begin{smallmatrix} \sigma & \sigma\\\sigma&\sigma \end{smallmatrix} \big]
F_{00} \big [ \begin{smallmatrix} \sigma & \sigma\\\sigma&\sigma \end{smallmatrix} \big]
\;+\;
\SC_{\sigma\sigma}^{~~\epsilon} \, \SC_{\mu\mu}^{~~\epsilon} \, 
F_{\epsilon\epsilon} \big [ \begin{smallmatrix} \sigma & \sigma\\\sigma&\sigma \end{smallmatrix} \big]
F_{\epsilon0} \big [ \begin{smallmatrix} \sigma & \sigma\\\sigma&\sigma \end{smallmatrix} \big]
\right)
\ ,
\\
\label{eq:FF-cross-d}
\big(\SC_{\sigma\mu}^{~~\psib} \big)^2 
&~=
- e^{\pi i/2}
\left(
\SC_{\sigma\sigma}^{~~1} \, \SC_{\mu\mu}^{~~1} \, 
F_{0\epsilon} \big [ \begin{smallmatrix} \sigma & \sigma\\\sigma&\sigma \end{smallmatrix} \big]
F_{00} \big [ \begin{smallmatrix} \sigma & \sigma\\\sigma&\sigma \end{smallmatrix} \big]
\;+\;
\SC_{\sigma\sigma}^{~~\epsilon} \, \SC_{\mu\mu}^{~~\epsilon} \, 
F_{\epsilon\epsilon} \big [ \begin{smallmatrix} \sigma & \sigma\\\sigma&\sigma \end{smallmatrix} \big]
F_{\epsilon0} \big [ \begin{smallmatrix} \sigma & \sigma\\\sigma&\sigma \end{smallmatrix} \big]
\right)
\ .
\end{align}
\end{subequations}
Using \eqref{eq:cyclic} we get the following relations between the structure constants appearing in the above constraints and those listed in \eqref{eq:FF-indep-odd-SC},
\begin{align}
\SC_{\mu\mu}^{~~\eps} &= \SC_{\mu\mu\eps} 
~,~~&
\SC_{\psi \sigma}^{~~\mu} &= \SC_{\sigma\mu}^{~~\psi} = \SC_{\mu\psi \sigma}
~,~~&
\SC_{\psi \mu}^{~~\sigma} &= e^{-\pi i /2} \, \SC_{\mu\psi \sigma }
~,
\nonumber\\
&&
\SC_{\psib \sigma}^{~~\mu}  &= \SC_{\sigma\mu}^{~~\psib} = \SC_{\mu\psib \sigma} 
~,~~&
\SC_{\psib \mu}^{~~\sigma} &= e^{\pi i /2} \, \SC_{\mu\psib \sigma } \ .
\end{align}
Then, combining this with \eqref{eq:FF-cross-a}, \eqref{eq:FF-cross-b} and \eqref{eq:Ising-p-pb-e} shows (in agreement with \eqref{eq:Ising-bulk-sse})
\be
	\SC_{\mu\mu}^{~~\eps}  = - \, \SC_{\sigma\sigma}^{~~\eps}  = -\frac12 \ .
\ee
Substituting this into \eqref{eq:FF-cross-c} and \eqref{eq:FF-cross-d} gives
$\big(\SC_{\sigma\mu}^{~~\psi} \big)^2 = \frac{i}{2}$ and
$\big(\SC_{\sigma\mu}^{~~\psib} \big)^2  = -\frac{i}{2}$.
Conditions \eqref{eq:FF-cross-a}--\eqref{eq:FF-cross-d} leave an overall sign in the normalisation of $\psi$ and $\psib$ undetermined, and we choose
\be
\SC_{\mu\psi\sigma} = \frac{e^{\pi i/4}}{\sqrt 2}
\;,\;\;
\SC_{\mu\psib\sigma} = \frac{e^{-\pi i/4}}{\sqrt 2}
\ .
\ee
This agrees with the constants $\SC_{\psi\sigma}^{~~\mu}$, $\SC_{\psib\sigma}^{~~\mu}$, $\SC_{\psi\mu}^{~~\sigma}$ and  $\SC_{\psib\mu}^{~~\sigma}$ given in~\cite[Eq.\,(4)]{Chatterjee:1993ca}
and~\cite[Eq.\,(12.68)]{YBk}, as well as with \eqref{eq:cdefs} 
(the normalisation constants $\lambda_a^{e/o}$ are given in appendix \ref{app:ising}).

\medskip

Consider for a moment the parity shifted theory
$\widetilde\FM(3,4)$. Its structure constants are related to the ones
given above as in \eqref{eq:SC-R-parity-shift}. This turns out to be
an equivalent theory, and a choice of parity-grading preserving isomorphism
from $\widetilde\FM(3,4)$ (whose fields we denote by
$\widetilde\sigma$, $\widetilde\psi$, etc.) to $\FM(3,4)$ is
\begin{align}
&\text{NS-sector:}
&&\text{R-sector:}
\nonumber\\
&
\widetilde\eps \mapsto -\eps
~,~~
\widetilde\psi \mapsto \psi
~,~~
\widetilde{\bar\psi} \mapsto -\bar\psi
&&
\widetilde\sigma \mapsto \mu
~,~~
\widetilde\mu \mapsto -i \, \sigma \ .
\end{align}
This can be interpreted as the action of an invertible interface and agrees with the solution to the (ungraded) fermionic defect classifying algebra called ``$\text{duality}_2$'' in table~\ref{tab:Dsols} below.

\subsection{Boundary classifying algebra and boundary conditions}

In the fermionic Ising model case, only the three spinless fields ($\epsilon,\sigma,\mu$)
can couple to a weight zero field on the boundary and of these, the
two even fields ($\epsilon,\sigma$) can only couple to a bosonic
field of weight zero while the odd field ($\mu$) can only couple to a
fermionic field of weight zero.

We shall denote a possible fermionic weight zero field by $a$ and normalise its
two point function to $\langle aa \rangle =1$, so that $c_{aa}^{~1}=1$. As a
consequence, we would like to find the sewing constraints on the
following set of bulk-boundary structure constants:
\be
B^\epsilon_1
\;,\;\;
B^\sigma_1
\;,\;\;
B^\mu_a
\;,
\ee
and we have six sewing constraints of the form \eqref{eq:BB},
namely
\begin{align}
B^\epsilon_1\,B^\epsilon_1
&~=~ \SC_{\epsilon\epsilon}^{~~1}\, F_{00}
\big[\begin{smallmatrix} \epsilon & \epsilon \\ \epsilon & \epsilon
\end{smallmatrix}\big]
~=~ 1 
\ ,
\nonumber\\
B^\epsilon_1\,B^\sigma_1
&~=~ \SC_{\epsilon\sigma}^{~~\sigma}\, F_{\sigma 0}
\big[\begin{smallmatrix} \sigma & \sigma \\ \epsilon & \epsilon
\end{smallmatrix}\big] \, B^\sigma_1
~=~ B^\sigma_1
\ ,
\nonumber\\
B^\epsilon_1\,B^\mu_a
&~=~ \SC_{\epsilon\mu}^{~~\mu}\, F_{\sigma 0}
\big[\begin{smallmatrix} \sigma & \sigma \\ \epsilon & \epsilon
\end{smallmatrix}\big] \, B^\mu_1
~= - B^\mu_a
\ ,
\nonumber\\
B^\sigma_1\,B^\sigma_1
&~=~ 
\SC_{\sigma\sigma}^{~~1}\, F_{00}
\big[\begin{smallmatrix} \sigma & \sigma \\ \sigma & \sigma 
\end{smallmatrix}\big] 
+
\SC_{\sigma\sigma}^{~~\epsilon}\, F_{\epsilon 0}
\big[\begin{smallmatrix} \sigma & \sigma \\ \sigma & \sigma 
\end{smallmatrix}\big] \, B^\epsilon_1
\,
~=~ \tfrac{1}{\sqrt 2} \left(1 + B^\epsilon_1 \right)
\ ,
\nonumber\\
B^\sigma_1\,B^\mu_a
&~=~ 0
\ ,
\nonumber\\
B^\mu_a\,B^\mu_a
&~=~ 
\SC_{\mu\mu}^{~~1} \, F_{00}
\big[\begin{smallmatrix} \sigma & \sigma \\ \sigma & \sigma 
\end{smallmatrix}\big] 
+
\SC_{\mu\mu}^{~~\epsilon}\, F_{\epsilon 0}
\big[\begin{smallmatrix} \sigma & \sigma \\ \sigma & \sigma 
\end{smallmatrix}\big] \, B^\epsilon_1
~=~ \tfrac{1}{\sqrt 2} \left(1 - B^\epsilon_1 \right)
\ .
\label{eq:FF-bulk-bnd-constraints}
\end{align}
These equations have four solutions,
\be
{\renewcommand\arraystretch{1.3}
\begin{array}{l|r|r|r}
	& B^\epsilon_1    & B^\sigma_1   & B^\mu_a \\ 
	\hline
	\hbox{fixed up}   &  1           &  2^{1/4}         &  0 \\
	\hline
	\hbox{fixed down} &  1           & -2^{1/4}         &  0 \\
	\hline
	\hbox{free}  & -1           &  0         &  2^{1/4} \\
	\hline
	\hbox{free}  & -1           &  0         & -2^{1/4} 
\end{array}
}
\label{eq:FF-bulkbndcross-sol}
\ee
but the last two can be identified by the change of normalisation $a
\to -a$, and so there are three inequivalent
solutions to the boundary sewing
constraints, two fixed boundary conditions which do not require a
fermionic weight zero field, and one free boundary condition which does
require a fermionic weight zero field.

From \eqref{eq:FF-bulk-bnd-constraints} we can also read off the fermionic boundary classifying algebra $\hat\cB$ in \eqref{eq:BB'} for the free fermion:
\begin{align}
B^\epsilon\,B^\epsilon
&= 1
\;,
&
B^\sigma\,B^\sigma
& = \frac{1}{\sqrt 2} \left(1 + B^\epsilon \right)
\;,
&
B^\mu\,B^\mu
&
= \frac{1}{\sqrt 2} \left(1 - B^\epsilon \right)
\;,
\nonumber \\
B^\epsilon\,B^\sigma
&= B^\sigma
\;,&
B^\epsilon\,B^\mu
&= - B^\mu
\;, &
B^\sigma\,B^\mu &= 0
\;.
\label{eq:BCA}
\end{align}
This is a super-algebra and the parity of each generator agrees with that of the corresponding bulk field: $B^\epsilon$, $B^\sigma$ are even and $B^\mu$ is odd.
Now we can more easily identify the elementary boundary conditions by expressing $\hat\cB$ as a direct sum of copies of $\mathbb{C}$ and the Clifford algebra $C\ell_1$ which correspond directly to the elementary boundary conditions without overcounting the free boundary condition (cf.\ the discussion in section~\ref{sec:ferm-class-alg}),
\be
	\hat\cB ~=~ 
	\mathbb{C} e_+ 
	\,\oplus\,
	\mathbb{C} e_-
	\,\oplus\,
	\big(\mathbb{C} f_f \oplus \mathbb{C} a_f\big) \ ,
\ee
where the indices $+$, $-$, $f$ correspond to fixed up, fixed down and free.
Explicitly,
\begin{align}
&e_{+} 
= \tfrac 14 ( 1 + B^\epsilon) + 2^{-3/4} B^\sigma
\quad , \qquad
e_{-} 
= \tfrac 14 ( 1 + B^\epsilon) - 2^{-3/4} B^\sigma \ ,
\nonumber
\\
&f_f 
= \tfrac 12 (1 - B^\epsilon)
\quad , \quad
a_f 
= 2^{-1/4} B^\mu \ .
\end{align}
One can check that these satisfy the algebra \eqref{eq:efa} and from
the action of the generators $B^\epsilon$, $B^\sigma$ and $B^\mu$,  we
can read off the representations: 
\be
{\renewcommand\arraystretch{1.3}
\begin{array}{l|c|c|c}
	& B^\epsilon    & B^\sigma   & B^\mu \\ 
	\hline
	\hbox{fixed up}   &  1           &  2^{1/4}         &  0 \\
	\hline
	\hbox{fixed down} &  1           & -2^{1/4}         &  0 \\
	\hline
	\hbox{free}  & 
	- \mathds 1 & 
	0 & 
	2^{1/4} \cdot \mathds A
\end{array}
}
\ee
where 
the $2\times 2$ matrices $\mathds 1$ and $\mathds A$ are
\be
\mathds 1 = \begin{pmatrix} 1 & 0 \\ 0 & 1 \end{pmatrix}
\;,\;\;
\mathds A = \begin{pmatrix} 0 & 1  \\ 1 & 0 \end{pmatrix}
\;.
\ee
It is easy to see that the matrix representation of the free boundary condition is reducible and (by diagonalising $\mathds A$) splits into exactly the two solution in \eqref{eq:FF-bulkbndcross-sol} but this does not respect the even-odd nature of the generators and leads to the over-counting of the free boundary condition in \eqref{eq:FF-bulkbndcross-sol}; only when ensuring that the representation of $\hat\cB$ respects the even-odd grading do we get the correct counting (cf.\ \eqref{eq:el-bc-vs-irrep} for the precise statement).

\subsection{Boundary field content and boundary states}

It is also instructive to consider the contribution of non-zero weight
boundary fields as this will determine the full boundary field content
of each boundary condition 
as well as give the gluing conditions for the fermion
fields on the boundaries.

\subsubsection{Boundary field content}

Given the bulk field content in the fermionic Ising model, one only
has to consider $h=\frac12$ boundary fields as no other couplings are
possible.

The first result is that there must always exist a fermionic weight
$\frac12$ field, which we denote $\muB$ and which could be identified with
the boundary disorder field. The reason is that the bulk fermionic
fields can only couple to a weight $\frac12$ field on the boundary. If we
denote the bulk-boundary coupling of $\psi$ as $B^\psi_{\muB}$ and if the
boundary field two point function normalisation is $c_{\muB
  \muB}{}^1$, then the sewing constraint \eqref{eq:BB2} requires 
$(B^\psi_{\muB})^2 \,c_{\muB \muB}{}^{\!1}=1$ and hence we see that not
only must such a field exist but that the bulk fermions must couple to
it. We shall not consider the possibility of two distinct boundary
fermions but instead suppose that both  bulk fields $\psi$ and $\psib$
couple to the same boundary field. 
 
The next question is whether there is also an even (bosonic) weight
$\frac12$ boundary field $\sigmaB$ (which could be identified with the boundary
spin). If there is such a field then its OPE with $\muB$ must
necessarily be to an odd weight 0 field and conversely the OPE of an
odd weight zero field with $\muB$ must be an even weight $\frac12$ field
$\sigmaB$. If we normalise $a^2=
c_{\sigmaB\sigmaB}{}^{\!1}=c_{\muB\muB}{}^{\!1}=1$ then the OPE algebra is forced by the boundary sewing constraints to have the form 
\be
a \sigmaB = \lambda \muB
\;,\;\;
a \muB = (1/\lambda) \sigmaB
\;,\;\;
\sigmaB a = (1/\lambda) \muB
\;,\;\;
\muB a = \lambda \sigmaB
\;,
\ee
where $\lambda^4=1$. The values $\lambda=\pm 1$ turn out to be
inconsistent with
the bulk-boundary sewing constraints and $\lambda=\pm i$ are
equivalent under a field redefinition, hence we make the choice
$\lambda=i$ from here on.
With this boundary field algebra and the seven structure constants
\be
  B^\epsilon_1
\;,\;\;
  B^\sigma_1
\;,\;\;
  B^\mu_a
\;,\;\;
B^\sigma_{\sigmaB}
\;,\;\;
  B^\mu_{\muB}
\;,\;\;
  B^\psi_{\muB}
\;,\;\;
  B^\psib_{\muB}
\;,
\ee
there are 28 sewing constraints coming from \eqref{eq:BB2}.
These amount to 28 multiplication rules for the structure constants, which are
given in table \ref{tab:28}.

\begin{table}[htb]
\[
{\renewcommand\arraystretch{1.5}
\begin{array}{c||c|c|c|c|c|c|c|}
 & B^\epsilon_1 & B^\sigma_1 & B^\sigma_\sigmaB &
   B^\mu_a & B^\mu_\muB & B^\psi_\muB & B^\psib_\muB 
\\
\hline
\hline
B^\epsilon_1 & 
1 & & & & & & 
\\
\hline
B^\sigma_1   & 
B^\sigma_1 & \frac{1}{\sqrt 2}({1 + B^\epsilon_1}) & & & & &
\\
\hline
B^\sigma_\sigmaB &
 - B^\sigma_\sigmaB & 0 & \frac{1}{2 \sqrt 2}({1 - B^\epsilon_1}) & & & &
\\
\hline
B^\mu_a &
- B^\mu_a & 0 & \frac{1}{2}(B^\psib_\muB - B^\psi_\muB) & 
  \frac{1}{\sqrt 2}({1 - B^\epsilon_1}) & & & 
\\
\hline
B^\mu_\muB &
B^\mu_\muB & \frac 12 (B^\psi_\muB + B^\psib_\muB) & 0 & 0 & \frac{1}{2
  \sqrt 2}(1 + B^\epsilon_1) & & 
\\
\hline
B^\psi_\muB &
B^\psib_\muB & \sqrt 2 B^\mu_\muB & -\frac{1}{\sqrt 2}B^\mu_a
& - \sqrt 2\, B^\sigma_\sigmaB & \frac{1}{\sqrt 2}\,{B^\sigma_1} & 1 &
\\ 
\hline
B^\psib_\muB &
B^\psi_\muB & \sqrt 2 B^\mu_\muB & \frac{1}{\sqrt 2}B^\mu_a
& \sqrt 2\, B^\sigma_\sigmaB & \frac{1}{\sqrt 2}\,{B^\sigma_1} & 
B^\epsilon_1 & 1
\\ 
\hline
\end{array}
}
\]
\caption{The sewing constraints from \eqref{eq:BB2}, rewritten in the from
$B_a^\alpha\, B_b^\beta = \sum_{\gamma,c} C_{\alpha a, \beta b}{}^{\gamma c} \, B_c^\gamma$ and stated explicitly as a multiplication table; 
the sewing constraints for the
standard Ising model are in the top left $3\times 3$ sub-table.}
\label{tab:28}
\end{table}

Viewed as equations, the sewing constraints have 8 
solutions, four fixed with $B^\epsilon_1=1$
and four free with $B^\epsilon_1=-1$. 
The eight solutions are given in table \ref{tab:8sols}.
In the four fixed solutions, there are no couplings to the fields $a$
and $\sigmaB$ and so these fields can be consistently excluded from the set
of boundary fields, as expected.   

\begin{table}[htb]
\[
{\renewcommand\arraystretch{1.3}
\begin{array}{c||c|c|c|c|c|c|c|}
 & B^\epsilon_1 & B^\sigma_1 & B^\sigma_\sigmaB &
   B^\mu_a & B^\mu_\muB & B^\psi_\muB & B^\psib_\muB 
\\
\hline
\hline
\hbox{fixed up} &
1 & 2^{1/4} & 0 & 0 & 2^{-1/4} & 1 &1
\\ 
\hline
\hbox{fixed up} & 
1 & 2^{1/4} & 0 & 0 &-2^{-1/4} &-1&-1
\\ 
\hline
\hbox{fixed down} & 
1 & -2^{1/4} & 0 & 0 &-2^{-1/4} & 1&1
\\ 
\hline
\hbox{fixed down} & 
1 & -2^{1/4} & 0 & 0 &2^{-1/4} &-1&-1
\\
\hline 
\hbox{free} &
-1 & 0& 2^{-1/4} & 2^{1/4} & 0&-1&1
\\
\hline
\hbox{free} &
-1 & 0& 2^{-1/4} & - 2^{1/4} &0&1&-1
\\
\hline
\hbox{free} &
-1 & 0& - 2^{-1/4} & 2^{1/4} &0&1&-1
\\
\hline
\hbox{free} &
-1 & 0& - 2^{-1/4} & - 2^{1/4} &0&-1&1
\\
\hline
\end{array}
}
\]
\caption{The eight solutions to the full sewing constraints with
  $\lambda=i$; the three solutions to the sewing constraints for the
  standard Ising model are given by the first three columns of the table}
 \label{tab:8sols}
\end{table}

Only three of the eight solutions are physically distinct. Namely, for
each of the sign choices $\zeta,\xi \in \{ \pm 1 \}$ we can redefine
the boundary fields as $a \mapsto \zeta \, a$, $\sigmaB \to \xi\,
\sigmaB, \muB \to \zeta\xi\, \muB$. This agrees with the three
physically distinct solutions found in
\eqref{eq:FF-bulkbndcross-sol}. 

What is perhaps surprising is that the full set of sewing constraints
in table~\ref{tab:28}
also defines a commutative algebra [with identity] with generators
$\{1,B^\epsilon_1,B^\sigma_1,B^\mu_a,B^\sigma_\sigmaB,B^\mu_\muB,B^\psi_\muB,B^\psib_\muB\}$,
with the boundary classifying algebra as a sub-algebra, and the solutions in table~\ref{tab:8sols} are the eight
one-dimensional representations of this algebra. 

As with the classifying algebra, this commutative algebra can also be
viewed as a super-algebra
with even generators 
$\{1,B^\epsilon_1,B^\sigma_1,B^\sigma_\sigmaB\}$,
and odd generators $\{B^\mu_a,B^\mu_\muB,B^\psi_\muB,B^\psib_\muB\}$, 
and the eight
one-dimensional representations in table \ref{tab:8sols} combine into four
representations of this super-algebra on $\mathbb{C}^{1|1}$, of which
the final two are again related by $\sigmaB \to - \sigmaB, \muB \to -
\muB$ and so are equivalent physically:
\be
{\renewcommand\arraystretch{1.3}
\begin{array}{c||c|c|c|c|c|c|c|}
 & B^\epsilon_1 & B^\sigma_1 & B^\sigma_\sigmaB &
   B^\mu_a & B^\mu_\muB & B^\psi_\muB & B^\psib_\muB 
\\
\hline
\hline
\hbox{fixed up} &
\mathds 1 & 
2^{1/4} \cdot \mathds 1 & 
0 & 0 & 
2^{-1/4} \cdot \mathds A&
\mathds A &
\mathds A
\\ 
\hline
\hbox{fixed down} &
\mathds 1 & 
- 2^{1/4} \cdot \mathds 1 & 
0 & 0 & 
2^{-1/4} \cdot \mathds A&
-\mathds A &
-\mathds A
\\ 
\hline
\hbox{free} &
- \mathds 1 & 
0 & 
2^{-1/4} \cdot \mathds 1& 
2^{1/4} \cdot\mathds A&
0 &
- \mathds A&
\mathds A
\\ 
\hline
\hbox{free} &
- \mathds 1& 
0 & 
- 2^{-1/4} \cdot \mathds 1 & 
2^{1/4} \cdot \mathds A &
0 &
\mathds A&
- \mathds A 
\\ 
\hline
\end{array}
}
\ee

\blank{\tiny

\be
\begin{array}{c||c|c|c|c|c|c|c|}
 & B^\epsilon_1 & B^\sigma_1 & B^\sigma_\sigmaB &
   B^\mu_a & B^\mu_\muB & B^\psi_\muB & B^\psib_\muB 
\\
\hline
\hline
\hbox{fixed up} &
\begin{pmatrix}1 & 0 \\ 0 & 1 \end{pmatrix} & 
\begin{pmatrix}2^{1/4} & 0 \\ 0 & 2^{1/4} \end{pmatrix}  & 
0 & 0 & 
\begin{pmatrix} 0 & 2^{-1/4}\\ 2^{-1/4} & 0 \end{pmatrix} &
\begin{pmatrix} 0 & 1\\ 1 & 0 \end{pmatrix} &
\begin{pmatrix} 0 & 1\\ 1 & 0 \end{pmatrix} 
\\ 
\hline
\hbox{fixed down} &
\begin{pmatrix}1 & 0 \\ 0 & 1 \end{pmatrix} & 
-\begin{pmatrix}2^{1/4} & 0 \\ 0 & 2^{1/4} \end{pmatrix}  & 
0 & 0 & 
\begin{pmatrix} 0 & 2^{-1/4}\\ 2^{-1/4} & 0 \end{pmatrix} &
\begin{pmatrix} 0 & -1\\ -1 & 0 \end{pmatrix} &
\begin{pmatrix} 0 & -1\\ -1 & 0 \end{pmatrix} 
\\ 
\hline
\hbox{free} &
\begin{pmatrix}-1 & 0 \\ 0 & -1 \end{pmatrix} & 
0 & 
\begin{pmatrix}2^{1/4} & 0 \\ 0 & 2^{1/4} \end{pmatrix}  & 
\begin{pmatrix} 0 & 2^{-1/4}\\ 2^{-1/4} & 0 \end{pmatrix} & 0&
\begin{pmatrix} 0 & -1\\ -1 & 0 \end{pmatrix} &
\begin{pmatrix} 0 & 1\\ 1 & 0 \end{pmatrix} 
\\ 
\hline
\hbox{free} &
\begin{pmatrix}-1 & 0 \\ 0 & -1 \end{pmatrix} & 
0 & 
-\begin{pmatrix}2^{1/4} & 0 \\ 0 & 2^{1/4} \end{pmatrix}  & 
\begin{pmatrix} 0 & 2^{-1/4}\\ 2^{-1/4} & 0 \end{pmatrix} & 0&
\begin{pmatrix} 0 & 1\\ 1 & 0 \end{pmatrix} &
\begin{pmatrix} 0 & -1\\ -1 & 0 \end{pmatrix} 
\\ 
\hline
\end{array}
\label{eq:8sols}
\ee
}

This table also shows that the fermions have opposite
gluing conditions on fixed and free boundary conditions. On the free
boundary condition with $B^\epsilon=-1$, the bulk fermions obey $\psi
= - \psib$ on the  boundary; on the the fixed boundary conditions they
obey $\psi = \psib$.

\subsubsection{Boundary states}

Since we are working with theories on a fixed spin structure, each
boundary condition $b$ will be associated to two boundary states, one
state 
$|b\rangle_\mathrm{NS}$ entirely in the NS sector and one state $|b\rangle_\mathrm{R}$ entirely in the R sector.
In our approach, correlators are overall even linear forms, so that a one-point correlator on a disc is an even linear map $\cH_F \to \mathbb{C}$. This implies that boundary states are purely even.

The bulk-boundary couplings to the identity field (cf.~\eqref{eq:FF-bulkbndcross-sol}) determine the overlap of the boundary states with each primary bulk field, normalised by the overlap with the vacuum. Namely, for primary bulk fields $\phi_i$ in the NS sector and $\phi_j$ in the R sector we have
\be
\frac{ {}_\mathrm{NS}\langle b | \phi_i \rangle }{ {}_\mathrm{NS}\langle b | 0 \rangle
} = B^i_1 
\qquad , \qquad
\frac{ {}_\mathrm{R}\langle b | \phi_j \rangle }{ {}_\mathrm{NS}\langle b | 0 \rangle
} = B^j_1 \ .
\ee
This fixes the expansion of $|b\rangle_{\mathrm{NS},\mathrm{R}}$ in terms of Virasoro-Ishibashi states $\kett{ 1 }$, $\kett{ \sigma }$, $\kett{ \epsilon }$ up to the overall constant $\langle 0 | b \rangle_{NS}$.

To determine $\langle 0 | b \rangle_\mathrm{NS}$ (up to a sign) 
we will make use of the fact that we know the boundary field content for each boundary condition:
\be
\fxed: 1, \muB
\qquad, \qquad
\fre: 1, a, \muB, \sigmaB
\ee
This means that the NS boundary states have to satisfy
\begin{align}
{}_\mathrm{NS}\langle \fxed|\, q^{L_0 + \bar L_0 - 1/24} \,|\fxed\rangle_\mathrm{NS}
&= \chi_0(\tilde q) + \chi_{1/2}(\tilde q)
\;,\;\;
\nonumber\\
{}_\mathrm{NS}\langle \fre|\, q^{L_0 + \bar L_0 - 1/24} \,|\fre\rangle_\mathrm{NS}
&= 2 \chi_0(\tilde q) + 2 \chi_{1/2}(\tilde q)
\;,
\end{align}
where as usual $q=\exp(2\pi i \tau), \tilde q = \exp(- 2 \pi i/\tau)$
and $\chi_h$ are the characters of the Virasoro highest weight
representations of weight $h$. 

Putting all this together, we arrive at 
(up to an overall undetermined sign for each boundary condition)
\begin{align}
 | \fxed+\rangle_\mathrm{NS} &~=~ \kett{ 0 } + \kett{ \epsilon }
\;,\;\;
&
 | \fxed+\rangle_\mathrm{R}  &~=~ 2^{1/4} \kett{ \sigma } \ ,
\nn\\
 | \fxed-\rangle_\mathrm{NS} &~=~  \kett{ 0 } + \kett{ \epsilon }
\;,\;\;
&
 | \fxed-\rangle_\mathrm{R} &~=  - 2^{1/4} \kett{ \sigma } \ ,
\nn\\
 | \fre\rangle_\mathrm{NS} &~=~ \sqrt 2 \left( \kett{ 0 } - \kett{ \epsilon } \right)
\;,\;\;
&
 | \fre\rangle_\mathrm{R} &~=~ 0 \ .
\label{eq:bss}
\end{align}
The overlaps between the R-sector boundary states must give
the supertrace over the field contents, and the above states
correctly give these supertraces,
\begin{align}
{}_{R}\langle \fxed|\, q^{L_0 + \bar L_0 - 1/24} \,|\fxed\rangle_{R}
&= \chi_0(\tilde q) - \chi_{1/2}(\tilde q)
\;,
\nonumber\\
{}_{R}\langle \fre|\, q^{L_0 + \bar L_0 - 1/24} \,|\fre\rangle_{R}
&= 0
\;.
\end{align}

As briefly mentioned in the end of section~\ref{sec:ferm-class-alg}, one can think of boundary states as coming in pairs which differ by an overall sign; the three boundary states presented above are then characterised by the condition 
$\langle 0 | b \rangle_\mathrm{NS} > 0$.

Note that the NS components in \eqref{eq:bss} agree with the
conjectures in \cite{mw} for the NS-sector of the boundary states,
although that paper does not correctly account for the fermionic
fields in this model nor the Ramond sectors.

Boundary states in the NS-sector of free fermions are also considered
in \cite{Tong2}, and the boundary states for a single Majorana fermion given
in Appendix D there agree with the NS-component in \eqref{eq:bss} up to
normalisation.

Finally, we note that our boundary states are related to the boundary
states for the purely bosonic Ising model by taking a superposition
of the NS- and R-sector boundary states,
\be
  \ket{a}_{\mathrm{Ising}} = \frac{1}{\sqrt 2}\Big(\, \ket{a}_\mathrm{NS} + \ket{a}_\mathrm{R}
  \,\Big)
\;.
\ee

\subsection{Defect classifying algebra and defect operators}
\label{ssec:dca}

\subsubsection{Classifying algebra and defect conditions}

We consider now the defect classifying algebra
of the standard Ising model and the fermionic
version, assuming that there is a single bosonic field of weight zero
(denoted 1) on the defect
and at most one fermionic defect field of weight zero
denoted $a$ which satisfies $a^2=1$.
The bulk-defect
structure constants for the non-identity bulk fields in
the Ising case are 
$\{D^{\epsilon\epsilon}_{~1},
D^{\sigma\sigma}_{~1}\}$, 
and for the fermionic case they are
$\{D^{\epsilon\epsilon}_{~1},
D^{\sigma\sigma}_{~1}, D^{\mu\mu}_{~1},D^{\sigma\mu}_{~a},
D^{\mu\sigma}_{~a},
D^{\psi\psi}_{~1},D^{\psib\psib}_{~1}
\}$.
The sewing constraints are given by the multiplication table
\ref{tab:Dtab}.

\begin{table}[htb]
\[
{\renewcommand{\arraystretch}{1.3}
\begin{array}{c||c|c|c|c|c|c|c|}
& 
D^{\epsilon\epsilon}_{~1} & D^{\sigma\sigma}_{~1} & 
D^{\mu\mu}_{~1} & D^{\sigma\mu}_{~a} & 
D^{\mu\sigma}_{~a} & D^{\psi\psi}_{~1} & D^{\psib\psib}_{~1}
\\ \hline \hline
D^{\epsilon\epsilon}_{~1} & 
1 & & & & & & 
\\ \hline
D^{\sigma\sigma}_{~1} & 
D^{\sigma\sigma}_{~1} & \frac 12 ( 1 + D^{\epsilon\epsilon}_{~1}) & & & & & 
\\ \hline
D^{\mu\mu}_{~1} & 
D^{\mu\mu}_{~1} & \frac 12 (D^{\psi\psi}_{1} + D^{\psib\psib}_1) &
\frac 12 (1 + D^{\epsilon\epsilon}_1) & & & & 
\\ \hline
D^{\sigma\mu}_{~a} & 
- D^{\sigma\mu}_{~a} & 0 & 0 & \frac 12 (-1 + D^{\epsilon\epsilon}_{1}) & & & 
\\ \hline
D^{\mu\sigma}_{~a} &
- D^{\mu\sigma}_{~a} & 0 & 0 &  - \frac i2(D^{\psi\psi}_1 -D^{\psib\psib}_1)
&\frac 12 (1 - D^{\epsilon\epsilon}_{1}) & & 
\\ \hline
D^{\psi\psi}_{~1} & 
D^{\psib\psib}_1 & D^{\mu\mu}_1 & D^{\sigma\sigma}_1 & - i D^{\mu\sigma}_{~a} &
i D^{\sigma\mu}_{~a} & 1 & 
\\ \hline
D^{\psib\psib}_{~1} &
D^{\psi\psi}_1 & D^{\mu\mu}_1 & D^{\sigma\sigma}_1 & i D^{\mu\sigma}_{~a} &
- i D^{\sigma\mu}_{~a} & D^{\epsilon\epsilon}_{~1} & 1 
\\ \hline
\end{array}
}
\]
\caption{}
\label{tab:Dtab}
\end{table}

Viewed as equations, there are eight solutions to the defect sewing
constraints, given in table \ref{tab:Dsols}.

\begin{table}[htb]
\[
\begin{array}{c||c|c|c|c|c|c|c|}
& 
D^{\epsilon\epsilon}_{~1} & D^{\sigma\sigma}_{~1} & 
D^{\mu\mu}_{~1} & 
D^{\psi\psi}_{~1} & D^{\psib\psib}_{~1} &
D^{\sigma\mu}_{~a} & D^{\mu\sigma}_{~a}  
\\ \hline \hline
\hbox{identity} & 
1 & 1 & 1& 1 & 1 & 0 & 0 
\\ \hline
\hbox{identity}' &
1 & 1 & -1& -1 & -1 & 0 & 0 
\\ \hline
\hbox{spin reversal}&
1 & -1 & -1& 1 & 1 & 0 & 0 
\\ \hline
\hbox{spin reversal}'&
1 & -1 & 1& -1 & -1 & 0 & 0 
\\ \hline
\hbox{duality}^1&
-1 & 0 & 0 & 1 & -1 & i & -1 
\\ \hline
\hbox{duality}^2&
-1 & 0 & 0 & 1 & -1 & -i & 1 
\\ \hline
\hbox{duality}^3&
-1 & 0 & 0 & -1 & 1 & -i & -1 
\\ \hline
\hbox{duality}^4&
-1 & 0 & 0 & -1 & 1 & i & 1 
\\ \hline
\end{array}
\]
\caption{}
\label{tab:Dsols}
\end{table}

We can view the bulk-defect structure constants  as algebra generators
of a super-algebra $\hat\cD$, the fermionic defect classifying
algebra, with odd generators $\{D^{\sigma\mu}_{~a}, D^{\mu\sigma}_{~a}
\}$ and the rest being even. As an ungraded algebra $\hat\cD$ is
commutative, and the solutions in table \ref{tab:Dsols} are its eight
one-dimensional representations. 
When viewed as representations of the
super-algebra, the first four are one-dimensional representations on
$\mathbb C^{1|0}$ while the second four combine to form two
two-dimensional representations on $\mathbb C^{1|1}$.
Accordingly, $\hat\cD$ 
decomposes  into four copies of $\mathbb C$ (with
generators $e_\alpha$) and two
copies of $C\ell_1$ (with generators $\{f_\alpha,a_\alpha\}$
satisfying the same algebra \eqref{eq:efa}. The explicit forms are:
\begin{align}
e_{id} &= \frac 18(1 + D^{\epsilon\epsilon}_1 + 2D^{\mu\mu}_1 +
2D^{\sigma\sigma}_1 + D^{\psib\psib}_1 +  D^{\psi\psi}_1 ) 
\ , \nonumber
\\
e_{id'} &= \frac 18(
1 + D^{\epsilon\epsilon}_1 - 2 D^{\mu\mu}_1 +  2 D^{\sigma\sigma}_1 - D^{\psib\psib}_1 -  D^{\psi\psi}_1)
\ , \nonumber
\\
e_{s} &= \frac 18 (1+ D^{\epsilon\epsilon}_1 - 2 D^{\mu\mu}_1 -  2D^{\sigma\sigma}_1 + D^{\psib\psib}_1 +  D^{\psi\psi}_1)
\ , \nonumber
\\
e_{s'} &=
\frac 18(1 + D^{\epsilon\epsilon}_1 + 2 D^{\mu\mu}_1 - 2 D^{\sigma\sigma}_1 - D^{\psib\psib}_1 - D^{\psi\psi}_1)
\ , \nonumber
\\
f_{d12} &= \frac 14 (1 - D^{\epsilon\epsilon}_1 -D^{\psib\psib}_1 +  D^{\psi\psi}_1)
~~,~~
a_{d12} = \frac 12(D^{\mu\sigma}_a + i D^{\sigma\mu}_a)
\ , \nonumber
\\
f_{d34} &= \frac 14 (1- D^{\epsilon\epsilon}_1 + D^{\psib\psib}_1 - D^{\psi\psi}_1)
~~,~~
a_{d34} = \frac 12( D^{\mu\sigma}_a - i D^{\sigma\mu}_a) \ .
\end{align}

\subsubsection{Defect field content and defect operators}

Topological defects can also be described by operators on the Hilbert
space of the theory which commute with the Virasoro algebra, and hence
are sums of intertwiners between equivalent Virasoro representations.
The bulk-defect structure constants are proportional to the
coefficients of these intertwiners and hence the defect operator is
defined, up to an overall scalar multiple, by the bulk-defect
structure constants. This overall multiple can itself be fixed (up to
a sign) by the requirement that it correctly determines the field
content on the defect, and in particular that it correctly determines
the dimension of the space of zero-weight fields on the defect.

We illustrate this in the case of the bosonic and fermionic Ising defects.
The bosonic Hilbert space is
\be
\cH_{\mathrm{Ising}} = 
(M_0\otimes \overline M_0 ) 
\oplus
(M_\epsilon\otimes \overline M_\epsilon ) 
\oplus
(M_\sigma\otimes \overline M_\sigma ) 
\ ,
\ee
where $M_a$ denotes the corresponding irreducible Virasoro representation.
This means a topological defect operator takes the form
\be
 \hat D = 
\alpha P_{00} + \beta P_{\epsilon\epsilon} + \gamma P_{\sigma\sigma}
\;.
\ee
The constants $D^{ij}_1$ are given by
\be
  D^{ij}_1 = \frac{ \langle j | \, \hat D \, | i \rangle }
                  { \langle 0 | \, \hat D \, | 0 \rangle }
\ee
and hence the defect is fixed up to a scalar,
\be
 \hat D = \lambda ( 
P_{00} + D^{\epsilon\epsilon}_1 P_{\epsilon\epsilon} 
        + D^{\sigma\sigma}_1 P_{\sigma\sigma}
          )
\;,
\ee
The field content on the defect is then given by the modular transform
of the trace on the cylinder,
\be
  Z(q,\bar q) 
= \mathrm{Tr}\left( q^{L_0 + \bar L_0 - c/12} D \cdot D^\dagger \right)
= \lambda^2 \Big( 
|\chi_0(q)|^2 + (D^{\epsilon\epsilon}_1)^2 |\chi_{1/2}(q)|^2
              + (D^{\sigma\sigma}_1)^2 |\chi_{1/16}(q)|^2 \Big)
\;.
\ee
The upshot is that $\lambda=\pm1$ for the identity and spin defects
and $\pm\sqrt 2$ for the duality defect,
cf.~\cite{Oshikawa:1996dj,Petkova:2000ip}.

The same ideas can be applied to the defects in the fermionic
model, with the observation that the defects come in two versions,
depending on the spin structure, so that there are separate defect
operators acting on the Neveu-Schwarz and Ramond sectors of the
Hilbert space. In the fermionic
model the Hilbert space is
\be
\cH_{F} = 
 \cH_{F}^{NS} + \cH_{F}^{R}
\;,\;\;\;\;
\ee\be
 \cH_{F}^{NS}
 = ( M_0 + M_{1/2}) \otimes ( \overline M_0 + \overline M_{1/2}) 
\;,\;\;
 \cH_{F}^{R}
 = \left. M_{1/16}\otimes \overline M_{1/16} \right|_{\sigma}
 \; \oplus\;
  \left. M_{1/16}\otimes \overline M_{1/16} \right|_{\mu}
\;.
\ee
The defect operators in the two sectors are then given by the
bulk-defect structure constants up to an overall constant
\be
  \hat D^{NS} = 
 \lambda \left( P_{0} + 
D^{\psi\psi}_1 P_{\psi} +
D^{\psib\psib}_1 P_{\psib} +
D^{\epsilon\epsilon}_1 P_{\epsilon}\right)
\;,\;\;
  \hat D^{R} = 
 \lambda \left( D^{\sigma\sigma}_1 P_{\sigma} + D^{\mu\mu}_1 P_{\mu}\right)
\;
\ee
The main difference is that the space of weight zero fields on the
defects can now be either one or two, depending on whether the defect
supports the fermionic weight zero field $a$ or not. 

When we perform the calculations we find that $\hat D^{NS}$ are in
fact identical with the defect operators proposed in \cite{mw}. We
give the explicit forms in table \ref{tab:FM34defects}. 

{\renewcommand{\arraystretch}{1.3}
\begin{table}[htb]
\[
\begin{array}{c||c|c|}
& \hbox{NS} & \hbox{R} 
\\ \hline \hline
\hbox{id} 
& P_0 + P_{\psi} + P_{\bar\psi} + P_{\eps}
& \phantom{-}P_{\sigma} + P_{\mu} 
\\ \hline
\hbox{id}'
& P_0 - P_{\psi} - P_{\bar\psi} + P_{\eps} 
&  \phantom{-}P_{\sigma} - P_{\mu} 
\\ \hline
\hbox{s}
& P_0 + P_{\psi} + P_{\bar\psi} + P_{\eps} 
&  -P_{\sigma} - P_{\mu} 
\\ \hline
\hbox{s}'
& P_0 - P_{\psi} - P_{\bar\psi} + P_{\eps}
& - P_{\sigma} + P_{\mu} 
\\ \hline
\hbox{d12}
& \sqrt2 \big(\,P_0+ P_\psi - P_{\bar\psi} - P_\eps \,\Big)
& 0
\\ \hline
\hbox{d34}
& \sqrt2 \big(\, P_0- P_\psi + P_{\bar\psi} - P_\eps \, \big)
& 0
\\ \hline
\end{array}
\]
\caption{}
\label{tab:FM34defects}
\end{table}
}

We note that the defects of the bosonic Ising model are given by the combinations
\be
D^{\text{Ising}}_{id} = \frac 12 \big( D_{id} + D_{id'} \big)
\;,\;\;
D^{\text{Ising}}_{spin} = \frac 12 \big( D_{s} + D_{s'} \big)
\;,\;\;
D^{\text{Ising}}_{duality} = \frac 12 \big( D_{d12} + D_{d34} \big)
\;.\;\;
\ee

\subsection{Interfaces between Ising and fermionic Ising}

As an example, consider interfaces between the bosonic Ising model and the fermionic version.
Let us assume that the Ising model is in the upper half plane and the
fermionic model in the lower half plane. This means that the possible
bulk-defect structure constants are $D^{ij}_x$ where $i$ takes 
values in $\{\epsilon,\sigma\}$, $j$ takes values in 
$\{\epsilon,\sigma,\mu\}$ and $x\in\{1,a\}$.

The interface sewing constraints are exactly the appropriate subset of the
fermionic defect sewing constraints in table \ref{tab:Dtab}, 
namely
\be
{\renewcommand{\arraystretch}{1.3}
\begin{array}{c||c|c|c|}
& 
D^{\epsilon\epsilon}_{~1} & D^{\sigma\sigma}_{~1} & 
D^{\sigma\mu}_{~a} 
\\ \hline \hline
D^{\epsilon\epsilon}_{~1} &  
1 & &
\\ \hline
D^{\sigma\sigma}_{~1} &  
D^{\sigma\sigma}_{1} & \frac 12 (1 + D^{\epsilon\epsilon}_1) & 
\\ \hline
D^{\sigma\mu}_{~a} & 
- D^{\sigma\mu}_{~a} & 0 & \frac 12 (-1 + D^{\epsilon\epsilon}_{1})
\\ \hline
\end{array}
}
\ee
There are four solutions to the sewing constraints, which constitute
the four one-dimensional representations of the commutative interface
classifying algebra and which also form two one-dimensional representations on
$\mathbb C^{1|0}$ and one two-dimensional representation on $\mathbb
C^{1|1}$ of the interface classifying super-algebra:
\be
\begin{array}{c||c|c|c|}
& D^{\epsilon\epsilon} & D^{\sigma\sigma} & D^{\sigma\mu} \\
\hline \hline
& 1 & 1 & 0 
\\ \hline
& 1 & -1 & 0
\\ \hline
& -\mathds 1 & 0 & \mathds A
\\ \hline
\end{array}
\ee

\section{Further Virasoro examples}\label{sec:furtherVir}

In this section we will consider a few examples of fermionic minimal models by increasing weight of the generator $G$ as listed in~\eqref{eq:models-by-weight}.
The smallest value
is $h_G=1/2$ in the single model $\FM(4,3)=\widetilde\FM(4,3)$ which is the free fermion and was already treated in section~\ref{sec:Ising}.

The next value is $h_G=3/2$, so that the fermionic theory has
super-Virasoro symmetry. There are two such examples, $\FM(4,5)$ which
is the fermionic tri-critical Ising model (section \ref{sec:tcim}) and
the non-unitary models $\FM(3,8)/\wtFM(3,8)$ (section \ref{sec:susyly}).

The $N=1$ superconformal
minimal model values of $c$ are $c(p,q) = 15/2 - 3 p/q - 3 q/p$,
parametrised by two integers $p,q$ with
$p-q$ even and $p,q\geq 2$ \cite{BKT,BFMRW}; we shall denote them generically by
$SM(p,q)$. The possible modular invariant partition functions for the
unitary models $|p-q|=2$ have been classified in~\cite{Cappelli1987} but, as noted
there, modular invariance does not fix the partition function
uniquely and it is determined only up to a constant which is fixed by
the parities of the states with $h=c/24$.

The final value we consider in any detail is 
$h_G=5/2$ in $\FM(4,7)$ in section~\ref{sec:M47}.
This final case was first noted in \cite{ZamW} and a longer list is
given in~\cite{BFKNRV} where chiral algebras which extend the Virasoro algebra
by a single fermionic field are considered.  $\FM(4,7)$ 
is a reduction the $\mathit{WB}(0,2)$ algebra (a.k.a.\ the
fermionic $\mathit{WB}_2$ algebra) at a value of $c$ at which the spin
4 field decouples.

These examples in fact all fit into two infinite series of fermionic models
with a current of spin $(2k-1)/2$, and these are $\FM(4,2k+1)$ and
$\FM(3,4k)$.  For $k\geq 3$, these two series are all  special
cases of the fermionic W-algebra $\mathit{WB}(0,k-1)$ (a.k.a.\ the
fermionic $\mathit{WB}_{k-1}$ algebra first introduced in \cite{FL}) in
which all but the fermionic W-algebra field decouple.

\subsection{Fermionic tri-critical Ising model}\label{sec:tcim}

The fermionic tri-critical Ising model is the second in the series of
fermionic extensions of the minimal models and some essential data is given in
appendix \ref{app:TCIM}.

\subsubsection{Fermionic TCIM boundary classifying algebra}

The boundary conditions of the tri-critical Ising model in both
bosonic and fermionic models have been studied before. For the bosonic
case see \cite{Affleck2000}, and for the fermionic case see
for example \cite{Nepomechie2001,Nepomechie2002}.
Here we will study boundary conditions of the fermionic model via the fermionic classifying algebra.

In the fermionic tri-critical Ising model, there are eight spinless
bulk fields which can hence
couple to a weight zero boundary field:
\begin{equation}
	\begin{array}{c|cccc|cc|cc}
		\text{spinless} & \multicolumn{6}{c|}{\text{even}} & \multicolumn{2}{c}{\text{odd}} 
		\\ 
		\text{fields} & \multicolumn{4}{c|}{\text{NS}} & \multicolumn{2}{c|}{\text{R}} & \multicolumn{2}{c}{\text{R}}
		\\ 
		& 1  & \eps & \eps' & \eps'' & \sigma & \sigma' & \mu & \mu' 
		\\[0.1em] \hline &&&&&&&& \\[-0.9em]
		\text{label} & (1,1) & (3,3) & (1,3) & (3,1) & (2,3) & (2,1) & (2,3) & (2,1)
		\\[0.1em] \hline &&&&&&&& \\[-0.9em]
h = \bar h & 0 & \tfrac1{10} & \tfrac3{5} & \tfrac3{2} & \tfrac3{80} & \tfrac7{16} & \tfrac3{80} & \tfrac7{16}
	\end{array}
\end{equation}
As
before, we assume that the only possible weight zero boundary fields
are the identity $1$ and an odd field $a$, and so the bulk-boundary
couplings fields are 
\be\label{eq:TCIM-bnd-coupligs}
  B^1_1 \equiv 1
  ~,~ 
  B^\epsilon_1
~,~ 
  B^{\epsilon'}_1
  ~,~ 
  B^{\epsilon''}_1
  ~,~ 
  B^\sigma_1
  ~,~ 
  B^{\sigma'}_1
  ~,~ 
  B^\mu_a
  ~,~ 
  B^{\mu'}_a
\;.
\ee
The sewing constraints can again
be considered as the relations in an 8
dimensional commutative algebra which also has the form of a
non-supercommutative super-algebra with even generators
$\{\,1, B^\epsilon, B^{\epsilon'}, B^{\epsilon''}, B^\sigma,
  B^{\sigma'}\}$ and odd generators
$\{ B^\mu, B^{\mu'}\}$.
It is also a graded algebra with respect to the spin grading, so it is
in fact a bi-graded algebra. 
Sorting the generators according to the
gradings $\phi$ and $\tilde\phi = \phi + \nu$ gives:
\be 
{\renewcommand{\arraystretch}{1.4}
\begin{array}{c|c|c}
& \phi=0 & \phi = 1 \\ \hline
 \tilde\phi=0 & 
 B^1 \equiv 1,\; B^\eps,\; B^{\eps'},\; B^{\eps''} & B^\mu,\; B^{\mu'}
\\ \hline 
  \tilde\phi = 1 & B^\sigma,\; B^{\sigma'}
& \hbox{---}
\end{array}
}
\ee

There are 28 sewing constraints for these structure constants which 
have eight solutions corresponding to the eight one-dimensional
representations of the commutative algebra with generators \eqref{eq:TCIM-bnd-coupligs}.

When viewed as super-algebra, the boundary classifying algebra decomposes into four copies of
$\mathbb C$ and two copies of $C\ell_1$, giving six boundary
conditions in all. This means the eight one-dimensional
representations combine to form 4 one-dimensional
representations on $\mathbb C^{1|0}$ and two two-dimensional
representations on $\mathbb C^{1|1}$, as shown in table \ref{tab:TCIMBCs}.

\begin{table}[htb]{\renewcommand{\arraystretch}{1.3}
\[
\begin{array}{c||ccc|cc|cc}
\hbox{boundary} & 
B^{\epsilon} & B^{\epsilon'} & B^{\epsilon''} &
B^{\sigma} &
B^{\sigma'} & B^{\mu} & B^{\mu'} 
\\ \hline\hline
(1,1) \equiv (-) 
 & 1 & 1 & 1 & 1 & 1 &  0 & 0
 \\ \hline
(3,1) \equiv (+)
& 1 & 1 & 1 & -1 & -1 & 0 & 0
\\ \hline
(1,2) \equiv(-0)
&  -\alpha & -\alpha  & 1 & \alpha & -1 & 0 & 0 
\\ \hline
(1,3) \equiv (0+)
&  -\alpha & -\alpha  & 1 & -\alpha & 1 & 0 & 0 
\\ \hline
(2,1) \equiv(0)
& -\mathds 1 & \mathds 1 & - \mathds 1 & 0 & 0 &   
\delta \mathds A & -\beta \mathds A
\\ \hline
(2,2) \equiv (d)
&
\alpha\mathds 1 & - \alpha\mathds 1 & - \mathds 1 & 0 & 0 & 
- \gamma \mathds A & \beta \mathds A 
\end{array}
\]
\caption{The boundary conditions of $\FM(4,5)$ at $c=7/10$; ~~ 
$\alpha = (3 - \sqrt 5)/2$, 
$\beta = (2/\sqrt 7)(1+i)$, 
$\gamma = \sqrt 7(3-\sqrt 5) (1+i)$, 
$\delta = 2\sqrt 7 (1+i) $.}
\label{tab:TCIMBCs}}
\end{table}

The boundary states in the NS and R sectors are given by taking linear
combinations determined by table \ref{tab:TCIMBCs} with an overall
normalisation,
\be
\ket{a}_\mathrm{NS} = \lambda_a( \ket 0 + B^\eps_1 \ket\eps +
B^{\eps'}_1\ket{\eps'} +  B^{\eps''}_1\ket{\eps''} )
\;,\;\;
\ket{a}_\mathrm{R} = \lambda_a(B^\sigma_1 \ket\sigma +
B^{\sigma'}_1\ket{\sigma'})
\;.
\ee
The normalisation is fixed (up to a sign) by the requirement that the number
of weight-zero fields is 1 on the boundaries corresponding to
one-dimensional representations and 2 on the boundaries corresponding
to two-dimensional representations. 

\subsubsection{Comparison to \cite{Nepomechie2001,Nepomechie2002}}

There has been a considerable amount of work on boundary conditions of
superconformal field theories. Here we compare our results
with those of Nepomechie in 
\cite{Nepomechie2002} which discusses boundary states in the
tri-critical Ising model $M(4,5)$ and which of these are ``supersymmetric''.
The boundary states discussed in that paper include both NS- and R-sectors and a boundary condition is said to be supersymmetric if the
partition function on a cylinder is a sum of characters of the super
Virasoro algebra, rather than simply a sum of characters of the
Virasoro algebra. 

Since the fermionic theory $\FM(4,5)$ we consider includes the
generators of the super-Virasoro algebra, the partition function
calculated using the overlaps of the boundary states in the NS-sector will always be sums of characters of the super-Virasoro
algebra. However, these are not the
partition functions discussed in \cite{Nepomechie2001} 
which are instead the average over the two
spin structures, that is the average of the overlaps between the NS-sectors and the R-sectors. Since the R-sectors will contribute the
supertrace over a super-Virasoro representation, not a trace, any
contribution from the R-sector will stop the partition function
being a sum of super-Virasoro characters. Hence we see that
``supersymmetric'' boundary conditions in the sense of \cite{Nepomechie2002}
correspond to boundary conditions with zero R-sector boundary state,
that is boundary conditions for which the algebra of weight zero
boundary fields is $C\ell_1$; if the weight zero boundary fields are
simply $\mathbb C$ then the boundary condition is
``non-supersymmetric''. 
The latter boundary conditions
fall into pairs related by the spin symmetry and which give
``supersymmetric'' boundary conditions when taken as a superposition.

In this sense we find, as in \cite{Nepomechie2002}, two
``supersymmetric'' boundary conditions and four ``non-supersymmetric''
ones. 

\subsubsection{Fermionic TCIM defect classifying algebra}
\label{tcimdefcla}

The defect classifying algebra of $\FM(4,5)$ has one generator
$D^{ij}$ for each pair of fields $\phi_i,\phi_j$ whose operator
product on the defect includes a field of weight zero. 
There are 6
such pairs where $i$ and $j$ are both even which generate the defect
classifying algebra of the bosonic
$M(4,5)$. There are 6 more pairs for which $i$
and $j$ are both odd, but for which $D^{ij}$ is again therefore even. 
Together these generate the 12 dimensional even subalgebra of $\hat\cD$.
There are further 4 pairs where one of $i$ and $j$ is even and the
other odd and these span the four dimensional odd part of $\hat\cD$:
\be\label{eq:TCIMDs}
{\renewcommand{\arraystretch}{1.4}
\begin{array}{c|c}
\hbox{Even generators} & 
D^{11}_1\equiv 1,\;
D^{\eps\eps}_1,\;
D^{\eps'\eps'}_1\!,\; 
D^{\eps''\eps''}_1\!,\;
D^{\sigma\sigma}_1, \;
D^{\sigma'\sigma'}_1\!, \;
D^{\mu\mu}_1, \;
D^{\mu'\mu'}_1 , \;
D^{GG}_1 , \;
D^{\bar G\bar G}_1 , \;
D^{\psi\psi}_1 , \;
D^{\bar \psi\bar \psi}_1 
\\ 
\hline
\hbox{Odd generators} & 
D^{\sigma\mu}_a,\; 
D^{\mu\sigma}_a, \;
D^{\sigma'\mu'}_a\!, \;
D^{\mu'\sigma'}_a 
\end{array}
}
\ee
As a vector space, $\hat\cD$ is equivalent to $\mathbb C^{12|4}$. 
As a  super-algebra, 
$\hat\cD$ splits into 8 copies of $\mathbb C$ and four
copies of $C\ell_1$ giving 12 defects in total, twice as many as the
purely bosonic defects of $M(4,5)$, with the differences occurring in
sectors that are not in $M(4,5)$.

If we restrict $i$ and $j$ to the NS sector alone, the algebra is instead
equivalent to $\mathbb C^{8|0}$, i.e. it is purely even and the
corresponding eight solutions are formally the same as those found in \cite{mw}.

Up to now, the question has been entirely one of calculating the
bulk-defect structure constants. 
The final step is to fix the normalisation of the defect operators (up
to a sign) to reproduce the correct counting
of weight zero fields from the torus expectation value of the defect
operator which is easily done.

\subsubsection{Fermionic TCIM interface classifying algebra}

The only interface we can consider is that between $M(4,5)$ and $FM(4,5)$.
Since the fields in $M(4,5)$ form a subalgebra of the fields in
$FM(4,5)$, the interface classifying algebra is a subalgebra of the
defect classifying algebra of $FM(4,5)$. From \eqref{eq:TCIMDs},
there are 6 surviving even generators and 2 surviving odd
generators. This means the interface classifying algebra is $\mathbb
C^{6|2}$ as a super-vector space, splitting into four copies of
$\mathbb C$ and two copies of $C\ell_1$ as
a graded algebra, giving 6 interfaces in total.

\subsection[$FM(3,8)/\wtFM(3,8)$, the supersymmetric Lee-Yang model]{$\boldsymbol{FM(3,8)/\wtFM(3,8)}$, the supersymmetric Lee-Yang model}\label{sec:susyly}

The case of $c=-21/4$ is the first where there is a half-integer spin
simple current allowing the extension of a bosonic Virasoro minimal
model to a fermionic model and there are also two bulk
invariants. There is the diagonal model $M(A_2,A_7)$ and the
D-invariant $M(A_2,D_5)$.

The representations and their weights are
as follows:
\be
{\renewcommand{\arraystretch}{1.4}
		\begin{array}{c|ccccccc}
		2 & \frac 32 & \frac{25}{32}& \frac 14& -\frac{3}{32}& -\frac 14&
		-\frac 7{32}& 0\\
		1 & 0 & -\frac 7{32} & - \frac 14 & - \frac 3{32} & \frac 14 & \frac{25}{32} & \frac 32 \\
		\hline
		\raisebox{.2em}{$r$} \large/ \raisebox{-.2em}{$s$}
		& 1 & 2 & 3 & 4 & 5 & 6 & 7
		\end{array}
}
\ee
As usual the Kac-table includes two copies of each representation and in this
case we remove the degeneracy by considering only the representations of type 
$(1,s)$. 

The value $c=-21/4$ corresponds to $SM(2,8)$ and has been looked at
before, see e.g.~\cite{Schoutens1990,Ahn:2000tj,Kormos2007}.
It has been
identified as the supersymmetric Lee-Yang model.
The issue of different models at the same central charge does
not seem to have been considered in these works --  they assume
that the bosonic projection is the diagonal invariant $M(3,8)$, and
hence in our language identify the superconformal theory as $\FM(3,8)$,
 
At $c=-21/4$, there are 4 relevant representations of the super-Virasoro
algebra with labels $(r,s)$, shown in table \ref{tab:Svir28}. Since
the Virasoro algebra is a subalgebra, each representation $\hat M_{r,s}$ of
SVir 
decomposes into a sum of one or more representations $M_{r,s}$ of Vir
and this information is included in this table along with their sector.

\begin{table}[htb]{\renewcommand{\arraystretch}{1.4}
		\[
		\begin{array}{c||ccc}
		(r,s) & h_{r,s} & NS/R & \hat M_{r,s}\\ \hline\hline
		(1,1) & 0      & NS   & \hat M_{1,1} = M_{1,1} \oplus M_{1,7} \\ \hline
		(1,2) & -\frac{7}{32}  & R   & \hat M_{1,2} = M_{1,2} \oplus M_{1,6}
		\\ \hline
		(1,3) &  - \frac 14   & NS  & \hat M_{1,3} = M_{1,3} \oplus M_{1,5}
		\\ \hline
		(1,4) &  -\frac 3{32} & R   & \hat M_{1,4} = M_{1,4} 
		\end{array}
		\]
		\caption{The representations of the super-Virasoro algebra at $c=-21/4$}
		\label{tab:Svir28}}
\end{table}

As there are two different invariants of the Virasoro algebra, there
are two different fermionic extensions. These are related by parity shift in the Ramond sector  and are the fermionic models
$\FM(3,8)$ and $\widetilde\FM(3,8)$ from section~\ref{sec:fermVir}.
Their field content is given
in table \ref{tab:fc}.

\begin{table}[htb]{\renewcommand{\arraystretch}{1.4}
\[
\begin{array}{cc}
\hbox{A invariant}&
\hbox{D invariant}
\\
\begin{array}{c|c|c|c}
\hbox{even, NS} & \hbox{even, R} & \hbox{odd, NS} & \hbox{odd, R}
\\ \hline
\phi^e_{(1,a)}   & \phi^e_{(1,a)}  & \phi^o_{(1,a)}  & \phi^o_{(1,a)}  \\ \hline
\text{\small (1,1)\,(1,1)}  & \text{\small (1,2)\,(1,2)} & \text{\small (1,1)\,(1,7)} & \text{\small (1,2)\,(1,6)} \\
\text{\small (1,3)\,(1,3)}  & \text{\small (1,4)\,(1,4)} & \text{\small (1,7)\,(1,1)} & \text{\small (1,4)\,(1,4)} \\
\text{\small (1,5)\,(1,5)}  & \text{\small (1,6)\,(1,6)} & \text{\small (1,3)\,(1,5)} & \text{\small (1,6)\,(1,2)} \\
\text{\small (1,7)\,(1,7)}  &            & \text{\small (1,5)\,(1,3)} &
\\
\end{array} 
&
\begin{array}{c|c|c|c}
\hbox{even, NS} & \hbox{even, R} & \hbox{odd, NS} & \hbox{odd, R} \\ \hline
\phi^u_{(1,a)}   & \phi^s_{(1,a)}  & \phi^s_{(1,a)}  & \phi^u_{(1,a)}  \\ \hline
\text{\small (1,1)\,(1,1)}  & \text{\small (1,2)\,(1,6)} & \text{\small (1,1)\,(1,7)} & \text{\small (1,2)\,(1,2)} \\
\text{\small (1,3)\,(1,3)}  & \text{\small (1,4)\,(1,4)} & \text{\small (1,7)\,(1,1)} & \text{\small (1,4)\,(1,4)} \\
\text{\small (1,5)\,(1,5)}  & \text{\small (1,6)\,(1,2)} & \text{\small (1,3)\,(1,5)} & \text{\small (1,6)\,(1,6)} \\
\text{\small (1,7)\,(1,7)}  &            & \text{\small (1,5)\,(1,3)} &
\\
\end{array}
\end{array}
\]
\caption{The field content of the $A$ and $D$ invariants of $M_{3,8}$
  (even part of the table) and their fermionic extensions (even
  and odd part). The fermionic extensions differ only in their Ramond
  sector parity. Listed are the parity and spin grade, the notation of the
  primary bulk field used in section~\ref{sec:fermVir}, and the
  Kac-label of the left/right representation of that field. 
}
\label{tab:fc}}
\end{table}

The representation content in the even and odd Ramond sector of
$\FM(3,8)$ and $\widetilde\FM(3,8)$ now differs, and so these models
cannot be graded-isomorphic. From the point of view of the
super-Virasoro algebra, the parity of the Ramond sector ground state
has changed between these two models, cf.\ Table~\ref{tab:Svir28},
where the R-ground state has label $(1,2)$. 

The structure of the boundary classifying algebras
also differs in the two models. 
According to the general theory in 
section~\ref{sec:class-parityshift}, there
are equal numbers of bulk-boundary
structure constants in the two models, but the parities are different.
In $\FM(3,8)$, where the Ramond ground state is even,
there are 7 even
and 1 odd generators.
This leads to a classifying algebra which
is $\mathbb C^{7|1}$ as a super-vector space and which decomposes into 
6 copies of $\mathbb C$ and 1 copy of $C\ell_1$ giving seven
fundamental boundary conditions.
From the point of view of the superconformal algebra, this means there
is 1 ``supersymmetric'' boundary condition in the sense of
\cite{Nepomechie2001} and 6 ``non-supersymmetric'' ones. 

In $\wtFM(3,8)$, where the Ramond ground state is odd,
there are 5 even
and 3 odd generators,
leading to a classifying algebra which
is $\mathbb C^{5|3}$ as a super-vector space and which decomposes into 
2 copies of $\mathbb C$ and 3 copies of $C\ell_1$ giving five
fundamental boundary conditions, 3 of which are ``supersymmetric'' and 2 are not.

As expected, the boundary conditions of $\FM(3,8)$ are in 1-1 correspondence with the 7 boundary conditions of $M(A_2,A_7)$, and those of $\wtFM(3,8)$ are in 1-1 correspondence with the 5 boundary conditions of $M(A_2,D_5)$ (see \cite{Behrend:1999bn} for the boundary conditions of minimal models).

The two boundary classifying algebras can be made identical [as
ungraded algebras] by a suitable rescaling of the fields in
$\wtFM(3,8)$ (or equivalently of the generators of the algebra).
The generators and their gradings are shown in table
\ref{tab:M38Bs} together with the rescaling of the generators that
makes the algebras identical.

{\renewcommand{\arraystretch}{1.4}
\begin{table}[htb]
\[
\begin{array}{|c||c|c|c|c|c|c|c|}
\hline
& \phi=0 & \phi = 1 \\ \hline
 \tilde\phi=0 & 
 B^{(1,1),e},\; B^{(1,3),e},\; B^{(1,5),e},\; B^{(1,7),e}
& B^{(1,4),o} 
\\ \hline 
  \tilde\phi = 1 & 
 B^{(1,2),e},\; B^{(1,4),e},\; B^{(1,6),e}
&  \hbox{---}
\\\hline
\end{array}
\]
\caption{The bi-grading of the generators of the boundary classifying
  algebras of $\FM(3,8)$ (parity $\phi$) and $\wtFM(3,8)$ (parity $\tilde\phi$), with the identification 
$B^{(1,a),u} = B^{(1,a),e}$, $B^{(1,4),s} = i B^{(1,4),o}$, 
cf.\ \eqref{eq:bnd-class-alg-shift-ident}.}
\label{tab:M38Bs}
\end{table}
}

There are now three different fermionic
defect classifying algebras, classifying
the defects in $\FM(3,8)$, defects in $\wtFM(3,8)$ and interfaces
between $\FM(3,8)$ and $\wtFM(3,8)$. Again, according to the general
theory in section~\ref{sec:class-parityshift}, these can be made equal as ungraded algebras by
a simple rescaling of the generators by phases, 
but they are not all equivalent as
as graded algebras. 
The two defect classifying algebras
have the same gradings, but the interface algebra has a different
grading. These are shown in table \ref{tab:M38Ds}. 
The two defect classifying algebras are $\mathbb C^{14|2}$ as vector
spaces, decomposing into 12 copies of $\mathbb C$ and two copies of
$C\ell_1$ as graded algebras, giving 14 defects in each of these two theories.
The interface algebra 
is $\mathbb C^{10|6}$ as a vector space and 4 copies $\mathbb C$ and 6
copies of $C\ell_1$ as
an algebra giving 10 interfaces between the two theories. 
When we descend to the bosonic $A$ and $D$ theories, the
defects/interfaces in the fermionic theories are identified in pairs
(as in subsection \ref{tcimdefcla}, the pairs only differ in sectors
that are not in the bosonic theories) and
give 7 defects in the each of the $A$ and $D$ theories and 5
interfaces between the $A$ and $D$ theories, 
in agreement with the computation in terms of traces of products
of bulk modular invariant matrices, see \cite{Petkova:2000ip}
and \cite[Rem.\,5.19]{FRS1}. 

{\renewcommand{\arraystretch}{1.4}
\begin{table}[htb]
\[
\begin{array}{|c||c|c|c|c|c|c|c|}
\hline
& \phi=0 & \phi = 1 \\ \hline
 \tilde \phi=0 & 
 D^{[(1,1),e][(1,1),e]},\; 
 D^{[(1,3),e][(1,3),e]},\; 
 D^{[(1,5),e][(1,5),e]},\; 
 D^{[(1,7),e][(1,7),e]},\; 
&
 D^{[(1,4),e][(1,4),o]},\; 
\\
& D^{[(1,1),o][(1,1),o]},\; 
 D^{[(1,3),o][(1,3),o]},\; 
 D^{[(1,5),o][(1,5),o]},\; 
 D^{[(1,7),o][(1,7),o]}\phantom{,\;} 
&
 D^{[(1,4),o][(1,4),e]} 
\\ \hline 
  \tilde \phi = 1 &
 D^{[(1,2),e][(1,2),e]},\; 
 D^{[(1,4),e][(1,4),e]},\; 
 D^{[(1,6),e][(1,6),e]},\; 
& 
 \hbox{---}
\\
&
 D^{[(1,2),o][(1,2),o]},\; 
 D^{[(1,4),o][(1,4),o]},\; 
 D^{[(1,6),o][(1,6),o]}\phantom{,\;} 
 &  
\\\hline
\end{array}
\]
\caption{The bi-grading of the generators of the defect and interface classifying
algebras of $\FM(3,8)$ and $\wtFM(3,8)$. The parity of a given generator in either of the two defect algebras is called $\phi$, and in the interface algebra it is called $\tilde\phi$, with the identifications used in~\eqref{eq:defect-class-ident}.} 
\label{tab:M38Ds}
\end{table}
}

\subsection[The $\FM(4,7) = \WB(0,2)_{5,7}$ example]{The $\boldsymbol{\FM(4,7) = \WB(0,2)_{5,7}}$ example}\label{sec:M47}

Finally, the only fermionic minimal model to have a weight $5/2$
current is $\FM(4,7)$ with central charge $c=-13/14$. This turns out
to be a restriction of the ``fermionic'' W-algebra $\WB(0,2)$. This
algebra, also known as the ``fermionic'' $\WB_2$ algebra, extends the
Virasoro algebra by primary fields of weights $4$ and $5/2$. Its
structure constants were worked out explicitly in 
\cite{FST} and it can be seen that the field of weight 4 decouples
from the algebra at this value of $c$. The minimal model of this
algebra has 6 representations labelled by $[rs;r's']$ where
$r,s,r',s'\geq 1$, $2r+s \leq 3$ and $2r'+s'\leq 5$. 
Three of these
are irreducible as Virasoro algebra representations and in the Ramond
sector and give the three R sector representations of $\FM(4,7)$; three are
reducible, splitting each into two Virasoro algebra representations
and these together are the six NS sector representations of
$\FM(4,7)$. The rest of the analysis is straightforward - there are 9
boundary conditions of $\FM(4,7)$ of which three are invariant under
the W-algebra automorphism $W \to - W$ and three are related in pairs.
This is exactly as in the fermionic
TCIM, where there are two ``supersymmetric'' boundary
conditions and two pairs which are related by $G \to -G$.

\section{Conclusions}

We have defined fermionic conformal field theories and their
classifying algebras, 
defined fermionic extensions of the Virasoro minimal models, found explicit expressions for all the bulk structure constants (of both Neveu-Schwarz and Ramond fields) of these models,  
and have given numerous examples in these cases.

We have found that it is natural for certain boundary conditions and
defects to support a weight zero fermionic field which has up to now
been introduced in an ad hoc manner. 

We have also found that there is a natural parity-shift operation
which can relate different theories. On the one hand, this relates
bosonic theories as the projections of parity-shifted fermionic
theories, and on the other hand this means that there are hitherto
unconsidered fermionic theories to be looked at.

There are quite a few questions that are unresolved and which suggest
new lines of enquiry. 

Firstly, we showed that the {\em full} set of bulk-boundary structure
constants in the Ising model defined an algebra, not just the
couplings to boundary fields of weight zero. This seems to  merit further
investigation, even in the purely bosonic case.

Secondly, we showed that the fermionic extensions of the $A$- and $D$-
invariant minimal models are related by a parity-shift operation. This
leads one to wonder if one can define extensions in other cases in
which the extended algebra would be bosonic, such as the $A$- and $D$-
invariants of $M(5,6)$, namely the tetra-critical Ising model and the
$3$-state Potts model respectively.
It would also be interesting to
investigate possible fermionic extensions of the exceptional invariants;
one of these, $M(A_4,E_6)$, is a product theory
$M(A_1,A_4)\times M(A_2,A_3)=M(2,5)\times M(3,4)$ \cite{QRW},
with the obvious proposal that $FM(A_4,E_6)= M(2,5)\times FM(3,4)$. It
would be good to have a general understanding of these models.

It would also be good to give
the structure constants for the
field theories on the boundaries, defects and interfaces that we have
found which would be a necessary first step to discuss their
perturbations and the resulting renormalisation group flows.

In \cite{Tong1}, the moduli space of $c=1$ CFTs with fermions was investigated (see
Figure 2 there), it would be interesting to look at this problem from our perspective.

Finally, one consequence of our construction relating $FM(p,q)$ to $\wtFM(p,q)$
is that when these are also superconformal field theories, their
partition functions differ in the sign of the ``bottom component'' of
the super-partition function, that is 
$Z_{RR}= \mathrm{TR}_R(\,(-1)^F\,)$ which is the trace of $(-1)^F$ on
the highest weight space of the Ramond fields \cite{CF}.
This cannot be
determined on the grounds of modular invariance. 
We have found that in
our construction, $Z_{RR} = +1$ for $\FM(3,8)$ and $Z_{RR} = -1$ for
$\wtFM(3,8)$ which are hence two inequivalent superconformal field theories
at $c=-21/4$.
It is a curious fact that this implies that in this model the Virasoro characters
satisfy $\chi_{1,2} = \chi_{1,6} + 1$, as is easily checked from the
character formulae in \cite{YBk}.
The value $Z_{RR}=0$ for the TCIM was already observed in \cite{CF}.
This leaves open now the question of investigating the superconformal
models in which this parity-shift relates two inequivalent field
theories, and how this might affect previous results on boundary
conditions, boundary perturbations, etc.

\medskip

\subsection*{Acknowledgements}

We would like to thank L.~Szegedy for helpful
discussions and careful reading of the manuscript.
GW thanks 
I.~Makabe for many discussions on boundary conditions in
superconformal field theories, P.~Mathieu for discussions on
superconformal models in general, 
as well as S.~Majid, A.~Recknagel and S.~Wood for further useful discussions.
IR thanks the Department of Mathematics at King's College London for
hospitality during a sabbatical in the first half of 2019 where this research was
completed. 
IR is partially supported by the Deutsche Forschungsgemeinschaft via
the Research Training Group RTG~1670 and the Cluster of Excellence
EXC~2121.

\appendix

\section{Appendix}

\newcommand{\Fmat}[6]{F_{#1#2}\big[ \begin{smallmatrix} #3&#4\\#5&#6 \end{smallmatrix} \big]}

\subsection{Ising data}
\label{app:ising}

We list here the data for the Ising model that is used in section \ref{sec:Ising}. The representations $(1,1),(1,2),(1,3)$ with conformal weights $0,\frac 1{16}$ and $\frac 12$ are denoted $1$, $\sigma$ and $\epsilon$.  

The bulk field content is as follows, giving both the conventional name from the free-fermionic extension of the Ising model and the names following the conventions of section \ref{ssec:fvmm}:
\begin{equation}
{\renewcommand{\arraystretch}{1.3}	
\begin{array}{cc|c|c|c}
& \hbox{Even} & \multicolumn{3}{c}{(r,s);h}\\
& (\bar r,\bar s);\bar h & (1,1);0 & (1,2);\frac 1{16} & (1,3);\frac 12 \\\hline
& (1,1);0 &    1 \equiv \phi_1^e   &   -               & - \\
&
  (1,2);\frac 1{16} & - & \sigma \equiv \phi_\sigma^e & - \\
& (1,3);\frac 12 & - & - & \epsilon \equiv \phi_\epsilon^e
\end{array}
\quad
\begin{array}{cc|c|c|c}
& \hbox{Odd}  & \multicolumn{3}{c}{(r,s);h}\\
& (\bar r,\bar s);\bar h  & (1,1);0 & (1,2);\frac 1{16} & (1,3);\frac 12 \\\hline
& (1,1);0 &    -    &  - &  \psi \equiv \phi_\epsilon^o              \\
&
  (1,2);\frac 1{16} & -  & \mu \equiv\phi_\sigma^o & - \\
& (1,3);\frac12 & \bar\psi \equiv \phi_1^o & - & -
\end{array}
}
\end{equation}

The values for the normalisation constants chosen in \eqref{eq:cdefs} to get the structure constants in section~\ref{sec:IsingBulk} are
\begin{align}
\lambda_1^e &= 1 
~,
&
\lambda_\eps^e &= 1
~,
&
\lambda_\sigma^e &= 2^{1/4}
~,
&
\lambda_1^o &= 1 
~,
&
\lambda_\eps^o &= i
~,
&
\lambda_\sigma^o &= e^{\pi i/4} \, 2^{-1/4}
\ .
\end{align}
The F-matrix entries we need are:
\begin{align}
\Fmat{0}{0}{\sigma}{\sigma}{\sigma}{\sigma} &= \frac{1}{\sqrt 2}
&
\Fmat{0}{\epsilon}{\sigma}{\sigma}{\sigma}{\sigma} &= \frac{1}{2 \sqrt 2}
&
\Fmat{\epsilon}{0}{\sigma}{\sigma}{\sigma}{\sigma} &= \sqrt 2
&
\Fmat{\epsilon}{\epsilon}{\sigma}{\sigma}{\sigma}{\sigma} &= \frac{-1}{\sqrt 2}
\nonumber \\
\Fmat{0}{\sigma}{\sigma}{\epsilon}{\sigma}{\epsilon} &= \frac{1}{2}
&
\Fmat{\sigma}{0}{\epsilon}{\epsilon}{\sigma}{\sigma} &= {2}
&
\Fmat{\sigma}{\sigma}{\epsilon}{\sigma}{\sigma}{\epsilon} &= {-1}
&
\Fmat{0}{0}{\eps}{\eps}{\eps}{\eps} &= 1
\end{align}
In the discussion of boundary states and defects we make use of the modular transformations of Virasoro characters, with $q = \exp(2 \pi i \tau)$ and $\tilde q = \exp(-2 \pi i /\tau)$,
\begin{align}
\chi_0(\tilde q) &~=~ \frac12 \chi_0(q) + \frac{1}{\sqrt{2}} \chi_\sigma(q) + \frac12 \chi_\epsilon(q) \ ,
\nonumber \\
\chi_\sigma(\tilde q) &~=~ \frac{1}{\sqrt{2}} \chi_0(q) - \frac{1}{\sqrt{2}} \chi_\epsilon(q) \ ,
\nonumber \\
\chi_\epsilon(\tilde q) &~=~ \frac12 \chi_0(q) - \frac{1}{\sqrt{2}} \chi_\sigma(q) + \frac12 \chi_\epsilon(q) \ .
\end{align}

\subsection{Tri-critical Ising data}
\label{app:TCIM}

The tri-critical Ising model is the Virasoro minimal model $M(4,5)$. 
The bulk model is described in \cite{LMC} and the conformal boundary conditions in \cite{Affleck2000}.

In table~\ref{tab:TCIM} we list the even and odd sectors in the same manner as for the Ising
model,  using mostly the naming conventions of
\cite{LMC} and \cite{Chim96} (\!\!\cite{FQS} use $t$ instead of
$\epsilon'$). The even sector is the field content of 
the bosonic tri-critical Ising model. Note that the field $G$ is
conventionally normalised to 
$\langle G | G \rangle = 3c/2=21/20$ 
(as opposed to our $\langle G | G \rangle = 1$)
and so we only have $G \propto
\phi_1^o$. Likewise, $\epsilon''\propto i G \bar G$.

\begin{table}[hbt]
{\renewcommand{\arraystretch}{1.3}	
\[
\begin{array}{c|c|c|c|c|c|c}
 \hbox{Even} & \multicolumn{6}{c}{(r,s);h}\\
(\bar r,\bar s);\bar h 
& (1,1);0           & (3,1);\frac 3{2} 
& (2,1);\frac 7{16} & (2,3);\frac 3{80} 
& (1,3);\frac 35    & (3,3);\frac 1{10} 
\\\hline
(1,1);0 & 
 1 \equiv \phi^e_1 & - & - & - & - & -
\\           
(3,1);\frac 3{2} &
- & \epsilon'' \equiv \phi_{\epsilon''}^e & - & - & - & -
\\
(2,1);\frac 7{16} &
- & - & \sigma' \equiv \phi_{\sigma'}^e & - & - & -
\\
(2,3);\frac 3{80} &
- & - & - & \sigma \equiv \phi_{\sigma}^e & - & - 
\\
(1,3);\frac 35 &
- & - & - & - & \epsilon' \equiv \phi_{\epsilon'}^e & - 
\\
(3,3);\frac 1{10} &
- & - & - & - & - & \epsilon \equiv \phi_{\epsilon}^e 
\end{array}
\]\[
\begin{array}{c|c|c|c|c|c|c}
 \hbox{Odd} & \multicolumn{6}{c}{(r,s);h}\\
(\bar r,\bar s);\bar h 
& (1,1);0           & (3,1);\frac 3{2} 
& (2,1);\frac 7{16} & (2,3);\frac 3{80} 
& (1,3);\frac 35    & (3,3);\frac 1{10} 
\\\hline
(1,1);0 & 
- & G \propto \phi^o_{\epsilon''} & - & - & - & - 
\\           
(3,1);\frac 3{2} &
\bar G \propto \phi_{1}^o & - & - & - & -
\\
(2,1);\frac 7{16} &
- & - & \mu' \equiv \phi_{\sigma'}^o & - & - & -
\\
(2,3);\frac 3{80} &
- & - & - & \mu \equiv \phi_{\sigma}^o & - & - 
\\
(1,3);\frac 35 &
- & - & - & - & - & \psib \equiv \phi_{\epsilon}^o 
\\
(3,3);\frac 1{10} &
- & - & - & - & \psi \equiv \phi_{\epsilon'}^o & - 
\end{array}
\]
}
\caption{}
\label{tab:TCIM}
\end{table}

As pointed out in \cite{FQS}, the tri-critical Ising model is related
to the first non-trivial unitary superconformal minimal model
$\mathit{SM}(3,5)$ which is the associated fermionic model.
There are four unitary highest weight representations $\hat M_{r,s}$
of the superconformal algebra at $c=7/10$. These each decompose as a finite sum of
Virasoro representations $M_{r,s}$ as shown in table \ref{tab:Svir35}.  

\begin{table}[H]{\renewcommand{\arraystretch}{1.4}
\[
\begin{array}{c||ccc}
(r,s) & h_{r,s} & NS/R & \hat M_{r,s}\\ \hline\hline
(1,1) & 0      & NS   & \hat M_{1,1} = M_{1,1} \oplus M_{3,1} \\ \hline
(1,2) & \frac{7}{16}  & R   & \hat M_{1,2} = M_{1,2}
\\ \hline
(1,3) & \frac 1{10}   & NS  & \hat M_{1,3} = M_{1,3} \oplus M_{3,3}
\\ \hline
(1,4) & \frac 3{80}   & R   & \hat M_{1,4} = M_{1,4} 
\end{array}
\]
\caption{The representations of the super-Virasoro algebra at $c=7/10$}
\label{tab:Svir35}}.
\end{table}

\newcommand\arxiv[2]      {\href{http://arXiv.org/abs/#1}{\tt #2}}
\newcommand\doi[2]        {\href{http://dx.doi.org/#1}{#2}}

\begingroup\raggedright

\setlength{\itemsep}{-2pt}
\raggedright

\endgroup

\end{document}